\documentclass[fleqn,usenatbib]{mnras}

\usepackage[T1]{fontenc}

\DeclareRobustCommand{\VAN}[3]{#2}
\let\VANthebibliography\thebibliography
\def\thebibliography{\DeclareRobustCommand{\VAN}[3]{##3}\VANthebibliography}

\usepackage{graphicx}	% Including figure files
\usepackage{amsmath}	% Advanced maths commands
\usepackage{amssymb}	% Extra maths symbols

\usepackage{pdflscape}
\usepackage{afterpage} % Prevent landscape from breaking text flow
\usepackage[export]{adjustbox}
\usepackage{upgreek}
\usepackage{multirow}

\usepackage{csvsimple}

\usepackage[separate-uncertainty=true]{siunitx}
\DeclareSIUnit{\parsec}{pc}
\DeclareSIUnit{\deg}{deg}
\DeclareSIUnit{\sky}{sky}
\DeclareSIUnit{\jansky}{Jy}
\DeclareSIUnit{\gauss}{G}
\DeclareSIUnit{\year}{yr}

\usepackage{tablefootnote}
\usepackage{hyperref}

\usepackage{caption}

\newcommand{\observed}{113}
\newcommand{\detected}{29}
\newcommand{\singlepulsedetected}{25}
\newcommand{\singlepulselofarnovel}{18} % single pulse detected - 7 in \citetalias{karako-argamanDiscoveryFollowupRotating2015}/\citetalias{michilliSinglepulseClassifierLOFAR2018}/\citetalias{sanidasLOFARTiedArrayAllSky2019b}
\newcommand{\periodicdetected}{14}
\newcommand{\periodicnovellofar}{7}
\newcommand{\consistentperiodicdetected}{6} % > 50% of observations had periodic emission
\newcommand{\totalnovel}{18} % single pulse + J0227
\newcommand{\timeable}{21} % > 0.5b/hr or snr > 10 after 2 hours
\newcommand{\pretimed}{10}
\newcommand{\wetimed}{8}
\newcommand{\tobetimed}{\the\numexpr\timeable-\pretimed-\wetimed-1\relax} % 21 - 9 - 8 - 1 = 4; 1 = 0317, upcoming lOFAR paper
\newcommand{\wecouldtime}{\the\numexpr\timeable-\pretimed\relax} % 21 - 9 = 12
\newcommand{\totalhours}{1408}
\newcommand{\surveyhours}{\the\numexpr495+\detected*5\relax}
\newcommand{\followhours}{\the\numexpr\totalhours-\surveyhours\relax}
\newcommand{\hmsangle}{\ang[angle-symbol-over-decimal,angle-symbol-degree=\textsuperscript{h},angle-symbol-minute=\textsuperscript{m},angle-symbol-second=\textsuperscript{s},minimum-integer-digits = 2,]}
\newcommand{\dmsangle}{\ang[angle-symbol-over-decimal,minimum-integer-digits = 2,]}
\newcommand{\galangle}{\ang[angle-symbol-over-decimal]}

\defcitealias{karako-argamanDiscoveryFollowupRotating2015}{K15}
\defcitealias{michilliSinglepulseClassifierLOFAR2018}{M18}
\defcitealias{sanidasLOFARTiedArrayAllSky2019b}{S19}
\defcitealias{goodFirstDiscoveryNew2020}{G21}
\defcitealias{dongCHIME2022}{D23}

\title[RRATs with I--LOFAR]
{A Census of Rotating Radio Transients at 150 MHz with the Irish LOFAR Station}
\author[D.~J.~McKenna et al.]{
D.~J.~McKenna\textsuperscript{1,2}\thanks{Email: \href{mailto:mckennd2@tcd.ie}{mckennd2@tcd.ie}},
E.~F.~Keane\textsuperscript{2},
P.~T.~Gallagher\textsuperscript{1,2},
J.~McCauley\textsuperscript{2}
\\
$^{1}$ School of Cosmic Physics, Dublin Institute for Advanced Studies, 31 Fitzwilliam Place, Dublin 2, Ireland.\\
$^{2}$ School of Physics, Trinity College Dublin, College Green, Dublin 2, Ireland.%\\
}

\date{Accepted XXX. Received YYY; in original form ZZZ}

\pubyear{2023}

\usepackage{newtxtext,newtxmath}

\begin{document}
\label{firstpage}
\pagerange{\pageref{firstpage}--\pageref{lastpage}}
\maketitle

\begin{abstract}
Rotating radio transients (RRATs) are neutron stars that emit detectable radio bursts sporadically. They are statistically distinct in the neutron star population, in many observable properties, but by their nature are practically difficult to study in depth. In this paper, we present the results from \SI{\totalhours}{\hour} of observations of RRAT candidates using the Irish station of the Low Frequency Array (LOFAR) at \SI{150}{\mega\hertz}. As of October 2022, this census involved observing $\observed$ sources, leading to $\detected$ detections which were then followed up systematically. Single-pulse emission was detected from $\singlepulsedetected$ sources, and periodic emission from $\periodicdetected$ sources. $\totalnovel$ sources were found to have emission behaviour that is not discussed in prior works using LOFAR instruments. Four novel or modified source periods have been determined, ranging from \SIrange{1.5}{3.9}{\second}, and $\wetimed$ new or updated phase-coherent pulsar timing ephemerides have been produced using detected bursts. One unexpected single-pulse with a clearly-Galactic dispersion measure was detected as a part of this work, but has not been re-detected in follow-up observations. Observations are ongoing to expand the number of observed sources and further characterise and improve ephemerides for the detected sources. This census has demonstrated the capability for international LOFAR stations to detect, monitor and characterise a significant fraction of these unique sources.

\end{abstract}


%%%%%%%%%%%%%%%%%%%%%%%%%%%%%%%%%%%%%%%%%%%%%%%%%%

%%%%%%%%%%%%%%%%% BODY OF PAPER %%%%%%%%%%%%%%%%%%

\section{Introduction}\label{sec:intro}
The majority of the over 3380 known radio pulsars\footnote{Pulsar catalogue v1.70, \citet{manchesterATNFPulsarCatalogue2005}} have been discovered through searches for periodic emission, and the majority of these through incoherent Fourier-based methods~\citep{lk12}. Although pulsars rotate in a reliable manner~\citep{hobbs+2012} they do not emit identical pulses each rotation and there is a large variety to the pulse amplitude distributions observed~\citep{burke-spolaorHighTimeResolution2011}. Rotating Radio Transient (RRAT) is the name generally given to any pulsar with a pulse amplitude distribution such that it is more significantly, or only, detectable through searches for individual bright pulses, as opposed to via periodicity-based searches~\citep{mclaughlinTransientRadioBursts2006a,keaneRotatingRadioTransients2011b}. Fourier-based searches select against long-period pulsars; if erratic emission in pulsars increases with age, then that would be an \textit{additional} compounding selection effect against long-period pulsars. Also, as real-world telescope data has a red noise characteristic~\citep{keane+2018}, this further compounds the selection effect against pulsars with periods longer than $\sim1$~s.

Understanding the evolution of pulsars, and the conditions of how they emit electromagnetic radiation throughout their lives, is a complex but fundamental problem of astrophysics. Much effort has rightly been dedicated to this problem~\citep{ridleyIsolatedPulsarSpin2010,vranesevicPulsarCurrentRevisited2011,johnstonPulsarBrakingPdotP2017} and work is ongoing, due to the large parameter space needed to describe the population. But in order to gain a complete understanding of the pulsar population, and its evolution, selection effects in the observed population must be understood and appropriately accounted for. To this end large-scale blind surveys across a broad spectral range are needed; these can counteract spectral selection effects which can be co-variant with other parameters~\citep{bates+2013}. Further, a wide range of search techniques must be applied to these data with single pulse searches~\citep{cm03}, and fast folding algorithms~\citep{morelloOptimalPeriodicitySearching2020b} acting as vital ingredients in these efforts. In recent years, this approach has been in action in many large-scale surveys undertaken with telescopes around the world~\citep{sanidasLOFARTiedArrayAllSky2019b,morelloSUrveyPulsarsExtragalactic2020a}. 

Many of these efforts have resulted in the detection of new transient/variable sources such as fast radio bursts (FRBs, \citealt{Lorimer2018}), nulling pulsars and other neutron stars with extreme properties~\citep{Ng2020}, a subsection of which fall into the definition of a RRAT. While the number of RRATs has been increasing in recent years, there has been a dearth of follow-up observations. Such follow-up timing is essential to properly characterise these sources, but due to the instantaneous sensitivity and observing time required to see sufficient individual pulses, only certain over-subscribed telescopes have been considered suitable. Consequently, as of early 2023, the vast majority of RRATs in the ATNF pulsar catalogue~\citep{manchesterATNFPulsarCatalogue2005} lack a precise rotation period, and only a third have a measurement of the first derivative of the period. Given these sources are outliers to the general pulsar population, both in detectability and often in their underlying properties, the lack of characterisation for the majority of these neutrons stars may be severely hampering efforts to understand the true pulsar population and its evolutionary track(s). Therefore, it might be expected that characterising RRATs would have a disproportionately large impact on disentangling of the pulsar evolution puzzle, given the small known population already appear statistically distinct from other pulsar populations \citep{2022arXiv220100295A}.

In this paper we present a systematic census of $\observed$ Northern Hemisphere RRATs using the Irish LOFAR station~\citep[I--LOFAR;][]{murphyFirstResultsREALtime2021}, a component of the International LOFAR Telescope~\citep[ILT;][]{haarlemLOFARLOwFrequencyARray2013a} with the aims of (i) determining if the sources can be profitably observed with a single international LOFAR station; and if so, (ii) to accurately quantify and characterise the nature of those sources that can be studied in this way through low-frequency observations and precision pulsar timing analyses. In~\S\ref{sec:sources} the source catalogues and selection constraints are described, while the observation and processing methodologies are described in~\S\ref{sec:obsback}. In~\S\ref{sec:results} the characteristics of the $\detected$ detected sources at \SI{150}{\mega\hertz} at the sensitivity of an international LOFAR station, from \SI{\surveyhours}{\hour} of initial census observations to determine the detectability of the sources and a further \SI{\followhours}{\hour} follow-up observations, are presented. These are further discussed in~\S\ref{sec:discuss}. The work is then concluded, with observation plans described, in~\S\ref{sec:conclusions}.

\section{Source Selection}\label{sec:sources}
%\section{Observation Background and Methodology}\label{sec:obsback}
%
%\subsection{Source Selection}\label{sec:sources}

There are a number of catalogues in the public domain which describe RRATs; each of these is detailed and updated heterogeneously. The sources chosen for this study come from three of these catalogues. (i) With a focus on RRATs detected in the Northern Hemisphere (at a central frequency of  \SI{600}{\mega\hertz}), the catalogues provided by the CHIME/FRB collaboration\footnote{\href{https://www.chime-frb.ca/galactic}{https://www.chime-frb.ca/galactic}} are referred to herein as the ``CHIME/FRB catalogue''. These are discussed in detail in \citealt{goodFirstDiscoveryNew2020} and \citealt{dongCHIME2022} (herein \citetalias{goodFirstDiscoveryNew2020} and \citetalias{dongCHIME2022}); (ii) the catalogue of the Pushchino Radio Astronomy Observatory\footnote{\href{https://web.archive.org/web/20220708232514/https://bsa-analytics.prao.ru/en/transients/rrat/}{https://bsa-analytics.prao.ru/en/transients/rrat/}} (at \SI{111}{\mega\hertz}), is referred to herein as the ``PRAO catalogue''. This has most recently been updated in \cite{samodurovDetectionStatisticsPulse2022}; and (iii) The West Virginia University RRATalog\footnote{\href{https://web.archive.org/web/20230210010446/http://astro.phys.wvu.edu/rratalog/}{http://astro.phys.wvu.edu/rratalog/}} (at various sites and frequencies), an extensive catalogue of sources combined from the announcements from surveys using several more `traditional' dish telescopes that reported RRAT detections prior to 2017. Sources included in these catalogues as accessed on May 1st 2022 were considered for inclusion in the census.

It is worth noting that several sources in these catalogues have been classified differently, following from further observations at different telescopes and observing frequencies (such as \cite{luStudyThreeRotating2019a}) or independent re-discovery during surveys~\citep[such as LOTAAS, described in][herein referred to as \citetalias{michilliSinglepulseClassifierLOFAR2018} and \citetalias{sanidasLOFARTiedArrayAllSky2019b}]{michilliSinglepulseClassifierLOFAR2018,sanidasLOFARTiedArrayAllSky2019b}. Given the `RRAT' label is an observation classification, and is not intrinsic to the source, it can initially be frequency or even telescope dependent due to the spectral and pulse-amplitude variability of pulsars. The extended family of reclassified pulsars were observed as a part of this census to determine if a single international LOFAR station has the sensitivity required to perform single-pulse detection and analysis of these extreme sources.

Some constraints were placed on each catalogue to limit the number of sources included in the census, with a goal to maximise the scientific potential of the allocated observing time. However, some sources that did not meet the typical observing requirements, but which were present at a celestial longitude where there is a dearth of sources, were additionally observed in time that otherwise would not have been allocated on the telescope.

%\subsubsection{The CHIME/FRB Galactic Catalogue}\label{sec:chime}
\subsection{The CHIME/FRB Galactic Catalogue}\label{sec:chime}
The CHIME/FRB catalogue contains sources discovered with the Canadian Hydrogen Intensity Mapping Experiment \citep{banduraCanadianHydrogenIntensity2014}, a zenith scanning array that it is focused on mapping hydrogen in the local universe. CHIME has additional back-ends for FRB and pulsar studies, which have helped it become a pivotal instrument in the recent developments in these fields \citep{amiriCHIMEPulsarProject2021,collaborationFirstCHIMEFRB2021}. These back-ends have allowed it to quickly become one of the most productive telescopes for single-pulse detection and folded pulsar observations due to its sensitivity, field of view, large fractional bandwidth and high-cadence observing across the entire Northern Hemisphere. The novel sources discovered by these backends are collected and published within this catalogue.

This catalogue contained 38 sources at the time of this census, 4 of which were presented with candidate plots for pulsars with periodic emission rather than sources of single-pulse emission. All of these sources were observed as a part of this campaign.

While the catalogue does not provide uncertainties on their measurements, they can be inferred from the results presented in \citetalias{goodFirstDiscoveryNew2020}. Depending on the activity of the sources in this catalogue, their positional uncertainties are mostly on the order of minutes and arcminutes (for both timed and untimed sources, e.g. J1931+4229 did not have a known rotation period and has a similar uncertainty on its parameters to the other sources presented), while 2 of the periodic sources have been localised to sub-arcsecond positions. The dispersion measures (DMs) presented in the catalogue have uncertainties on the order of \SIrange{0.1}{1}{\parsec\per\cubic\centi\metre}, reduced from the underlying $2^i\times$\SI{1.62}{\parsec\per\cubic\centi\metre} CHIME/FRB trial sampling, and, where provided, the periods were found to be accurate.

%\subsubsection{The Pushchino Radio Astronomy Observatory Catalogue}\label{sec:push}
\subsection{The Pushchino Radio Astronomy Observatory Catalogue}\label{sec:push}
The PRAO catalogue focuses on sources detected with the Big Scanning Array of the Lebedev Physical Institute (BSA LPI, \cite{tyulbashevDetectionNewPulsars2016}), a zenith scanning array at the Pushchino Radio Observatory outside Moscow, Russia.

The BSA LPI continually monitors declinations between \SIrange{-6}{42}{\degree} using 96 beams (48 of which are currently monitored), with \SI{12.5}{\milli\second} sampling across 32 channels between \SIrange{109}{111.5}{\mega\hertz}, which is contained within the bandwidth observed during each observation taken with I--LOFAR. While lacking the incredible \SI{45000}{\metre\squared} collecting area of the BSA LPI, the relatively wide beam and \SI{96}{\mega\hertz} usable bandwidth of an international LOFAR station makes these arrays a strong candidate for follow-up observations on the brighter sources contained in the PRAO catalogue.

With the narrow bandwidth and relatively low sampling rate of the BSA LPI, the sources reported within the catalogue have wide uncertainties on their dispersion measure, typically from \SIrange{2}{6}{\parsec\per\centi\metre\cubed}, due to the coarse dispersion measure trials at these channels widths and sampling rates. Similarly, frequently detected sources have periods reported to $\sim$\SI{}{\milli\second} precision.

Assuming that the sources in this catalogue have a bias towards having a steeper spectral index than typical pulsars, sources reported to have had their brightest pulse at a peak flux density below \SI{5}{\jansky} were not observed. Sources below this peak brightness are suspected to not be visible to I--LOFAR, as an international LOFAR station observing a $12.5$-ms
%\SI{12.5}{\milli\second}
pulse, integrated over a \SI{10}{\mega\hertz} bandwidth at \SIrange{108}{118}{\mega\hertz} results in an expected signal-to-noise ratio (SNR) of only $3$. While the wider bandwidth available would improve the significance of such a pulse, it is unexpected that I--LOFAR would be able to detect such quiet pulses. It is expected that only sources with a reported peak flux density of between \SIrange{11}{60}{\jansky} would be detectable, with the range depending on the sky temperature in the direction of the source (see~\S\ref{sec:sensitivity} for further details). Accounting for this, $41$ sources from this catalogue (and the four sources discussed in \S~\ref{sec:overlap}) were observed as a part of the census, with one source, J2018$-$07, not observed due to its low declination.

Several sources from the PRAO catalogue were detected during the LOTAAS survey, and were reported to have both primarily periodic (J0317+1328, J1132+2513, J1404+1159) and single pulse emission (J0139+3336). These sources were observed as a part of the census to further gauge the capabilities of an international LOFAR station as compared to the 8 LOFAR core stations that were coherently beam-formed as a part of the LOTAAS survey~\citep{sanidasLOFARTiedArrayAllSky2019b}.

%\subsubsection{The RRATalog}\label{sec:rratalog}
\subsection{The RRATalog}\label{sec:rratalog}
The RRATalog is a catalogue maintained by Cui and McLaughlin at West Virginia University, last updated in 2017, of previously announced RRAT sources from a wide variety of surveys. It contains sources detected during the multiple surveys performed over the past two decades using several telescopes, such as those at Arecibo, Parkes, and Green Bank. Some sources detected using the Green Bank telescope have previously been followed up at the LOFAR core \citep[][herein referred to as K15]{karako-argamanDiscoveryFollowupRotating2015}.

With the large number of `traditional' surveys consisting of tessellated pointings with narrow beams forming the foundation of this catalogue, it contains significantly lower uncertainties on positions and dispersion measures as compared to the catalogues discussed previously. Further, many of the sources have received full follow-up campaigns, allowing for their parameters to be constrained even further.

A large fraction of the sources in the catalogue below a declination of \SI{15}{\degree} were not observed as a part of this census. The majority of these excluded sources are grouped around a small region of the sky near the Galactic plane, and have relatively low peak flux densities. This region of the sky already represents the location where the sensitivity of I--LOFAR is reduced due to the low peak elevations of the sources during their transits, and the scaling of the sky temperature of the Galactic disk to low frequencies (generally discussed in \S~\ref{sec:sensitivity}). It was concluded that it would be unlikely these sources would be detected in the several months of observations it would take to follow up the 18 sources that were not observed as a result of this constraint. 

After this filtering, 39 sources were observed from this catalogue as a part of this census.

This catalogue acts as a strong foundation for observers looking to dive into the world of RRATs, however there are several errata regarding entries in the catalogue, while some RRATs reported by the referenced surveys with discovery dates as far back as 2014 were found to be missing from the database. These were noted as this census progressed, with some additional sources from these surveys being added to the census under the ``RRATalog'' label. A full list of the updated sources is provided in~\S\ref{ap:extendedrratalog}.

%\subsubsection{Catalogue Overlap}\label{sec:overlap}
\subsection{Catalogue Overlap}\label{sec:overlap}
Several sources are suspected to be duplicated in the aforementioned catalogues. 

J1130+09 of the CHIME/FRB catalogue and J1132+0921 of the PRAO catalogue have an angular separation of \SI{0.2}{\degree} with overlapping positional uncertainties, and have observed dispersion measures of 22.4 and \SI{22(2)}{\parsec\per\cubic\centi\metre}. Due to the reasonably large separation between the reported positions and low demands for telescope time at this source's right ascension, both pointings were observed, though neither resulted in a detection. 

J1848+1518 of the PRAO catalogue and J1849+15 of the RRATalog (originally detected by the Green Bank Telescope) are suspected to be the same source for similar reasons \citep[see further discussion in][]{tyulbashevDetection25New2018}. The ephemeris produced from a timing campaign performed at the LOFAR core (J1848+1516) was used for observing this source.

During the later stages of the census, the PRAO catalogue was updated with results from \cite{samodurovDetectionStatisticsPulse2022} to include two more sources, J1929+42 and J2214+45, that closely matched the location and dispersion measure properties of sources previously announced by CHIME, J1931+4229 and J2215+45. These sources had been monitored with I--LOFAR for several months at the time of the announcements using the original CHIME candidate parameters, consequently the PRAO reported positions were not specifically observed as a result.

\section{Observations and Methodology}\label{sec:obsback}

\subsection{Census Strategy}
Each source was observed for a minimum of \SI{5}{\hour}, typically in observations that ranged from \SIrange{30}{90}{\minute} centred near their transit time. When studying RRATs, longer total observation times increase the chance of brighter (detectable) pulses occurring, with the exact improvement dependent on the (typically unknown) pulse amplitude distributions. Five hours was chosen as a meaningful compromise between (a) probing sufficiently deep into the pulse-amplitude distribution of the sources given the sensitivity of the instrument, (b) investigating potentially low burst-rate sources emitting in a manner following Poisson statistics, and (c) the desire to complete the census in a reasonable time.

\subsection{Observing with I--LOFAR}\label{sec:observing}
I--LOFAR is one of 14 stations that make up the extended baselines of the ILT, which stretches from this westernmost site in Birr, Co. Offaly, Ireland to Irbene, Latvia in the East with a dense core of 38 stations with different configurations in the Netherlands \citep{haarlemLOFARLOwFrequencyARray2013a}. 

For \SIrange{32}{48}{\hour} per week, it is operated by the Irish LOFAR Consortium\footnote{\url{https://lofar.ie/}} in `local mode', whereby astronomers can apply to use the station for standalone or coordinated observations. It is during this time that this census was performed, with data processed on the  Real-time Transient Acquisition compute cluster \citep[REALTA;][]{murphyFirstResultsREALtime2021}.

The observations for this census were performed using the 96 High Band Antenna tiles (HBAs), operating in their lower frequency range of \SIrange{102}{197}{\mega\hertz} with \SI{195.3125}{\kilo\hertz} channel bandwidths (subbands 12 to 499 inclusive) and an underlying \SI{5.12}{\micro\second} sampling frequency. Raw voltages were written to disk, which are then processed to the needs of the current project (the details for this census are discussed further in \S\ref{sec:processing}). This process allows for corrections in the dispersion measure used for coherent dedispersion~\citep{1975MComP..14...55H} in the case that a source is detected with a significant deviation from the expected dispersion measure, re-sampling or extending the output data products in cases of particular interest and further analysis of unexpected features of the data.

I--LOFAR is located in a relatively radio-quiet zone as compared to the rest of the stations spread across Europe. The low local population density alongside the lack of digital audio broadcast (DAB) radio in Ireland significantly reduced the radio frequency interference (RFI) present during observations, though typically 7--20 per cent of the total bandwidth is still flagged during observations due to transient RFI, the FM-radio band and to reduce noise contributions from filtered regions near the edge of the observed bandwidth.

% The HBA tiles were arranged with the intention of several stations being used for (in)coherent beamforming simultaneously rather than single station use. As a result of the tile and antenna pattern being placed on a regular grid, side-lobes (peaks in the beam response outside of the main lobe) can often reduce the sensitivity of the instrument at higher frequencies, particularly when the radio-bright A--team sources, such as Cas A or Cyg A, align with these side lobes. These contributions are not considered when determining source flux densities, and may result in sources having higher spectral indices in this work as a result.

\subsection{Sensitivity and Flux Calibration}\label{sec:sensitivity}
The underlying sensitivity and resulting flux density ($S_{\text{emission}}$) measurements are calculated through a modified form of the methodology used by \cite{kondratievLOFARCensusMillisecond2016a}, based on the radiometer equation and the additional factors needed to accurately characterise the properties of a low-frequency, wide-bandwidth interferometer. Further modifications to the methodology have been made for this work to account for the use of a single LOFAR station, resulting in a set of equations, with a focus on single-pulses in Eq.~\ref{eq:radpulse}, and on periodic emission in Eq.~\ref{eq:radperiodic}, for a given observing frequency ($f$), and pointing direction ($l$, $b$).

\begin{equation}\label{eq:radpulse}\centering
S_\text{pulse}(f, l, b) = \text{S/N}  \frac{2k_b  m_\text{II}\left(T_\text{sky} + T_\text{ant}\right)}{A_\text{eff}\beta \sqrt{n_p  \Delta \nu_\text{eff} w_\text{pulse}}} \hfill\left[\text{Jy}\right]
\end{equation}

\begin{equation}\label{eq:radperiodic}\centering
S_\text{mean}(f, l, b) = \text{S/N} \frac{2k_b  m_\text{II}\left(T_\text{sky} + T_\text{ant}\right)}{A_\text{eff}\beta \sqrt{n_p  \Delta \nu_\text{eff} t_\text{obs}}} \sqrt{\frac{W}{P - W}}\hfill\left[\text{Jy}\right]
\end{equation}

These equations differ only in the fraction of time integrated over, the pulse width $\sqrt{w_\text{pulse}}$ against the integrated periodic emission for the given observation time, $\sqrt{t_\text{obs}\left(P - W\right) / W}$, based upon the observing time $t_{\text{obs}}$, pulsar period $P$ and emission width $W$. The remaining components describe the sensitivity of the station for a given SNR, which include standard parameters covering the antenna polarizations ($n_p$, set to 2 for the two polarizations of the antennas) and the effective observing bandwidth after performing RFI flagging ($\nu_\text{eff}$).

The gain of the telescope ($A_\text{eff} / \left(2k_bm_\text{II}\right)$) depends on the effective area of the telescope, which can be found for an international station in \cite{haarlemLOFARLOwFrequencyARray2013a}. The base value of \SIrange{1150}{2400}{\metre\squared} was modified to account for 2 of the 96 HBA tiles not being used in the majority of the observations during this census due to hardware issues. As a LOFAR station is an interferometric array, the effective area varies both in frequency, with an upper limit of the area reached at \SI{135}{\mega\hertz} due to mutual coupling of the densely packed antenna, and in pointing, which requires a correction in the form of a Mueller matrix, which is calculated from a Jones matrix generated by dreamBeam \citep{carozzi2baOrNot2baDreamBeam2020} using the Hamaker beam model \citep{hamaker2011mathematical} for each pulse or folded sub--integration.

The $\beta$ correction factor of 0.94 in the single-pulse brightness equation is to account for the losses while processing the signal. The 8-bit digitisation of the signal is an extremely minor contribution, while the majority is to account for the loss in signal when attempting to integrate over Gaussian-like signals with a boxcar~\citep{morelloOptimalPeriodicitySearching2020b}. While not all pulses take the shape of a Gaussian function, the majority do, with the remaining pulses having multiple components, or appear as scattered delta functions, both of which would see further losses beyond this correction factor (and are brighter than reported as a result).

The antenna and electronics temperatures ($T_\text{ant}$) are sampled based upon measurements by \cite{wijnholdsSituAntennaPerformance2011}, and are averaged across \SI{5}{\mega\hertz} bandwidth blocks.

The sky temperatures ($T_\text{sky}$) are calculated by performing a convolution between a 2D Gaussian function representing the approximated central component of the telescope beam, including a width correction factor of 1.02 \citep[Table~B.1]{haarlemLOFARLOwFrequencyARray2013a}, with the region of the Low Frequency Sky Model \citep[LFSM]{2017MNRAS.469.4537D}, as sampled using \texttt{pygdsm}~\citep{2016ascl.soft03013P}. The LOFAR beam width and sky temperature are sampled in \SI{5}{\mega\hertz} blocks, which are then fit by a power law that is sampled to determine the sky temperature for the frequency-variable beam size. The $T_\text{sky}$ variable was found to have a minimum, median, and maximum of \SI{205}{\kelvin}, \SI{444}{\kelvin} and \SI{2006}{\kelvin} at \SI{150}{\mega\hertz} for the corpus of sources observed during the census. The corresponding sensitivity limits for the I--LOFAR telescope for single pulses and periodic emission at these temperatures are given in Table~\ref{tab:sensitvity_pulse} and~\ref{tab:sensitvity_pulsar}.

While the recent work of \cite{priceGlobalSkyModels2021} has indicated the advantages of using diffuse global sky models to determine more accurate measurements of $T_\text{sky}$ for pulsar and FRB studies, Price notes that across the bandwidth used in this work the uncertainty is kept within a few per cent as the spectral index will only vary between $\alpha = - 2.43, -2.53$ near the Galactic plane. However, it was noted that the extra Gaussian convolution step, introduced to account for the wide and variable beam width of a single international LOFAR station, had a significantly larger effect on the effective spectral index associated with $T_\text{sky}$ than expected. After fitting the modelled sky temperatures to a power law, the index $\alpha$ was found to vary significantly more than expected, with a maximum, median and minimum fit of $\alpha = -2.29, -2.58, -2.83$ observed across the sampled sources. The difference in the $T_\text{sky}$ measurement at \SI{110}{\mega\hertz} of~$3$ per cent,~$1$ per cent,~$11$ per cent as compared to sampling the observed coordinate in the LFSM as a result of this modification.

This effect is due to the beam widening at lower frequencies which will amplify any anisotropic region in the field of view, causing $\alpha$ to be more shallow where a source is located near a local maximum on the sky temperature max, and steeper in the case that is it near a local minimum. However, the median effect on the combined $T_\text{sky} + T_\text{ant}$ term is found to be below~$1$ per cent across the observed bandwidth, as a result of the large contributions of $T_\text{ant}$ from each antenna that comprises the LOFAR interferometer.

For single pulses, the pulse width used is determined through SNR optimisation for the bandwidth-flattened 1D pulse profile that has been dedispersed to the optimal dispersion measure determined across all pulses observed at the end of the census. 

SNRs are calculated in windows of \SI{10}{\mega\hertz}, with the upper and lower components of the observation bandwidth discarded due to the sensitivity fall-off due to the pass-band of the electronics and the large amount of RFI due to the FM radio band near \SI{100}{\mega\hertz}. Consequently, statistics are generated over the bandwidth between \SIrange{116}{186}{\mega\hertz}, centred on \SI{151}{\mega\hertz}. An equivalent flux density is then calculated for each segment of the bandwidth. These frequency dependant flux densities were fit to a power law to determine the spectral indices for detected sources. This fit is performed across the full time-averaged folded profile for periodic emission, while for single-pulses the presented value is the best fit of a Gaussian function to a histogram of spectral indices fit to each observed pulse.

Continuing the method used by \citeauthor{kondratievLOFARCensusMillisecond2016a}, a 50 per cent uncertainty is introduced on the results of these calculations, to account for compounding uncertainties on the measured station parameters, potential failures in the beam model, side-lobe contributions (which are particularly strong when using a single international LOFAR station) and the transient background sky due to ionospheric scintillation at low frequencies (see \citealt[\S~3.2]{kondratievLOFARCensusMillisecond2016a} for further details). 

\begin{table}
    \centering
    \begin{tabular}{llccc}\hline\hline
    Samples & Width & S$_{T_\text{sky, min}}$ & S$_{T_\text{sky, median}}$ & S$_{T_\text{sky, max}}$  \\
     & \multicolumn{1}{c}{\SI{}{\milli\second}} & \SI{}{\jansky} & \SI{}{\jansky} & \SI{}{\jansky} \\
   \hline
1 & 0.655 & 27 & 38 & 110 \\
2 & 1.31 & 19 & 27 & 77 \\
4 & 2.62 & 14 & 19 & 55 \\
16 & 10.5 & 6.8 & 9.5 & 27 \\
64 & 41.9 & 3.4 & 4.8 & 14 \\
256 & 168 & 1.7 & 2.4 & 6.8 \\

   \hline\hline
    \end{tabular}
    \caption{The bandwidth-averaged minimum flux density, in jansky, of a pulse of varying width required to be detected with an international LOFAR station, with a signal-to-noise ratio of 7.5.}
    \label{tab:sensitvity_pulse}
\end{table}

\begin{table}
    \centering
    \begin{tabular}{llccc}\hline\hline
    \multicolumn{2}{c}{Duty Cycle} & S$_{T_\text{sky, min}}$ & S$_{T_\text{sky, median}}$ & S$_{T_\text{sky, max}}$  \\
    & per cent & \SI{}{\milli\jansky} & \SI{}{\milli\jansky} & \SI{}{\milli\jansky} \\
   \hline
0.001 & 0.1 & 0.27 & 0.38 & 1.1 \\ % These values differ from the single pulse/1 sample values by both an order of magnitude, and 1-3\%. This is not a data-production error.
0.003 & 0.3 & 0.48 & 0.67 & 1.9 \\
0.01 & 1 & 0.87 & 1.2 & 3.5 \\
0.03 & 3 & 1.5 & 2.1 & 6.2 \\
0.1 & 10 & 2.9 & 4.1 & 12 \\
0.3 & 30 & 5.7 & 8 & 23 \\
   \hline\hline
    \end{tabular}
    \caption{The folded average flux density, in millijansky, for a one-hour observation of a pulsar of varying duty cycles required to be detected with an international LOFAR station, with a signal-to-noise ratio of 6.}
    \label{tab:sensitvity_pulsar}
\end{table}

\subsection{Observation Processing and Archiving}\label{sec:processing}
The raw data from the station is processed into Stokes $I$ SigProc filterbanks \citep{lorimerSIGPROCPulsarSignal2011} using a modified version\footnote{Modified to accept raw voltages from an international station through udpPacketManager~\citep{McKenna2023} rather than COBALT(2) correlated H5 files, see \url{https://github.com/David-McKenna/cdmt}} of CDMT \citep{bassaEnablingPulsarFast2017}, to perform coherent dedispersion, channelisation, temporal downsampling and detection to Stokes $I$ on the Nvidia Tesla V100s present in the REALTA compute cluster.

Coherent dedispersion reduces the effect of temporal smearing of an incoming pulsar signal by performing a convolution between the station voltages and the predicted inverse transfer function for the interstellar medium at a given dispersion measure. Initially, the dispersion measure is taken from the source catalogue, but where observations with I--LOFAR constrained a source dispersion measure further, the dispersion measure is updated and used for re-processing voltages and future observations.

Channelisation was performed to reduce the bandwidth of the subbands by a factor of 8, increasing the number of channels in the output data and allowing for a more accurate measurement of dispersion measure when a pulse is detected.

The data is down-sampled in time by a factor of 16 as a compute and storage saving procedure. This is not considered to have a negative effect on the data due to three main considerations, (a) the predicted underlying pulse widths, (b) the expected scattering timescales and (c) the data is coherently dedispersed prior to this step, causing any smearing to be on timescales far below the output time resolution. As discussed in~\S\ref{sec:intro}, RRATs typically have longer periods than the general pulsar population due to selection effects. As a result, given a typical duty cycle, the underlying width of a pulse can be expected to be measured on a scale of \SI{}{\milli\second} to tens of \SI{}{\milli\second}. Further, at low frequencies these pulses are heavily scattered, with scattering behaviour often scaling strongly with frequency, roughly following $\nu^{-4}$ \citep[see][]{langInterstellarScintillationPulsar1971,krishnakumarMultifrequencyScatterBroadening2017a}, further broadening the pulses. As a result, the sub-millisecond sampling available after this down-sampling step is still more than sufficient to detect and analyse the morphology of the pulses observed when a sufficiently bright signal is detected.

The resulting filterbank is a 32-bit floating point file with 3904 channels, with bandwidths of \SI{24.41}{\kilo\hertz} each, and \SI{655.36}{\micro\second} temporal sampling. However, in the case that a source has been observed as a part of the census and after the \SI{5}{\hour} of observations there has been no detections, either single pulse or periodic, the filterbank is re-sampled to reduce the number of channels by a factor of 8, returning them to the original resolution of the telescope, resulting in channel widths of \SI{195.3126}{\kilo\hertz}. In all cases, the final filterbank is compressed using zstandard\footnote{\url{https://github.com/facebook/zstd}} for a further 5--10 per cent space-saving and archived.

Prior to processing, the 32-bit filterbanks are re-sampled using DSPSR's \texttt{digifil} tool~\citep{vanstratenDSPSRDigitalSignal2011}, removing the bandpass contributions across the default $10$-s %\SI{10}{\second}
re-scale interval and reducing the data to 8--bit samples, which are computationally easier to analyse. The reduced bit depth further dampens any excess contributions of spurious RFI samples to dedispersed time series.

All observations undergo a single-pulse search using the GPU-accelerated \texttt{heimdall} software~\citep{barsdellAdvancedArchitecturesAstrophysical2012}, with a more sensitive search (with a 0.3 per cent loss of signal per dispersion measure trial) over a range of \SIrange{10}{20}{\parsec\per\cubic\centi\metre} centred on the known dispersion measure of the source, and a less sensitive search (with a 3 per cent loss of signal per trial) across the \SIrange{10}{500}{\parsec\per\cubic\centi\metre} range to search for unexpected signals in the data. The lower limit of \SI{10}{\parsec\per\cubic\centi\metre} was chosen to discard numerous spurious candidates from RFI and ionospheric scintillation, while the upper limit was initially set to \SI{200}{\parsec\per\cubic\centi\metre} to consider the limitations associated with observations of high dispersion measure sources at low frequencies (see \S~\ref{sec:discussdm}), but was expanded to \SI{500}{\parsec\per\cubic\centi\metre} due to the minimal additional compute cost.

The remaining \texttt{heimdall} configuration options were not modified, resulting in a candidate threshold and $6~\sigma$ and baseline smoothing across \SI{2}{\second} windows.

It is noted that the recent work of \cite{quiFredda2023}
%keaneHeimdallDMTol???}(this is intentional, unsure if the second ref was released) 
highlights some systemic errors and pitfalls in the \texttt{heimdall} processing pipeline. The dense sampling of dispersion measure trials used for this work reduces the effect of the errors described by \citeauthor{quiFredda2023}, however 
%\citeauthor{keaneHeimdalDMTol???} 
%indicates 
it seems that the spacing of the dispersion measure trials is insufficient for low--frequency observation, and will result in a true loss of signal higher than the intended 0.3 per cent and 3 per cent defined above between dispersion measure trials.

Any candidates above a SNR of $7.5$ are used to generate a candidate plot, which is inspected by eye to determine if a candidate is a true pulse detection. In the case of a detection, the candidate is processed with the methodology discussed in~\S\ref{sec:singlepulse} and~\S\ref{sec:timing}.

Given that \texttt{heimdall} performs a search for single-pulses on a series of dispersed 1-dimensional time series, this methodology results in a reduced sensitivity towards pulses that are only visible in a fraction of the observed bandwidth. Consequently, pulses observed in a fraction of the bandwidth, either due to their intrinsic spectral index, scintillation, or other phenomena that cause sharp spectral variability are less likely to be reported in the results of this census, but may be detected in future processing of the census archive (further discussed in~\S\ref{sec:frbdiscuss}).

If the period of a source is known, the data is also folded with PRESTO's \texttt{prepfold} tool \citep{ransomPRESTOPulsaRExploration2011} at the given period and dispersion measure to search for periodic emission. Blind periodicity searches were not a component of the main processing pipeline for the census observations (but see~\S\ref{sec:J1329},~\S\ref{sec:conclusions}).

\subsection{Single Pulse Analysis}\label{sec:singlepulse}
Single pulses are extracted and pre-processed using a combination of DSPSR and PSRCHIVE \citep{hotanPSRCHIVEPSRFITSOpen2004}. DSPSR's \texttt{digifil} is used to extract a small block of time around a candidate arrival time\footnote{The amount of padding time varies depending on the source period due to the requirement of PSRCHIVE's \texttt{pat} tool, which is used for producing time of arrival (TOA) measurements, whereby input time series must be powers of two in length, but is typically on the order of half a rotation period on each side of a pulse.}, and re-sampled from 32-bit floats to 8-bit unsigned chars, removing the bandpass offset in the process. \texttt{dspsr} then dedisperses this filterbank, performs spectral kurtosis for RFI flagging and writes an output in the TIMER format. This is initially only performed for visually approved candidates with a SNR above 7.5, but once a source has been well constrained through detections in multiple observations, this criterion is reduced to a SNR of 7.

PSRCHIVE's \texttt{paz} tool is then used to perform further RFI zapping based on outlying channels in a windowed scan across the bandpass of the 32-bit data, to ensure no badly contaminated channels are included in the output pulse archive.

The overall pulse populations are then analysed to determine their properties. Namely, spectral brightness, pulse widths, burst rates and pulse amplitude distribution. Where noted, fitting is performed in Python using the least-squares method implemented by the \texttt{lmfit} \citep{newvilleLmfitLmfitpy2021} module, which is also the source of any associated uncertainties (which are 1~$\sigma$). These results can be found in Table~\ref{tab:obs_summary}. Pulse population analysis that includes model fitting was only performed on sources with more than 40 pulses, as sources with less observed pulses were not found to be well described by the models used for fitting.

The spectral properties of the pulses are calculated by fitting a power law to the spectral flux densities of each pulse determined using Eq.~\ref{eq:radpulse}, and a Gaussian is fit to the overall distribution of power law exponents across the observed pulses. The mean term of the Gaussian and its uncertainty is reported as the single-pulse spectral index in Table~\ref{tab:obs_summary}.

Pulse amplitude distributions were analysed by fitting multiple models (power law, broken power law, log--normal, powerlaw--log--normal and log--normal--powerlaw) to the distribution of frequency-averaged flux densities of observed pulses. From these models, the small-N corrected Akaike information criterion (AICc) was used to determine the best fit model. The fit values for these properties are presented accordingly in Fig.~\ref{fig:spectralmodfits} and Table~\ref{tab:pulseamplitudetab}.

Pulse widths and dispersion measures were determined with PSRCHIVE's \texttt{pdmp}. The measured dispersion measures were well-fit by a Gaussian distribution, however pulse widths did not form a smooth distribution. As a result, the mean and standard deviation of the widths has been presented, rather than parameters from a Gaussian fit.

\subsection{Timing}\label{sec:timing}
Using the single-pulse archive produced by \texttt{dspsr} as described in the previous section, PSRCHIVE's \texttt{pat} generates a per-pulse time of arrival (TOA) measurement using an analytical profile generated using the brightest pulse and the \texttt{paas} tool as a standard reference. 

\texttt{tempo2} \citep{hobbsTEMPO2NewPulsartiming2006} is then used to model and update the source ephemeris using all times of arrival of a source. Some TOAs of low-significance pulses (SNR ranging from 7 to 7.5) are often rejected at this stage due to unexpectedly large uncertainties, large phase offsets or other unexpected phenomena. These cases are often caused by unflagged RFI or unexpected behaviour in the underlying dataset, most of which can be fixed through manually re-processing the pulse archives.

Where sources have not yet been published in \texttt{psrcat}, the provided source catalogues were used to generate an ephemeris for each source, which was then used as a basis for timing. When a source with only a known period (and no derivative) was observed to have more than 3 pulses in an observation, the times of arrival were used to brute force a new period near the previously published result\footnote{The \href{https://github.com/evanocathain/Useful_RRAT_stuff}{\texttt{getper.py} script} from the SUPERB survey \citep{keane+2018} was used for the brute-force periodicity search.}, which was then used for timing.

When more than 6 pulses were seen for a source without a period, it is possible to use the same brute force approach to determine an underlying periodicity of the pulses. With 6 pulses, each generated candidate has a $2~\sigma$ likelihood of representing the source's rotation, assuming the pulses fall within a narrow phase window, with 9 pulses reaching a $3~\sigma$ likelihood \citep[Fig.~2]{2017ApJ...840....5C}. The candidate periods were then tested and refined by using prior TOAs to find the period that represents the greatest common divisor across all observed TOAs, which was then used for timing the source with \texttt{tempo2}.

\subsection{Periodic Emission Analysis}
In the case that a source was found to emit periodic emission with a SNR greater than 7 in the previously described PRESTO \texttt{prepfold} search, the observations of the source were folded and analysed using \texttt{DSPSR} and \texttt{PSRCHIVE} to prepare folded archives with integration times of \SI{30}{\second} and 256 bins. Observations are combined and integrated into a single-frequency time series for final analysis. To reduce the effects of RFI on the folded data and improve the accuracy of periodic emission SNRs, the data in each observation were flagged using \texttt{clfd} \citep{morelloHighTimeResolution2019} prior to analysis.

\subsection{Archival Data Follow-up}
In the case that a source was detected during the census, pointings from the LOTAAS survey~\citep{sanidasLOFARTiedArrayAllSky2019b} were downloaded from the LOFAR Long Term Archive\footnote{\url{https://lta.lofar.eu/Lofar}} (LTA) and processed using the same methodology described in this section to search for single-pulses and periodic emission. Due to the large positional uncertainty associated with many sources in the census, typically at least one `ring' of nearby beams were downloaded about the reported pointing of the observation.

This allowed for a potential detection of these sources at an earlier epoch and potentially allows for changes in source properties to be tracked from an earlier epoch.

\afterpage{\begin{landscape}
\begin{table}
\centering
    \caption{A summary of the properties of the $\detected$ sources detected as a part of this census, based on both single-pulse and periodic emission. Data across \SIrange{116}{186}{\mega\hertz}, centered on \SI{151}{\mega\hertz}, are analysed and presented from observations that started in August 2020 and ended in September 2022. Each source notes their source catalogue (\textbf{C}HIME, \textbf{P}ushchino, or \textbf{R}RAatalog), the reference to any previous work related to the source with LOFAR, the known rotation period  of the source, best-fit dispersion measure (from single-pulse emission, or periodic emission if single-pulses were not observed) and the number of hours each source was observed for. Basic parameters regarding the single-pulse emission and periodic emission during the observations are provided (notably $w_{50}$, the 50 per cent emission brightness width, and $\alpha_{sp}$ or $\alpha_f$, the spectral index for single-pulse or periodic emission). Additionally, the single-pulse emission provides the number of pulses observed, the ratio between the brightest and quietest observed pulses (SR), and the effective burst rate for each source.}
	\label{tab:obs_summary}
\begin{tabular}{lccccc|ccccccc|cccc}
\hline\hline
& & & & & & \multicolumn{7}{c|}{Single Pulse} & \multicolumn{4}{c}{Periodic Fold} \\
Source & Cat. & Prev. & Period & DM & T\textsubscript{obs} & N\textsubscript{pulses} &  w$_{50}$ & Duty Cycle & S$_{150}^{\text{peak}}$\textsuperscript{a} & SR &$\alpha_{\text{sp}}$\textsuperscript{b} & Burst Rate\textsuperscript{c} &  w$_{50}$ & Duty Cycle & S$_{150}^{\text{mean}}$\textsuperscript{a} &  $\alpha_{\text{f}}$\textsuperscript{b}\\
   & & & (\SI{}{\second}) & (\SI{}{\parsec\per\cubic\centi\metre})  & (\SI{}{\hour}) & & (\SI{}{\milli\second}) & per cent & (\SI{}{\jansky}) & & & (\SI{}{\per\hour}) & (\SI{}{\milli\second}) & per cent & (\SI{}{\milli\jansky}) & \\
\hline
J0054+6650 & R & -- & 1.3902 & 14.560(2) & 31.5 & 282 & 14(6) & 1.0(4) & 45 &  10 & $-$1.24(7) & 9.0(5) & -- & -- & -- & -- \\
J0102+5356 & R & \citetalias{karako-argamanDiscoveryFollowupRotating2015} & 0.3543 & 55.6200(8) & 27.3 & 88 & 4.0(1.4) & 1.1(4) & 60 &   8 & 0.13(13) & 3.2(3) & -- & -- & -- & -- \\
J0139+3336 & P & \citetalias{michilliSinglepulseClassifierLOFAR2018} & 1.2480 & 21.223(4) & 16.6 & 157 & 13(6) & 1.0(5) & 43 &   9 & $-$1.65(9) & 9.4(8) & -- & -- & -- & -- \\
J0201+7005 & R & \citetalias{karako-argamanDiscoveryFollowupRotating2015} & 1.3492 & 21.047(2) & 10.9 & 9 & 2.6(8) & 0.19(6) & 37 &   3 & -- & 0.8(3) & 12 & 0.9 & 2.5 & $-$0.2(5) \\
J0209+5759\textsuperscript{d} & C & -- & 1.0639 & 55.855(4) & 26.6 & 43 & 13(7) & 1.2(7) & 29 &   5 & $-$0.25(11) & 1.6(2) & 25 & 2.4 & 25.1 & $-$0.6(3) \\
J0226+3356 & C & -- & 1.2401 & 27.397(14) & 18.0 & -- & -- & -- & -- & -- & -- & -- & 43 & 3.5 & 3.3 & $-$0.9(3) \\
J0317+1328 & P & \citetalias{michilliSinglepulseClassifierLOFAR2018} & 1.9742 & 12.7452(10) & 10.1 & 59 & 3.2(1.3) & 0.16(7) & 41 &   4 & $-$5.3(8) & 5.9(8) & 20 & 1.1 & 3.8 & $-$1.8(4) \\
J0332+79 & R & \citetalias{karako-argamanDiscoveryFollowupRotating2015} & 2.0562 & 16.589(7) & 15.4 & 1 & 5.9 & 0.29 & 15 &   1 & -- & (0.065) & -- & -- & -- & -- \\
J0348+79 & C & -- & -- & 26.09(3) & 26.7 & 6 & 36(15) & -- & 9 &   3 & -- & 0.22(9) & -- & -- & -- & -- \\
J0746+5514 & C & -- & 2.8938 & 10.318(7) & 43.2 & 48 & 32(20) & 1.1(7) & 35 &  13 & $-$0.13(16) & 1.11(16) & -- & -- & -- & -- \\
J0854+5449 & C & -- & 1.2330 & 18.843(3) & 24.2 & 2 & 2.29(98) & 0.19(8) & 19 &   1 & -- & 0.08(6) & 11 & 0.9 & 1.1 & 0.7(2) \\
J0939+45 & P & -- & -- & 17.45(4) & 18.5 & 3 & 19(5) & -- & 7 &   2 & -- & 0.16(9) & -- & -- & -- & -- \\
J1006+3015 & P & -- & 3.0664 & 18.085(4) & 95.6 & 166 & 33(16) & 1.1(5) & 50 &  16 & $-$2.37(12) & 1.74(13) & -- & -- & -- & -- \\
J1132+2513 & P & \citetalias{sanidasLOFARTiedArrayAllSky2019b} & 1.0021 & 23.716(7) & 11.2 & -- & -- & -- & -- & -- & -- & -- & 17 & 1.8 & 7.6 & $-$1.3(4) \\
J1218+47 & P & -- & -- & 20.144(11) & 34.7 & 13 & 16(9) & -- & 13 &   4 & -- & 0.37(10) & -- & -- & -- & -- \\
J1329+13 & P & -- & -- & 12.367(13) & 49.6 & 5 & 23(11) & -- & 11 &   1 & -- & 0.10(5) & -- & -- & -- & -- \\
J1336+3414 & P & -- & 1.5066 & 8.4688(11) & 79.0 & 132 & 5(3) & 0.31(18) & 48 &  11 & 0.12(18) & 1.67(15) & -- & -- & -- & -- \\
J1400+2125 & P & -- & 1.8555 & 11.214(3) & 86.7 & 43 & 17(6) & 0.9(3) & 29 &   4 & $-$3.3(6) & 0.50(8) & -- & -- & -- & -- \\
J1404+1159 & P & \citetalias{michilliSinglepulseClassifierLOFAR2018} & 2.6504 & 18.53(3) & 5.7 & -- & -- & -- & -- & -- & -- & -- & 62 & 2.3 & 19.1 & $-$1.13(13) \\
J1538+2345 & R & \citetalias{karako-argamanDiscoveryFollowupRotating2015} & 3.4494 & 14.89814(98) & 38.3 & 371 & 9(6) & 0.25(18) & 73 &  22 & $-$1.60(12) & 9.7(5) & 121 & 3.5 & 4.3 & $-$0.51(17) \\
J1848+1516 & R & \citetalias{michilliSinglepulseClassifierLOFAR2018} & 2.2338 & 77.488(11) & 32.5 & 67 & 41(12) & 1.8(6) & 16 &   3 & $-$0.2(4) & 2.1(3) & 52 & 2.3 & 4.5 & 0.52(14) \\
J1931+4229 & C & -- & 3.9210 & 50.987(6) & 53.7 & 133 & 49(20) & 1.2(5) & 22 &   9 & $-$1.06(7) & 2.5(2) & -- & -- & -- & -- \\
J2105+19 & P & -- & 3.5298 & 34.47(3) & 7.7 & -- & -- & -- & -- & -- & -- & -- & 187 & 5.3 & 3.8 & $-$1.4(9) \\
J2108+4516\textsuperscript{e} & C & -- & 0.5774 & 82.520(7) & 6.2 & 75 & 24(12) & 4(2) & 10 &   3 & 1.91(14) & 12.0(1.4) & 96 & 16.8 & 231.2 & $-$0.4(4) \\
J2138+69 & C & -- & -- & 46.530(3) & 30.8 & 9 & 7(2) & -- & 20 &   2 & -- & 0.292(97) & -- & -- & -- & -- \\
J2202+2134 & P & \citetalias{sanidasLOFARTiedArrayAllSky2019b} & 1.3573 & 17.7473(19) & 18.1 & 11 & 3.0(7) & 0.22(5) & 25 &   3 & -- & 0.61(18) & 11 & 0.8 & 4.8 & $-$1.0(9) \\
J2215+4524 & C & -- & 2.7231 & 18.5917(16) & 37.2 & 121 & 9(5) & 0.32(18) & 19 &   4 & $-$1.76(17) & 3.3(3) & 21 & 0.8 & 1.7 & $-$1.45(12) \\
J2325$-$0530 & R & -- & 0.8687 & 14.9580(8) & 7.0 & 322 & 10(3) & 1.2(4) & 131 &  11 & $-$3.66(5) & 46(3) & 12 & 1.4 & 7.5 & $-$3.7(5) \\
J2355+1523 & C & -- & 1.0964 & 26.924(16) & 51.9 & 14 & 17(8) & 1.5(7) & 15 &   3 & -- & 0.27(7) & -- & -- & -- & -- \\

\hline\hline
\end{tabular}

\bigskip

{\footnotesize \textsuperscript{a} The standard uncertainty on all LOFAR flux density measurements is 50 per cent, see~\S\ref{sec:sensitivity}.}

{\footnotesize \textsuperscript{b} Spectral indices are expected to be biased towards more negative values due to the positional uncertainty of the sources, as the telescope beam FWHM will increase the effective gain for emission at lower frequencies, see~\S\ref{sec:sensitivity}.}
	
{\footnotesize \textsuperscript{c} Provided burst rate uncertainties are Poissonian.}

{\footnotesize \textsuperscript{d} Periodic emission statistics for J0209+5759 are based on emission during the active phases of the observations, representing \SI{1.3}{\hour} of data, see~\S\ref{sec:periodicres}.}
 
{\footnotesize \textsuperscript{e} Single pulse and periodic emission statistics for J2108+4516 are based on the single \SI{1}{\hour} observation where it was detectable.}

\end{table}

\end{landscape}}

\afterpage{\begin{landscape}
\begin{table}
	\centering
    \caption{Summary of the 8 new source ephemerides, and their derived quantities, determined through pulsar timing performed with single-pulse times of arrival at I--LOFAR.}
	\label{tab:newephemerides}
\begin{tabular}{lcccccccc}

\hline\hline
Source & J0054+6650 & J0102+5356 & J0746+5514 & J1006+3015 & J1336+3414 & J1400+2125 & J1931+4229 & J2215+4524 \\
\\
Catalogue & RRATalog & RRATalog & CHIME/FRB & PRAO & PRAO & PRAO & CHIME/FRB & CHIME/FRB \\
Catalogue Source Name & J0054+66 & J0103+54 & J0746+55 & J1005+3015 & J1336+3346 & J1400+2127 & J1931+4229 & J2215+45 \\
\hline
Right Ascension (hms) & \hmsangle{00;54;55.412}(14) & \hmsangle{01;02;57.786}(12) & \hmsangle{07;46;47.4}(2) & \hmsangle{10;06;34.44}(7) & \hmsangle{13;36;33.953}(18) & \hmsangle{14;00;14.19}(3) & \hmsangle{19;31;10.87}(7) & \hmsangle{22;15;46.847}(13) \\
Declination (dms) & \dmsangle{+66;50;23.81}(8) & \dmsangle{+53;56;11.86}(15) & \dmsangle{+55;14;37}(2) & \dmsangle{+30;15;46}(2) & \dmsangle{+34;14;37.8}(2) & \dmsangle{+21;25;40.0}(5) & \dmsangle{+42;29;16.1}(1.0) & \dmsangle{+45;24;43.3}(2) \\
Galactic Longitude ($^\circ$) & \galangle{123.27546}(6) & \galangle{124.64827}(5) & \galangle{162.5359}(8) & \galangle{197.9536}(3) & \galangle{71.97329}(8) & \galangle{16.66008}(11) & \galangle{75.1545}(3) & \galangle{96.39171}(5) \\
Galactic Latitude ($^\circ$) & \galangle{3.97052}(2) & \galangle{-8.89671}(4) & \galangle{29.7344}(6) & \galangle{53.9052}(6) & \galangle{77.98183}(7) & \galangle{73.33655}(15) & \galangle{11.2299}(3) & \galangle{-9.28754}(6) \\
Dispersion Measure  (\SI{}{\parsec\per\centi\metre\cubed}) & 14.560(2) & 55.6200(8) & 10.318(7) & 18.085(4) & 8.4688(11) & 11.214(3) & 50.987(6) & 18.5917(16) \\
Distance\textsuperscript{a} (\SI{}{\parsec}) & 878 & 2020 & 400 & 2120 & 715 & 990 & 4240 & 1140 \\
\\
Period (\SI{}{\second}) & 1.390218110074(8) & 0.354299198996(4) & 2.8936675025(4) & 3.0663651860(2) & 1.50660326853(3) & 1.85546492484(6) & 3.9210375131(2) & 2.72313605772(5) \\
Period Derivative (\SI{e-15}{\second\per\second}) & 5.5532(8) & 0.5203(5) & 10.19(5) & 5.52(2) & 0.111(3) & 1.358(7) & 31.14(5) & 5.223(5) \\
Characteristic Age (\SI{}{\mega\year}) & 3.97 & 10.8 & 4.5 & 8.8 & 216 & 21.7 & 2 & 8.27 \\
Magnetic Field (\SI{e12}{\gauss}) & 2.81 & 0.434 & 5.49 & 4.16 & 0.413 & 1.61 & 11.2 & 3.82 \\
\\
Timing Start (MJD) & 59108 & 59240 & 59268 & 59123 & 59206 & 59142 & 59191 & 59184 \\
Timing End (MJD) & 59849 & 59850 & 59848 & 59830 & 59848 & 59820 & 59849 & 59822 \\
Reference Epoch (MJD) & 59478 & 59545 & 59558 & 59476 & 59527 & 59481 & 59520 & 59534 \\
N\textsubscript{TOAs} & 279 & 88 & 48 & 163 & 123 & 43 & 133 & 183 \\
RMS Residuals (\SI{}{\micro\second}) & 1663 & 1250 & 11702 & 10701 & 2844 & 2942 & 9262 & 2536 \\
\hline\hline

\end{tabular}
	
		\vspace{0.25cm}
	{\footnotesize \textsuperscript{a} The YMW16 model~\citep{yaoDensityDistance2017} was used to determine the provided source distances.}
	
\end{table}

\begin{figure}
    \centering
    \begin{tabular}{cc}
    \begin{minipage}[t]{10.5cm}
    \centering
    \adjincludegraphics[height=5cm,trim={0 0 {0.85\width} 0}]{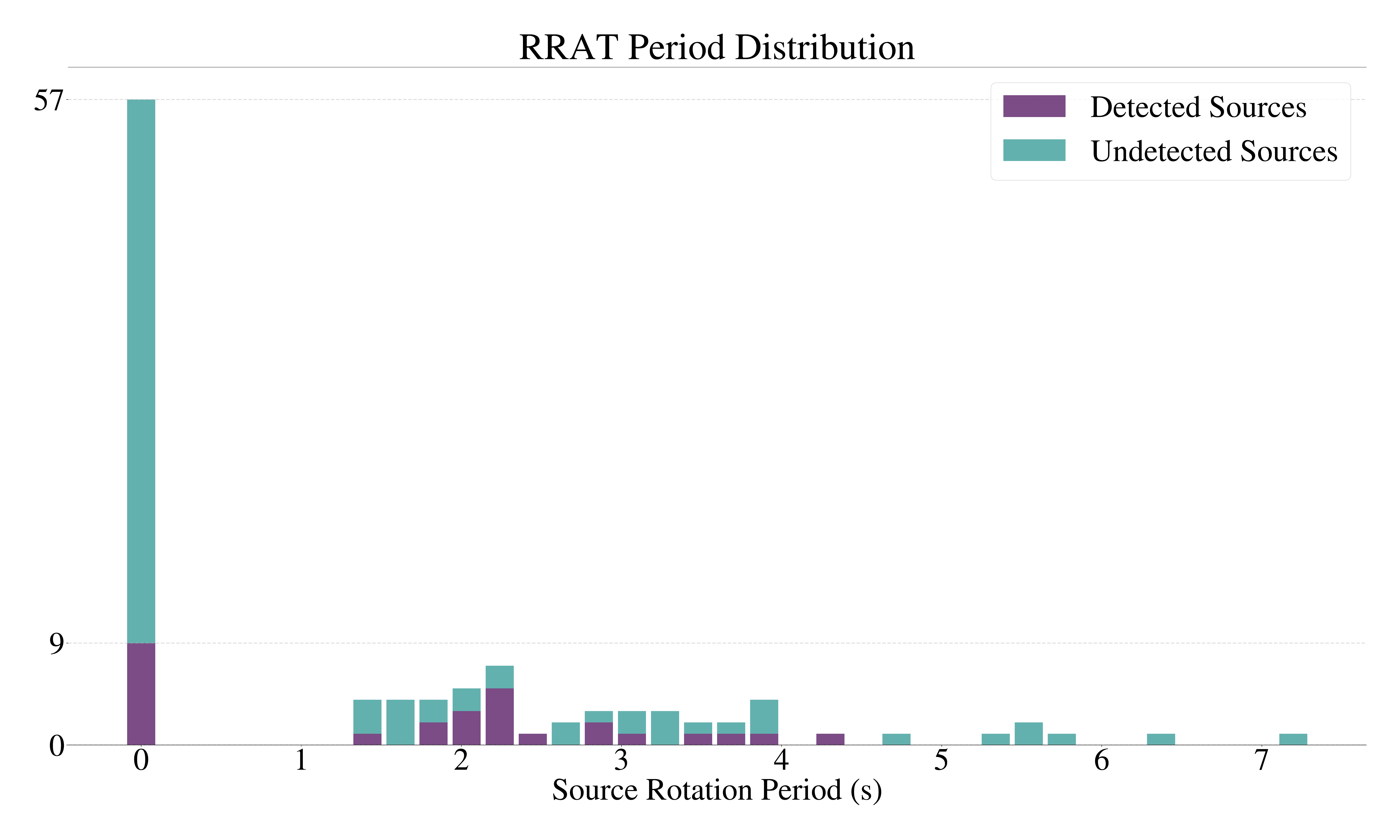}
    \includegraphics[height=5cm]{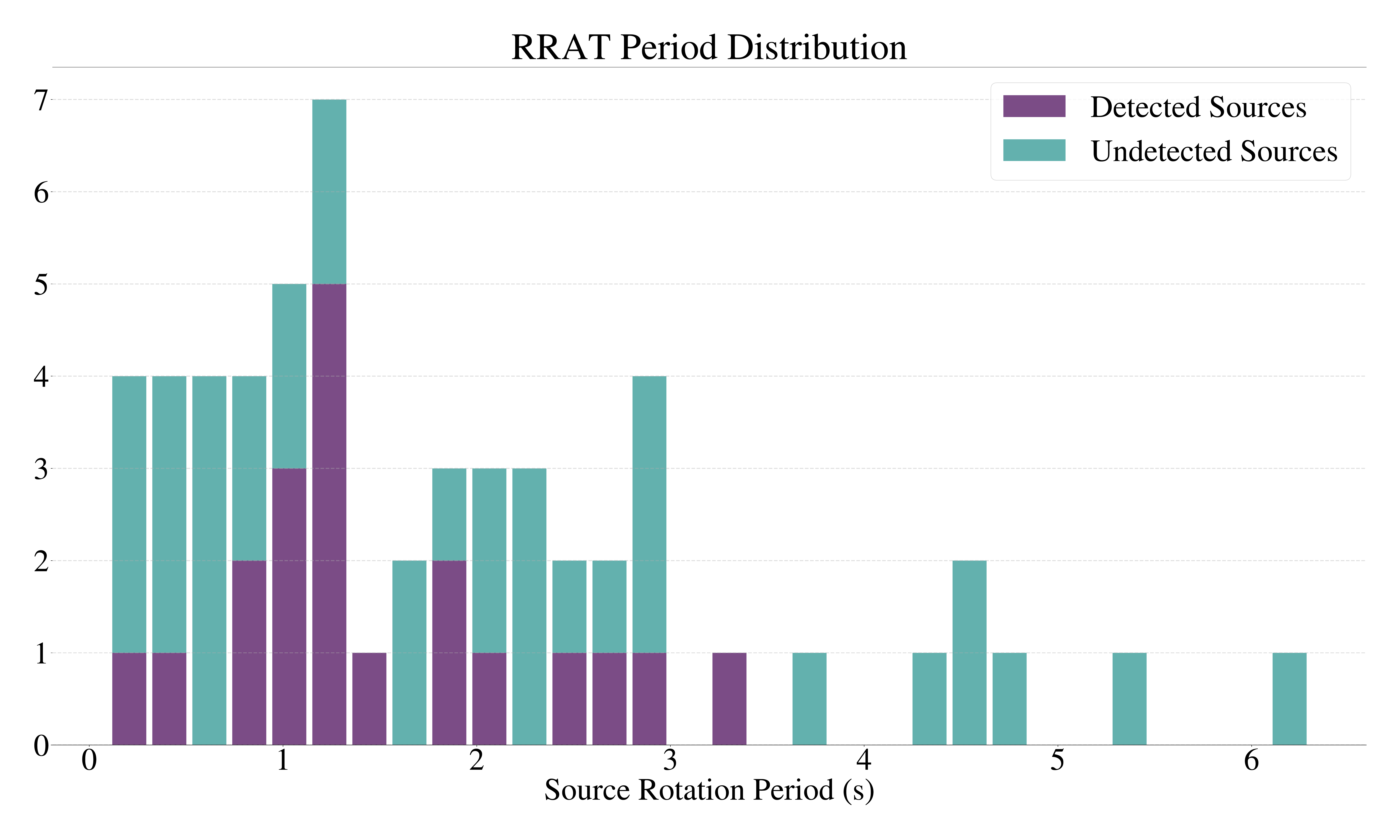}
    \caption{A histogram of source rotation periods, separating the sources that were detected (\SIrange{0.22}{3.45}{\second}) and not detected (\SIrange{0.15}{6.4}{\second}) as a part of this census. The group of sources without a known period has been placed on a separate axis on the left due to the order of magnitude difference between this bin and the remaining sources. The 0-period bin also contains sources that periods were determined for as a part of this census. See further discussion in~\S\ref{sec:discussproperties}.}
    \label{fig:rrat_periods}
    \end{minipage}\hspace{0.9cm}
    \begin{minipage}[t]{10.5cm}
    \centering
    \includegraphics[height=5cm]{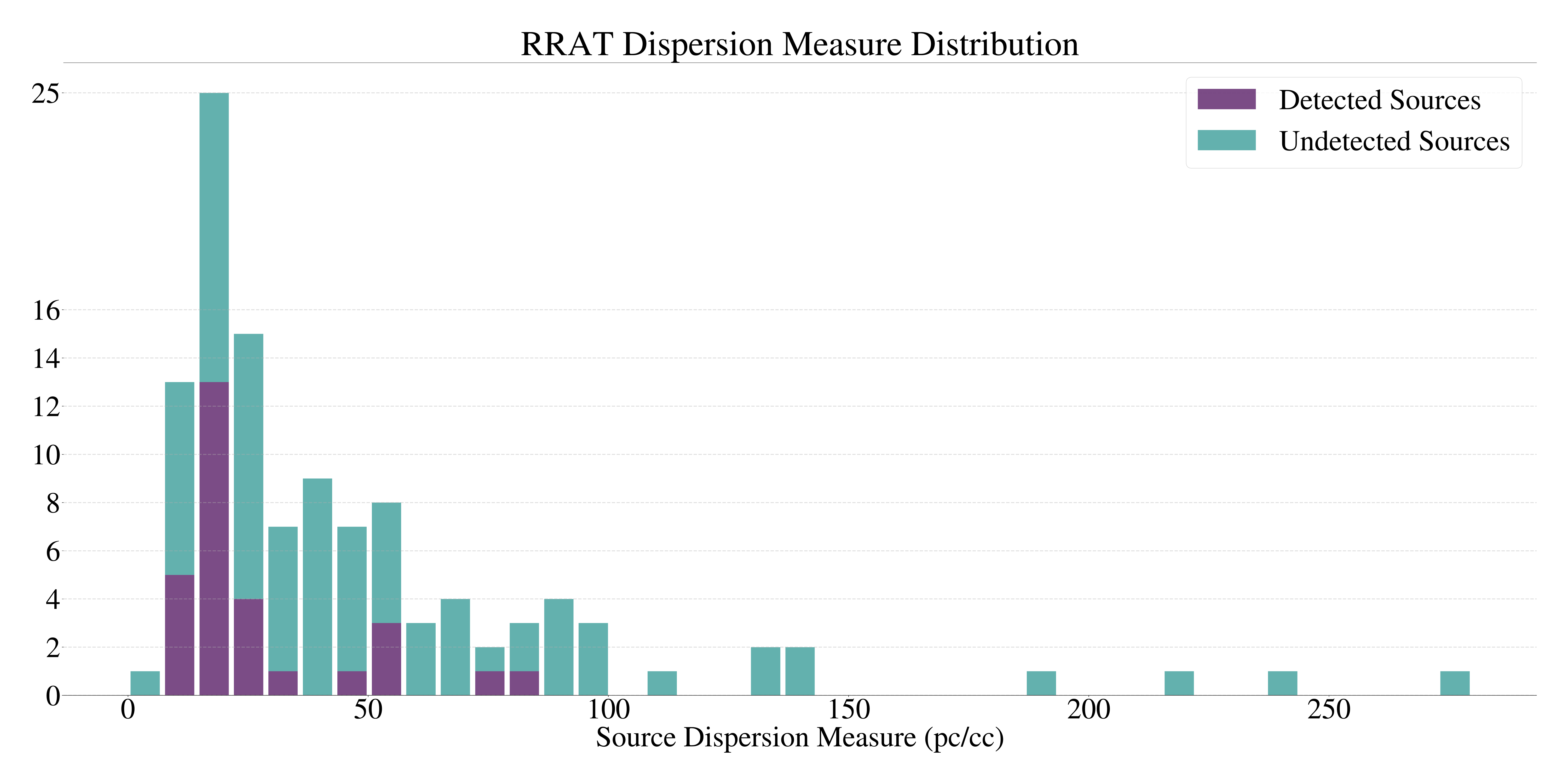}
    \caption{A histogram visualising the distribution of dispersion measures of the sources observed (\SIrange{5}{280}{\parsec\per\centi\metre\cubed}) and detected (\SIrange{8.5}{82.5}{\parsec\per\centi\metre\cubed}) as a part of the census. See further discussion in~\S\ref{sec:discussproperties}.}
    \label{fig:rrat_dms}
    \end{minipage}
    \end{tabular}
\end{figure}
\end{landscape}}

\section{Results}\label{sec:results}

From the $\observed$ sources in this census, $\detected$ sources were detected in some manner. $\singlepulsedetected$ sources were detected through single-pulse emission, while $\periodicdetected$ produced sufficient emission to be detected as a periodic source in at least one observation during the census or further follow-up observations. Only $\consistentperiodicdetected$ sources were found to produce consistent periodic emission during the census.

The detected sources cover dispersion measures from \SIrange{8.5}{82.5}{\parsec\per\centi\metre\cubed} and rotation periods over the range of \SIrange{0.354}{3.92}{\second}. The ratio of the maximum to the minimum observed peak brightness of the pulses is found to exceed an order of magnitude in 6 sources. For all the sources detected in this census, there are likely to be dimmer pulses, undetectable below the noise floor of the instrument. Thus, all the single-pulse-detected pulsars likely meet the criterion to be considered giant-pulse emitters~\citep{2004IAUS..218..315J}.

The source names used in this work may differ from their catalogue entries. The names used in this section and beyond will refer to their names in \texttt{psrcat}, or the updated names as a result of timing with I--LOFAR (see Table~\ref{tab:newephemerides}) or further cited publications. A map between names used can be found in Table~\ref{tab:sourcemappings}.

The single-burst behaviour of detected sources is discussed in ~\S\ref{sec:singlepulseres} and periodic emission is discussed in ~\S\ref{sec:periodicres}. Novel source periods are noted in ~\S\ref{sec:newperiod}. An overview of new properties determined from performing pulsar timing on these sources to-date will be described in ~\S\ref{sec:timnigres}. Some notes regarding inspection of LOTAAS archival pointings are discussed within each relevant section.

\subsection{Sources with Single-Pulse Emission}\label{sec:singlepulseres}
From the $\singlepulsedetected$ sources with single-pulse emission, $\singlepulselofarnovel$ are novel to LOFAR detections, having not been discussed in prior literature. 

Single-pulses from one source previously reported to have single-pulse emission during the LOTAAS survey (\citetalias{michilliSinglepulseClassifierLOFAR2018}), J1404+1159, was not detected as a part of the census. From the 7 sources detected with the full LOFAR core in \citetalias{karako-argamanDiscoveryFollowupRotating2015}, only J0054+69 and J2105+6223 were not detected during this census.

\subsubsection{LOTAAS Archive Reprocessing}
While a number of the detected sources were previously detected and discussed in \citetalias{karako-argamanDiscoveryFollowupRotating2015}, \citetalias{michilliSinglepulseClassifierLOFAR2018} and \citetalias{sanidasLOFARTiedArrayAllSky2019b}, all but one of the $\singlepulselofarnovel$ novel to LOFAR detected sources were within the field of the sky surveyed as a part of the LOTAAS survey (J2325$-$0530, too low in declination), while pointings could not be found for two sources (J0102+5356 and J0348+79). The remainder of the \SI{60}{\minute} observations for each source were downloaded and re-analysed following the methodology described in~\S\ref{sec:processing}. The burst rates of detected sources can be found in column three of Table \ref{tab:lotaas_summary}.

From the novel detections, pointing near J0054+6650, J0746+5514, J1006+3015, J1336+3414, J1400+2125 and J2215+4524 were found to contain at least 1 pulse at the previously described detection criteria. Apart from J1400+2125, these sources were observed to have burst rates in excess of \SI{1}{\per\hour} at I--LOFAR, making the probability of their detection more favourable during the LOTAAS survey. From the non-detected sources, only J0209+33 and J1931+4229 met the same burst rate criteria, with many of the other sources producing bursts extremely infrequently. 

Notably, while J2202+2134 was previously only reported as a source of periodic emission in \citetalias{sanidasLOFARTiedArrayAllSky2019b}, 1 pulse was detected in the archival pointing.

\subsection{Sources With Periodic Emission}\label{sec:periodicres}
Of the $\periodicdetected$ sources that were detected through periodic emission, $\periodicnovellofar$ sources (J0209+5759, J0226+3356, J0854+5449, J2105+19, J2108+4516, J2215+4524, J2325$-$0530) have not previously been discussed in any LOFAR works, while J1538+2345 has only been discussed in a single-pulse context by \citetalias{karako-argamanDiscoveryFollowupRotating2015} (this source is not discussed in \citetalias{sanidasLOFARTiedArrayAllSky2019b} as the survey pointing had not been performed at the time of publication). The remaining 6 sources were reported as novel detections as a part of the LOTAAS survey, or re-detections of known sources in \citetalias{sanidasLOFARTiedArrayAllSky2019b}.

J0209+5759 is a nulling pulsar reported by \citetalias{goodFirstDiscoveryNew2020}, which placed a lower limit on the nulling fraction of the source of 0.21 due to the limited observing windows of the CHIME instrument. I--LOFAR is able to detect blocks of emission after periodic folding at the nominal frequency that align with observed single-pulses. Using PSRSALSA~\citep{WeltevredePSRSALSA2016} to analyse 26 hours of data in 30-second sub-integrations, the nulling fraction can be estimated to be 0.953(4) for a 3-sigma detection threshold, or 0.971(3) for a 4-sigma detection threshold. Poissonian uncertainties are calculated from the number of active mode integrations.

\subsubsection{LOTAAS Archive Reprocessing}\label{sec:archivalmining}
From the 6 sources not previously detected with LOFAR instruments that have nearby LOTAAS pointings, the survey pointings for J0226+3356, J0854+5449 and J2215+4524 show evidence of weak periodic emission, though at signal-to-noise levels that were below the detection threshold for the survey. The signal-to-noise ratios of detected sources can be found in column four of Table \ref{tab:lotaas_summary}.

J2108+4516 was not detected in the LOTAAS pointing despite significant brightness reported in this work, though this is to be expected given \citetalias{goodFirstDiscoveryNew2020} reported this source is an eclipsing binary pulsar, and is only visible during narrow windows of the system's orbit.

The LOTAAS pointing containing CHIME source PSR J0209+5759 showed similar behaviour to observations performed at I--LOFAR, with a nulling fraction of 0.95(2) for a 3-sigma cut-off.

\begin{table}
\centering
    \caption{Inspected LOTAAS pointing properties. In the case that emission is detected in some manner, the highest SNR beam has been noted, otherwise the closest sub-array-pointing is provided. Sources with a note of 'R' indicate that we detected the source in the pointing in a manner that was not previously reported (novel emission is emboldened), 'P' notes there were no nearby pointing in the LOFAR LTA, sources with 'E' were previously reported as detected in a given pointing, but that pointing was not available to be downloaded, and 'N' highlights non-detections.}
	\label{tab:lotaas_summary}
\begin{tabular}{llccl}\hline\hline
Source     & Pointing          & \multicolumn{1}{l}{Bursts} & Periodic      & Note \\ 
 & ID/SAP/BEAM & \SI{}{\per\hour} & $\sigma$ \\
\hline
J0054+6650 & L625106/2/18 & \textbf{12} & No & R \\ % no report
J0102+5356 & -- & -- & -- & \hphantom{R}P \\
J0139+3336 & L526427/0/61 & 17 & No & \\
J0201+7005 & L642095/1/65 & \textbf{7} & 10.1 & R \\ % only periodic reported
J0209+5759 & L441006/0/72 & 0 & \textbf{Nulling} & R \\ % no report
J0226+3356 & L560221/0/59 & 0 & \textbf{5.4} & R \\ % no report
J0317+1328 & L215823/1/15 & -- & -- & \hphantom{RP}E \\ 
J0332+79 & L441036/2 & 0 & No & \hphantom{RPE}N \\
J0348+79 & -- & -- & -- & \hphantom{R}P \\
J0746+5514 & L687604/2/67 & \textbf{4} & No & R \\ % no report
J0854+5449 & L769469/2/40 & 0 & \textbf{4.4} & R \\ % no report
J0939+45 & L468036/0 & 0 & -- & \hphantom{RPE}N\\
J1006+3015 & L452116/1/44 & \textbf{1} & No & R \\ % no report
J1132+2513 & L347106/1/47 & 0 & 18.1 & \\
J1218+47 & L644719/1 & 0 & -- & \hphantom{RPE}N\\
J1329+13 & L568513/1 & 0 & -- & \hphantom{RPE}N\\
J1336+3414 & L347300/0/28 & \textbf{3} & No & R \\ % no report
J1400+2125 & L646389/2/21 & \textbf{1} & No & R \\ % no report
J1404+1159 & L521602/0/54 & 8 & 76.8 & \\
J1538+2345 & L773549/0/32 & \textbf{47} & \textbf{12.5} & R \\ % obs after \citetalias{sanidasLOFARTiedArrayAllSky2019b}
J1848+1516 & L337774/0/17 & 51 & 20.4 & \\
J1931+4229 & L543465/1 & 0 & No & \hphantom{RPE}N\\
J2105+19 & L640743/1 & 0 & No & \hphantom{RPE}N\\
J2108+4516 & L651504/0 & 0 & No & \hphantom{RPE}N\\
J2138+69 & L646349/1 & 0 & -- & \hphantom{RPE}N\\
J2202+2134 & L663798/2/70 & \textbf{1} & 21.6 & R \\ % only periodic reported
J2215+4524 & L599673/0/58 & \textbf{19} & \textbf{8.6} & R \\ % no report
J2325$-$0530 & -- & -- & -- & \hphantom{R}P \\
J2355+1523 & L691558/2 & 0 & No & \hphantom{RPE}N\\ \hline\hline

\end{tabular}
\end{table}

\begin{figure*}
    \centering
    \begin{tabular}{cc}
\includegraphics[width=3.62cm]{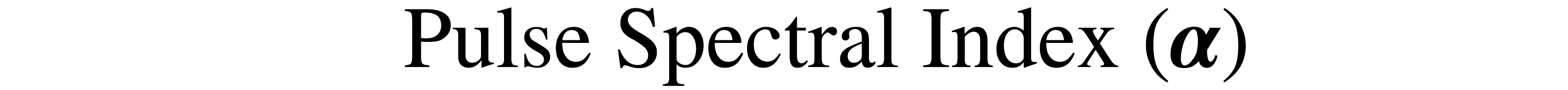}\hspace{0.1cm} \includegraphics[width=3.62cm]{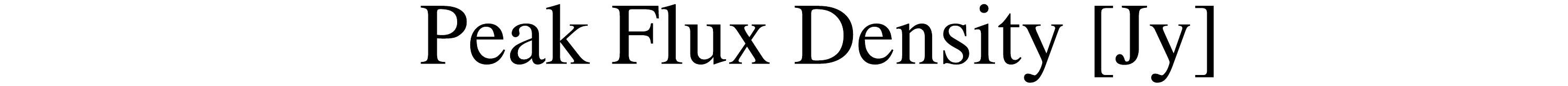} &   \includegraphics[width=3.62cm]{Plots/spectralFits/spectralFitAxis.pdf}\hspace{0.1cm} \includegraphics[width=3.62cm]{Plots/spectralFits/fluxDensityAxis.pdf} \\

\includegraphics[width=3.62cm]{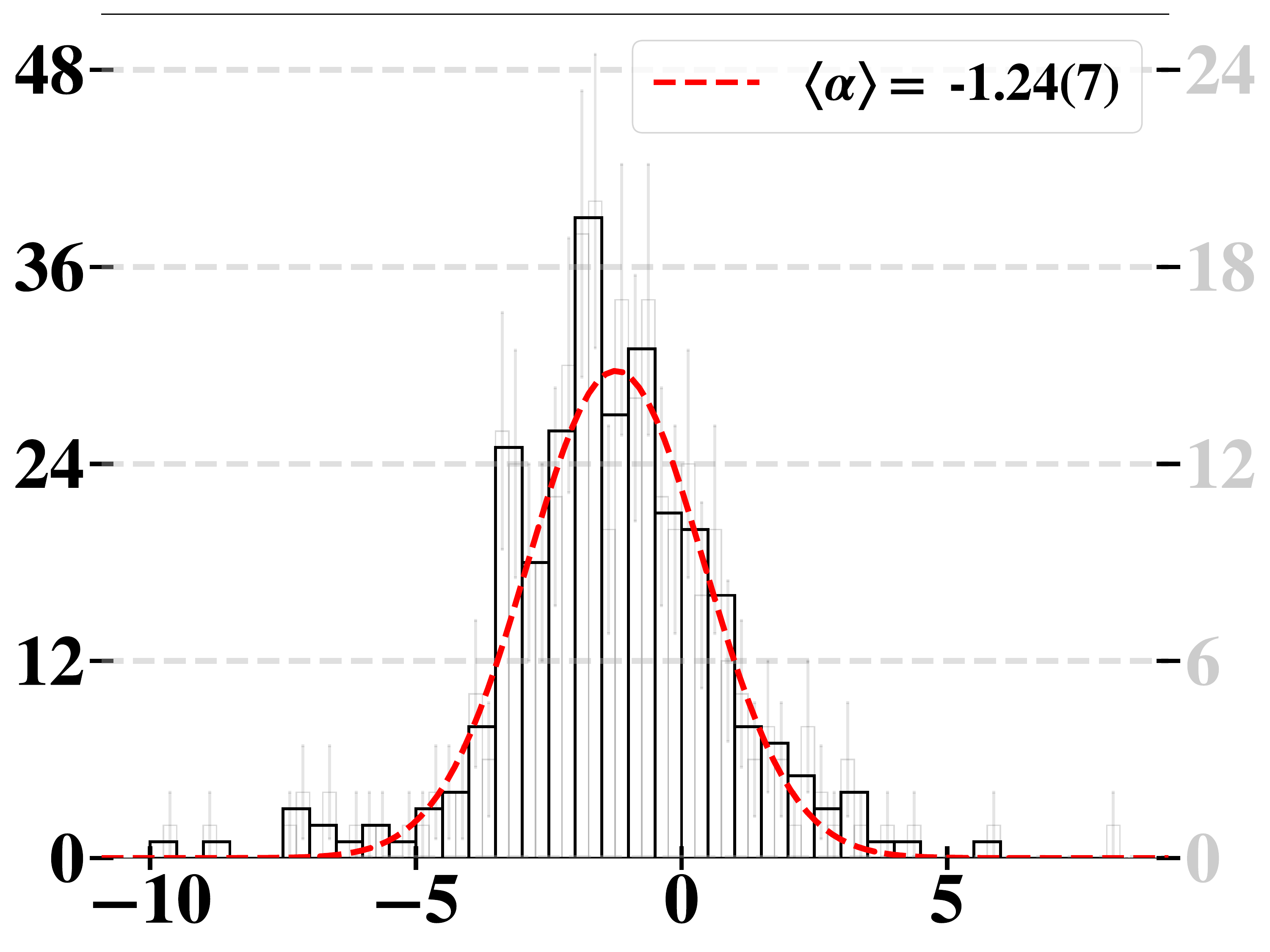}\hspace{0.1cm} \includegraphics[width=3.62cm]{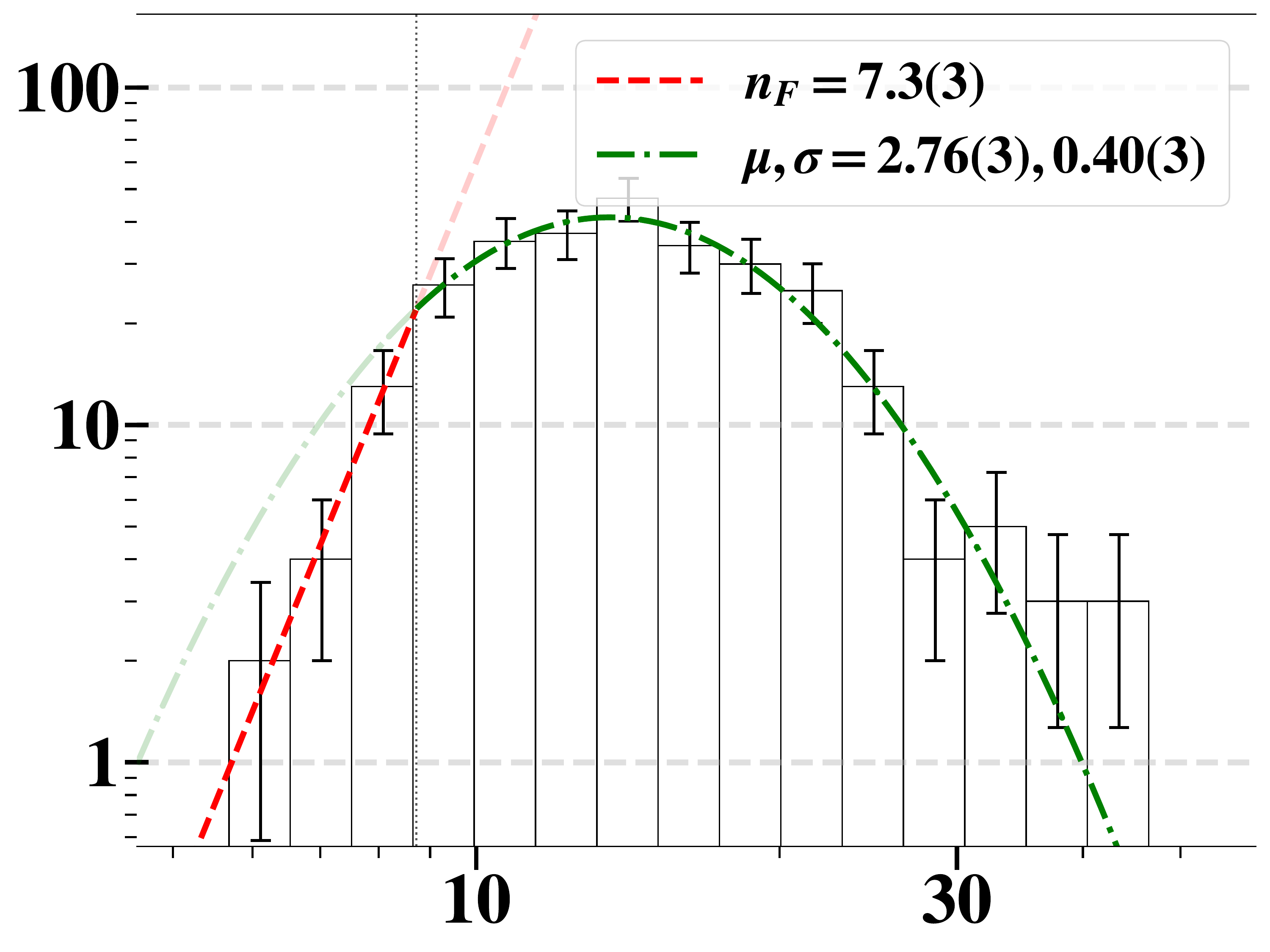} &   \includegraphics[width=3.62cm]{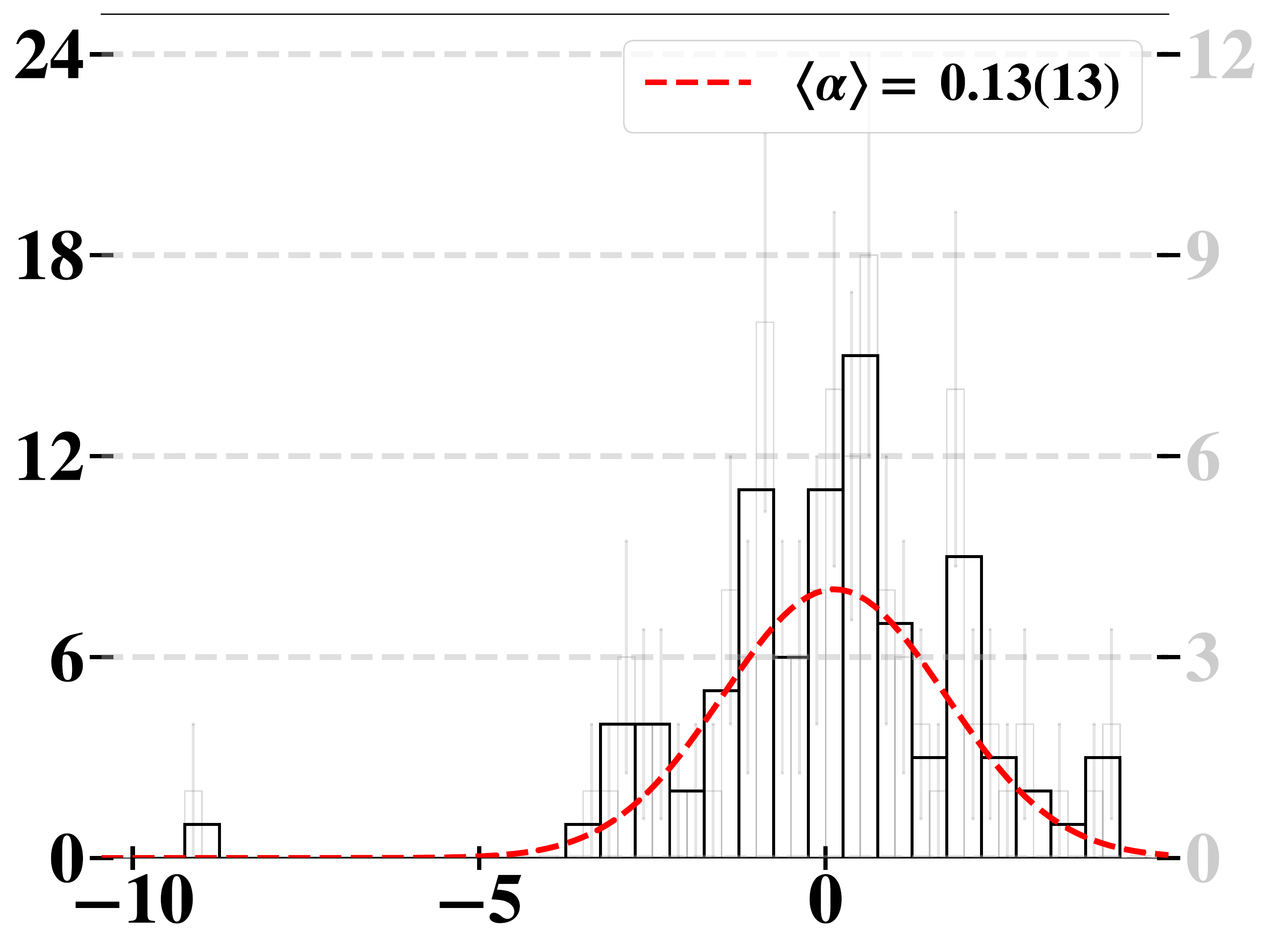}\hspace{0.1cm} \includegraphics[width=3.62cm]{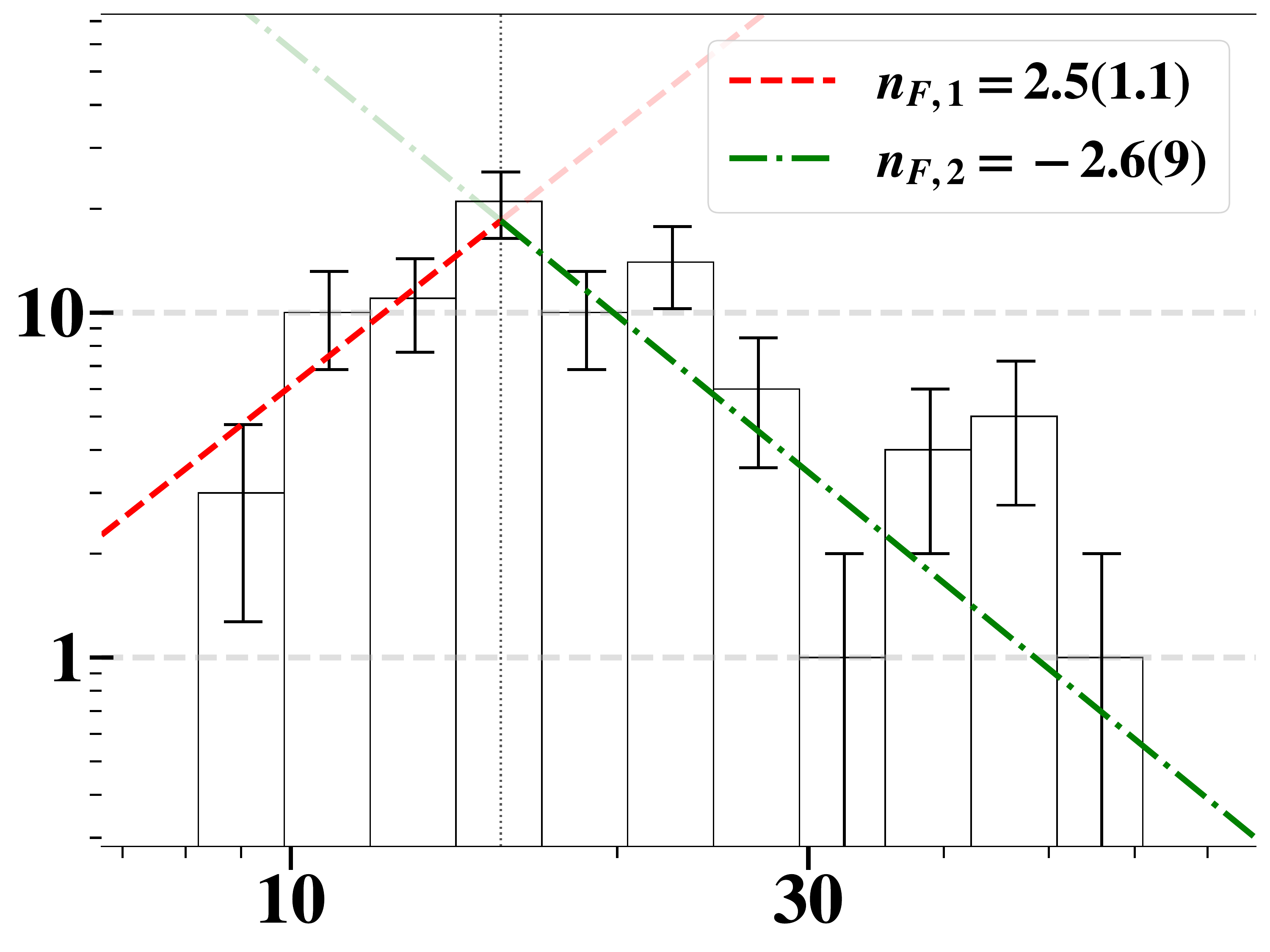}\vspace{-0.15cm}\\

(a) J0054+6650 & (b) J0102+5356 \vspace{0.13cm}\\

\includegraphics[width=3.62cm]{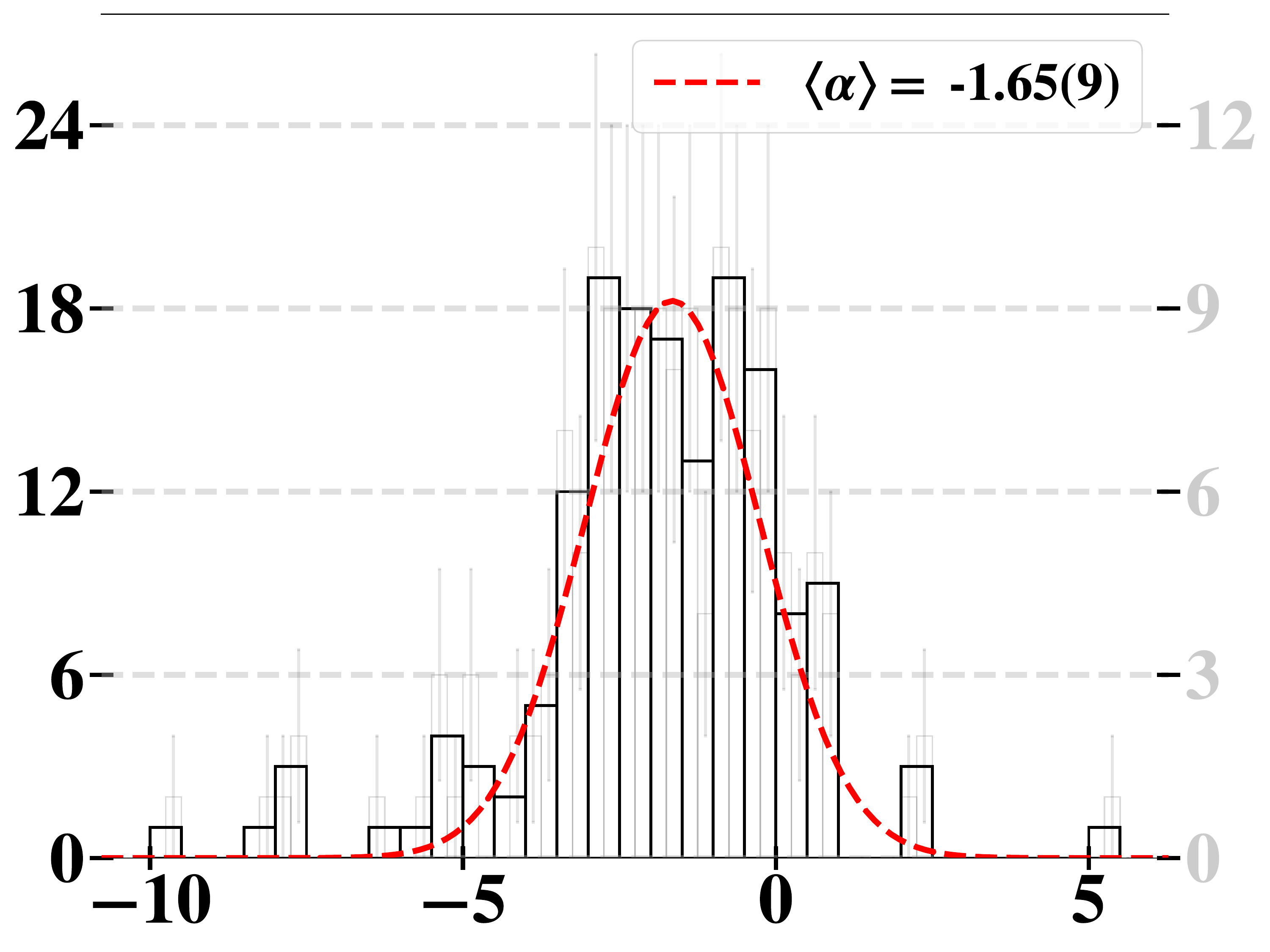}\hspace{0.1cm} \includegraphics[width=3.62cm]{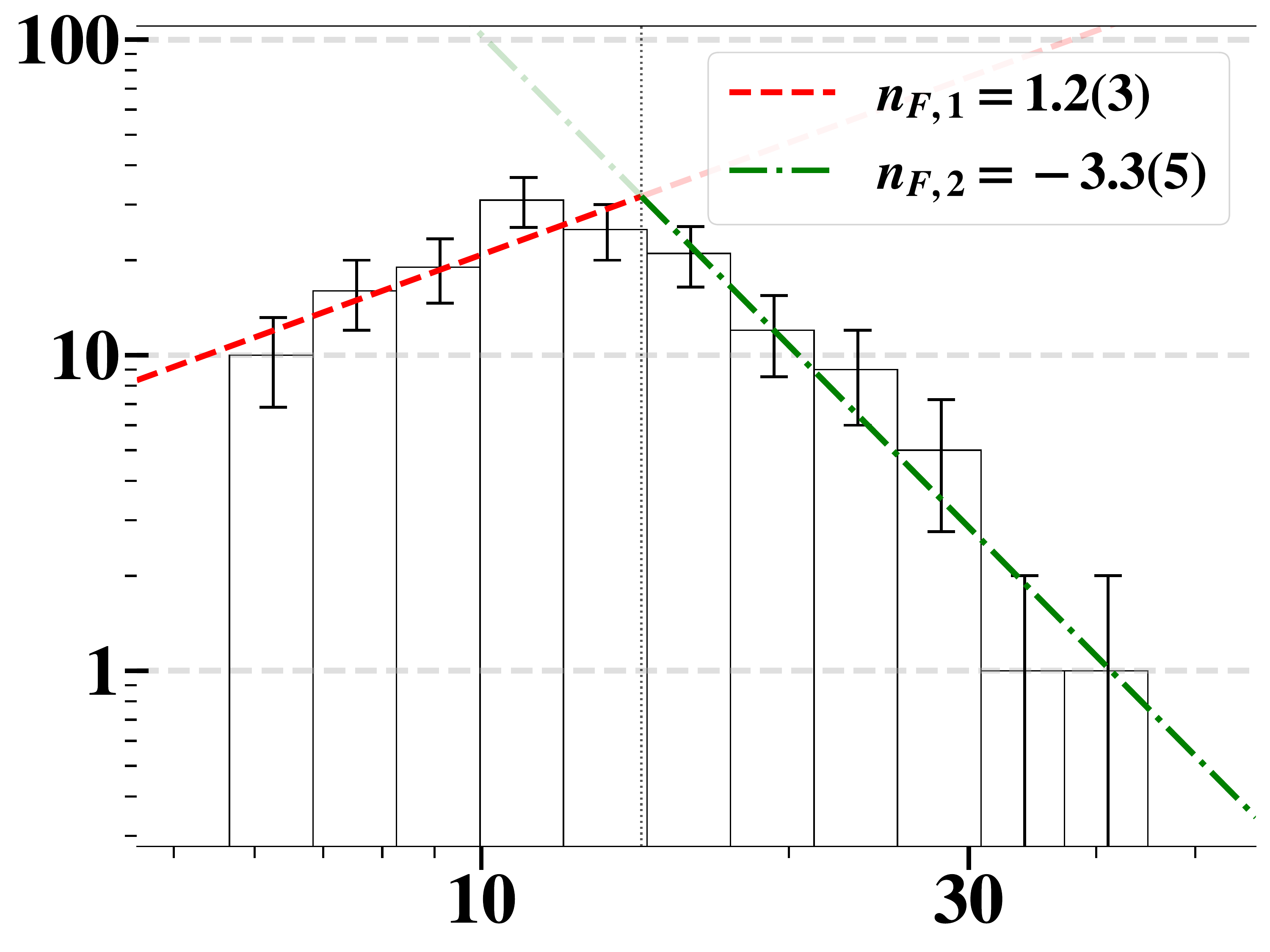} &   \includegraphics[width=3.62cm]{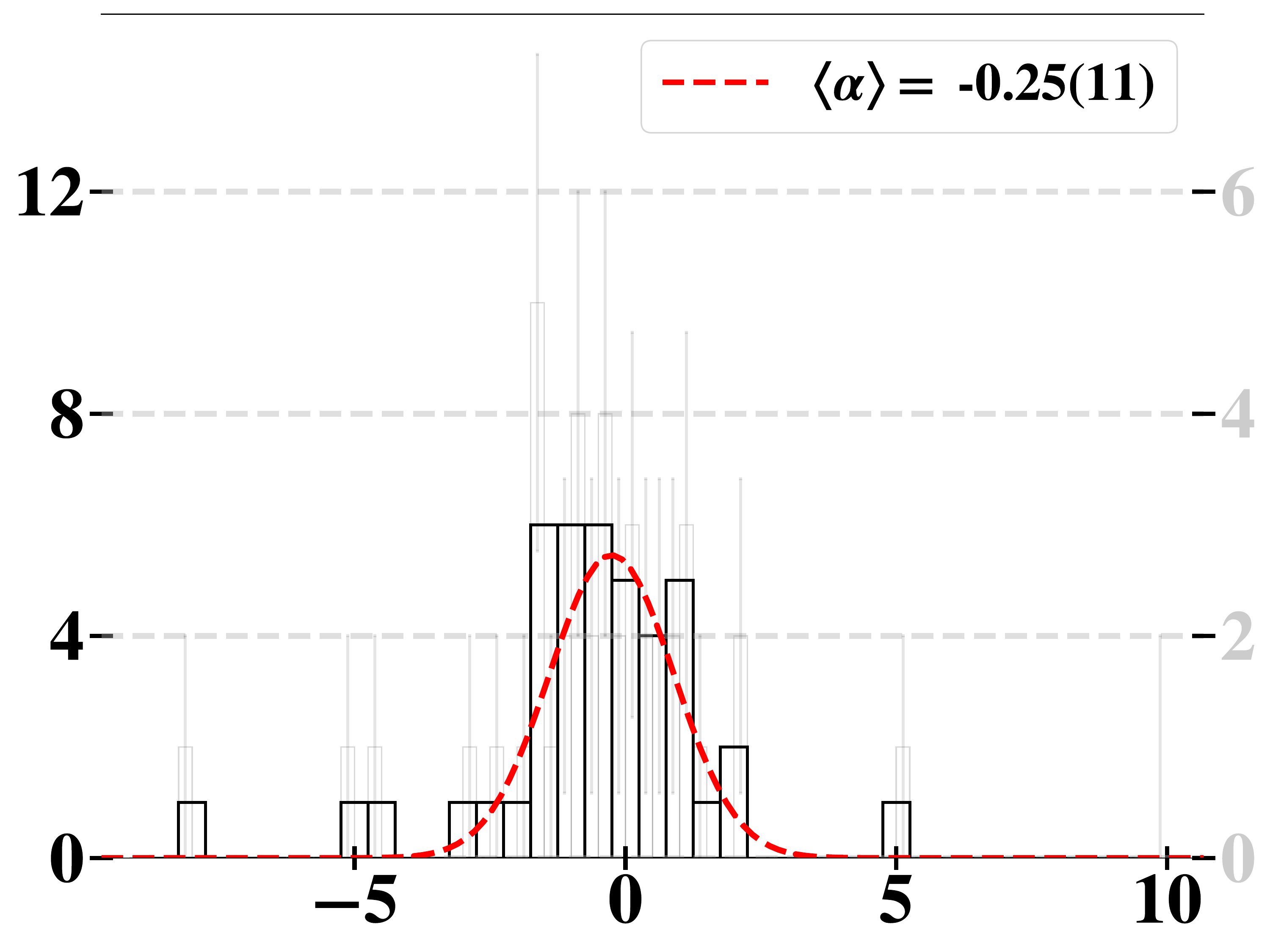}\hspace{0.1cm} \includegraphics[width=3.62cm]{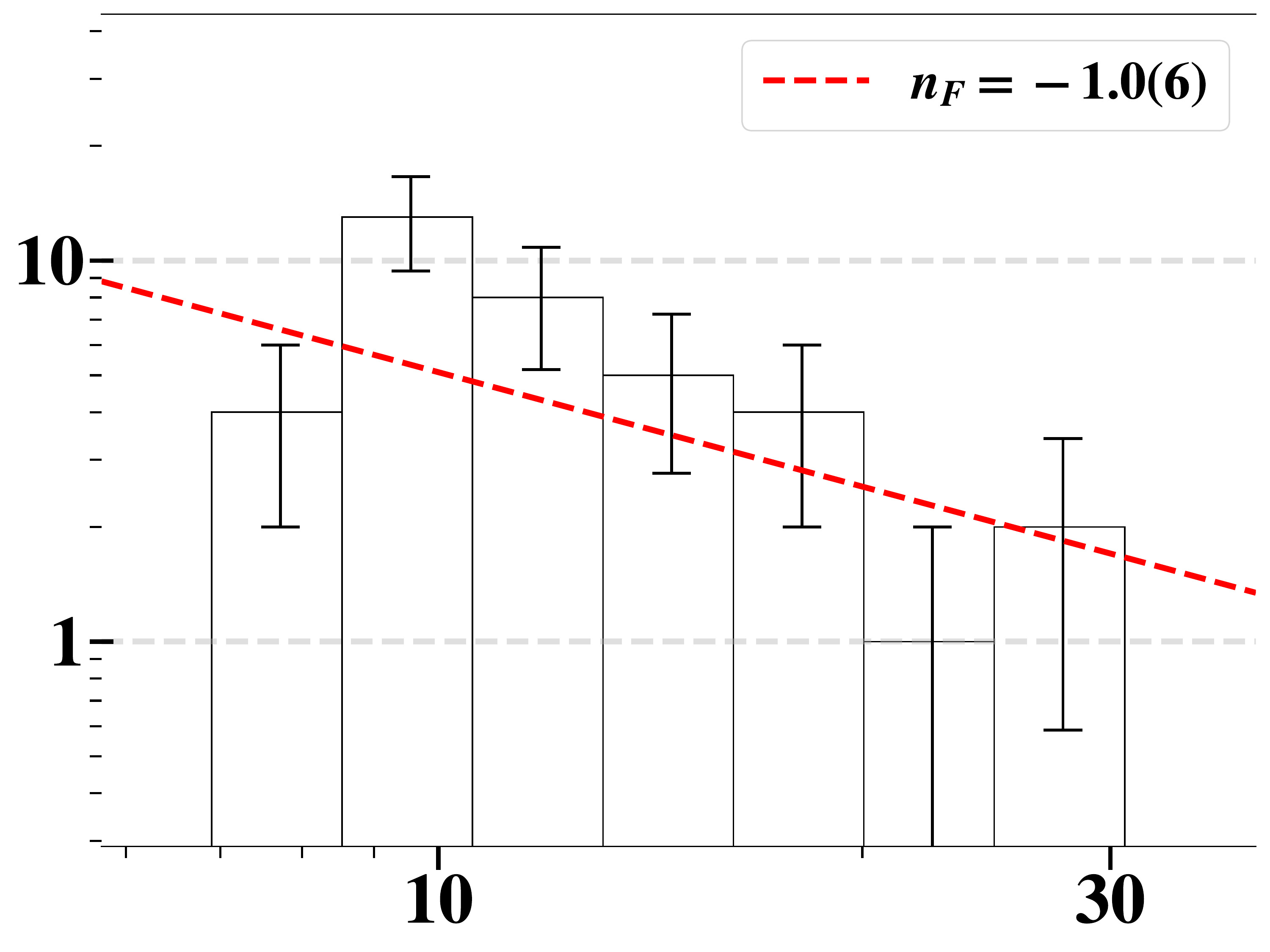}\vspace{-0.15cm}\\

(c) J0139+3336 & (d) J0209+5759 \vspace{0.13cm}\\

\includegraphics[width=3.62cm]{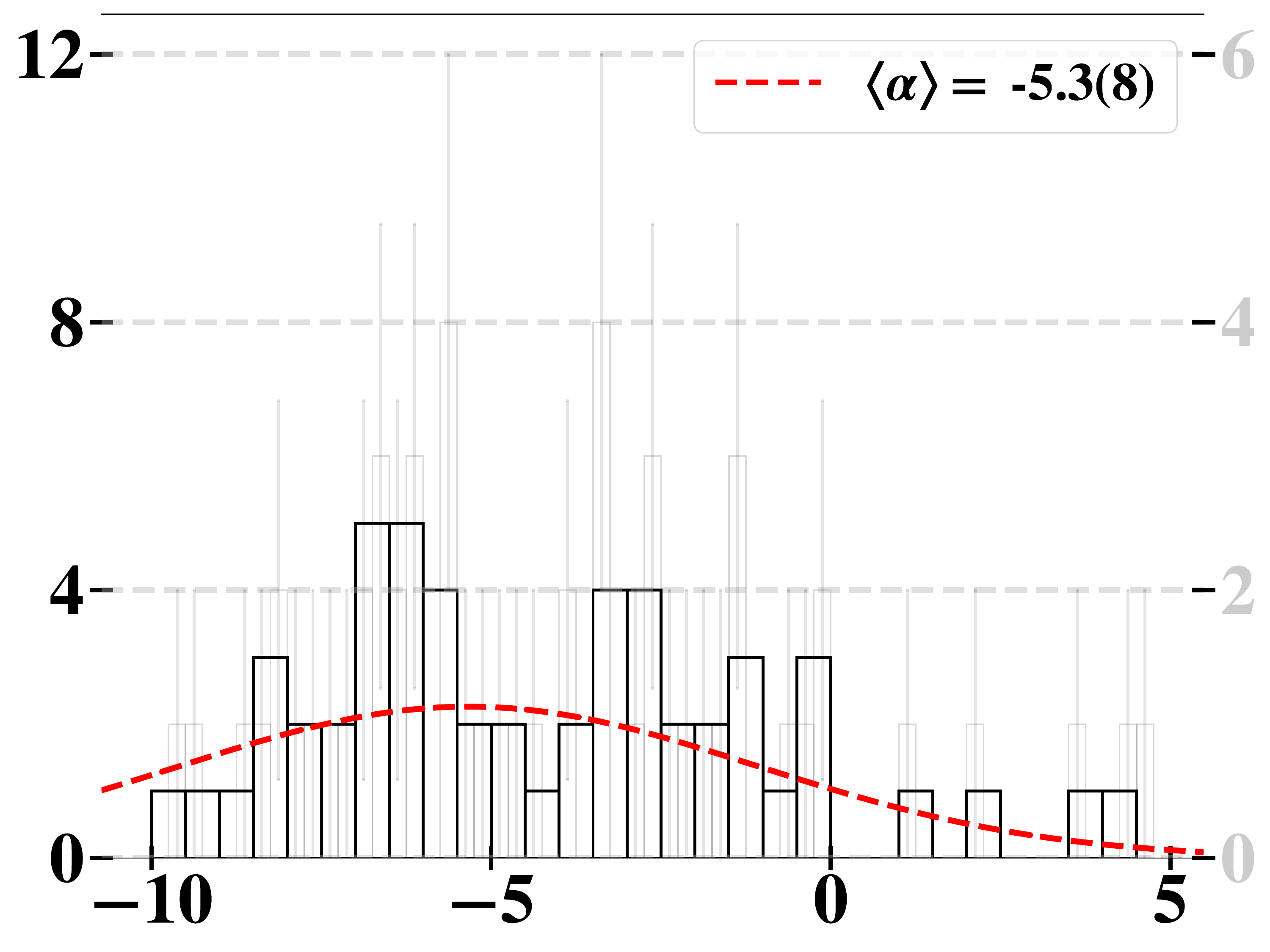}\hspace{0.1cm} \includegraphics[width=3.62cm]{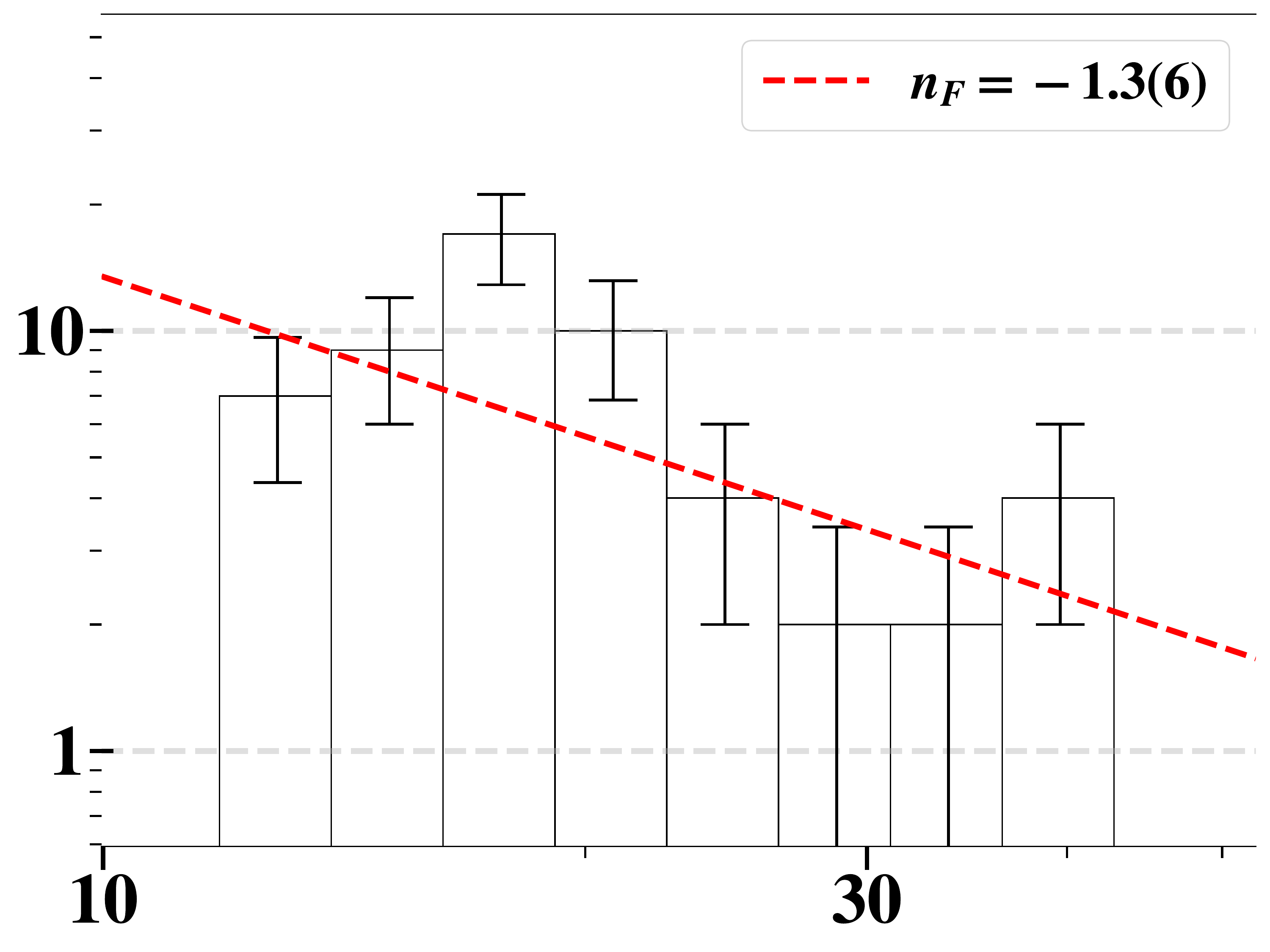} & \includegraphics[width=3.62cm]{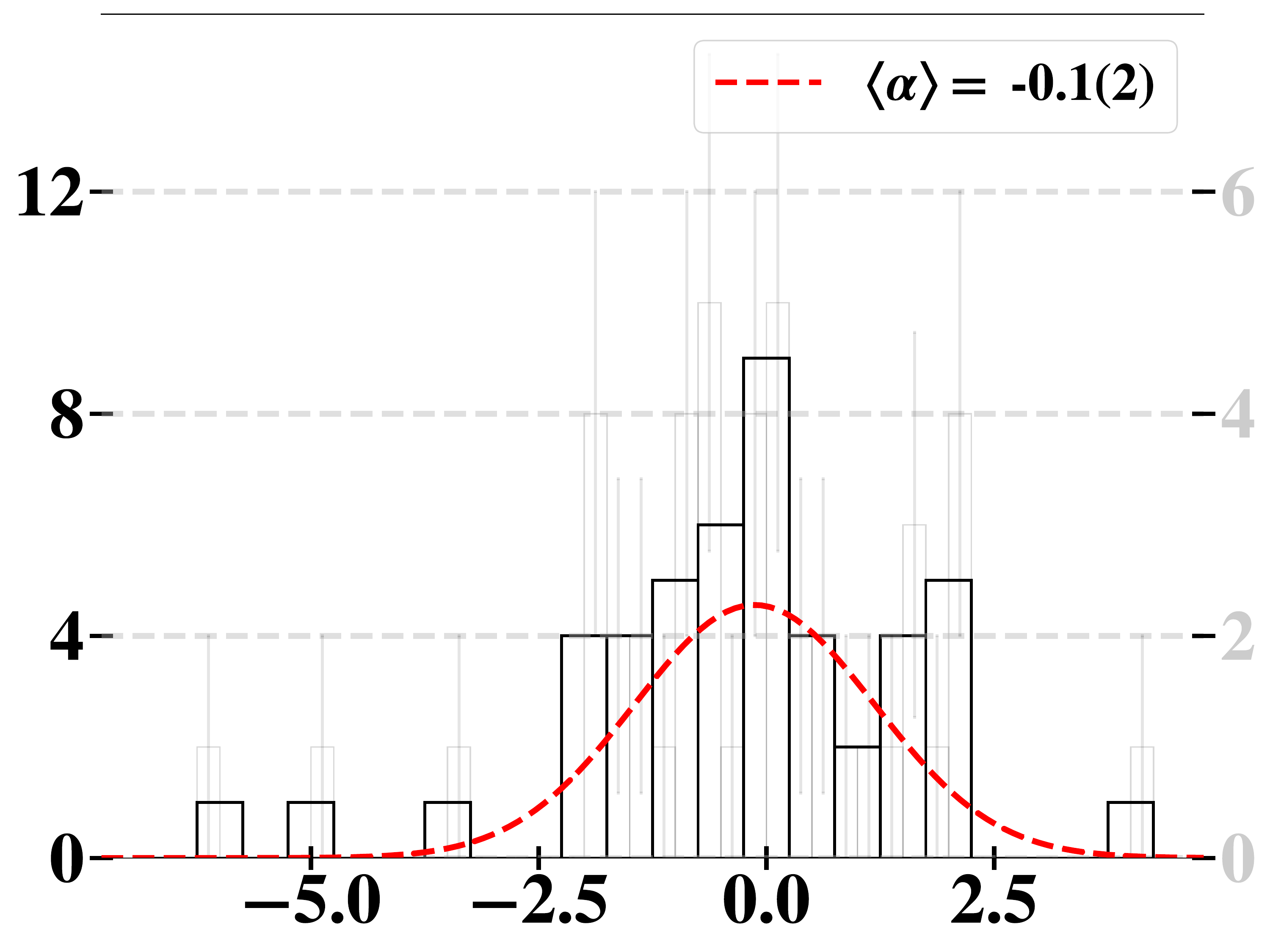}\hspace{0.1cm} \includegraphics[width=3.62cm]{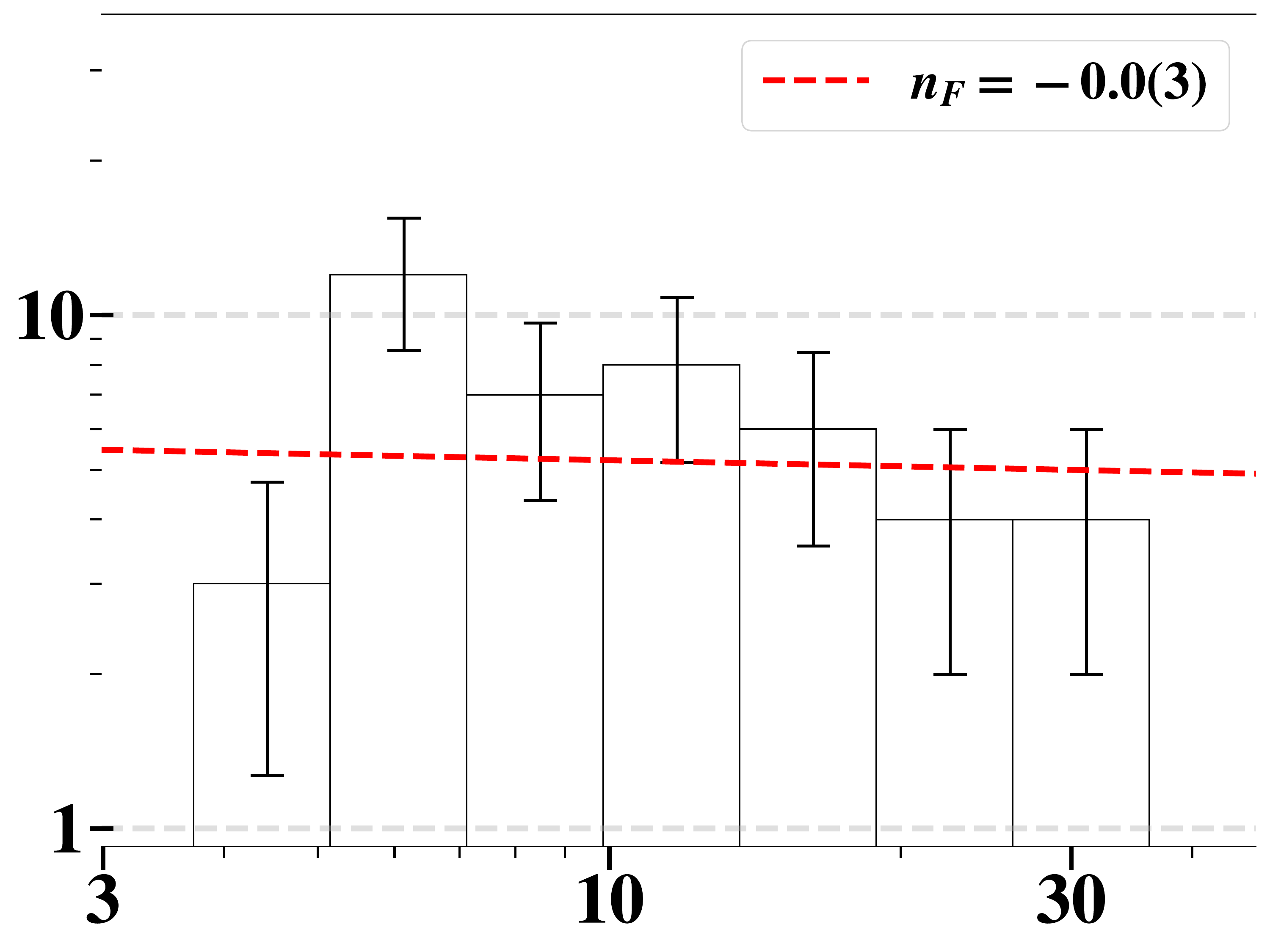}\vspace{-0.15cm}\\

(f) J0317+1328 & (g) J0746+5514 \vspace{0.13cm}\\

\includegraphics[width=3.62cm]{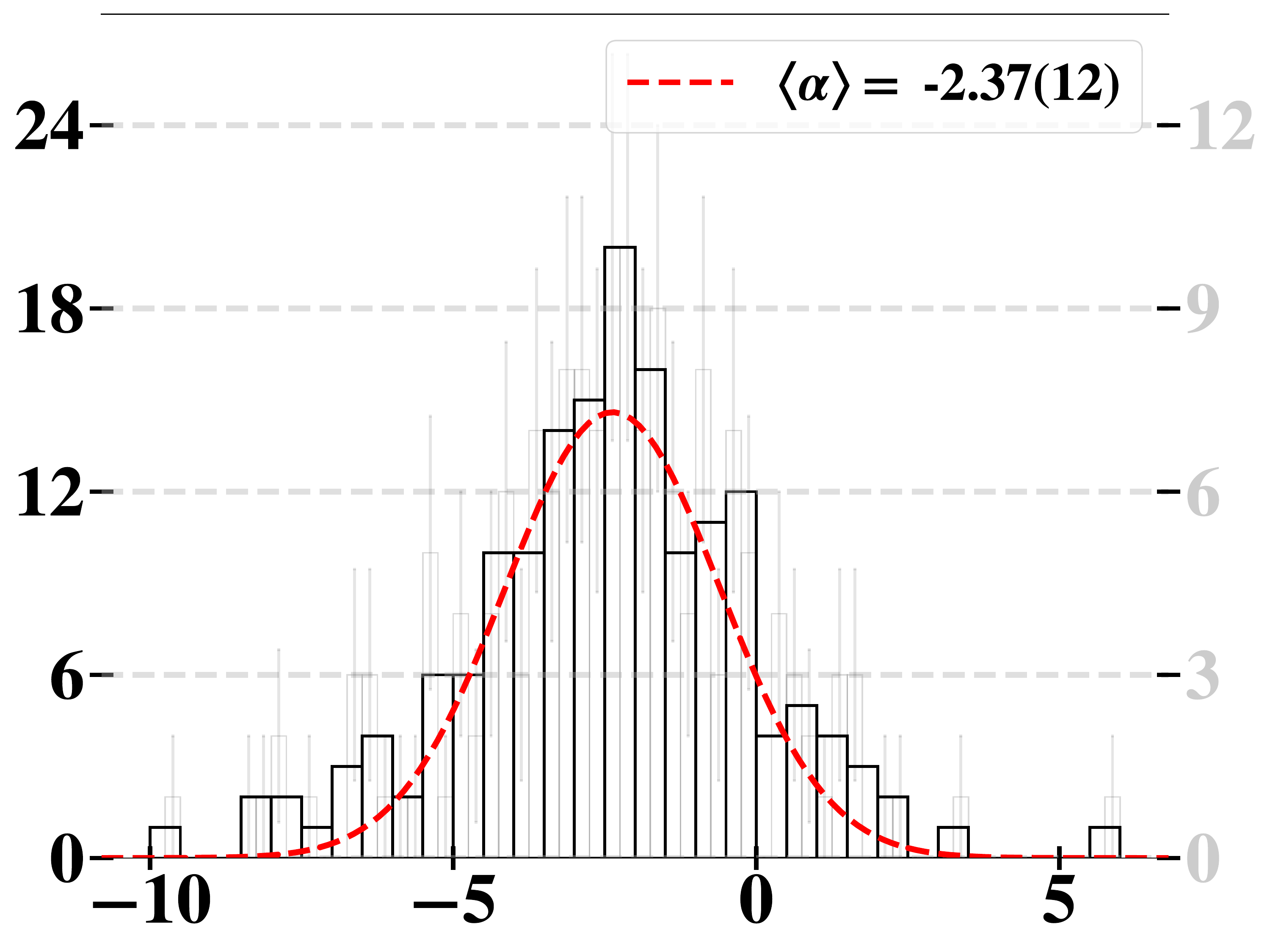}\hspace{0.1cm} \includegraphics[width=3.62cm]{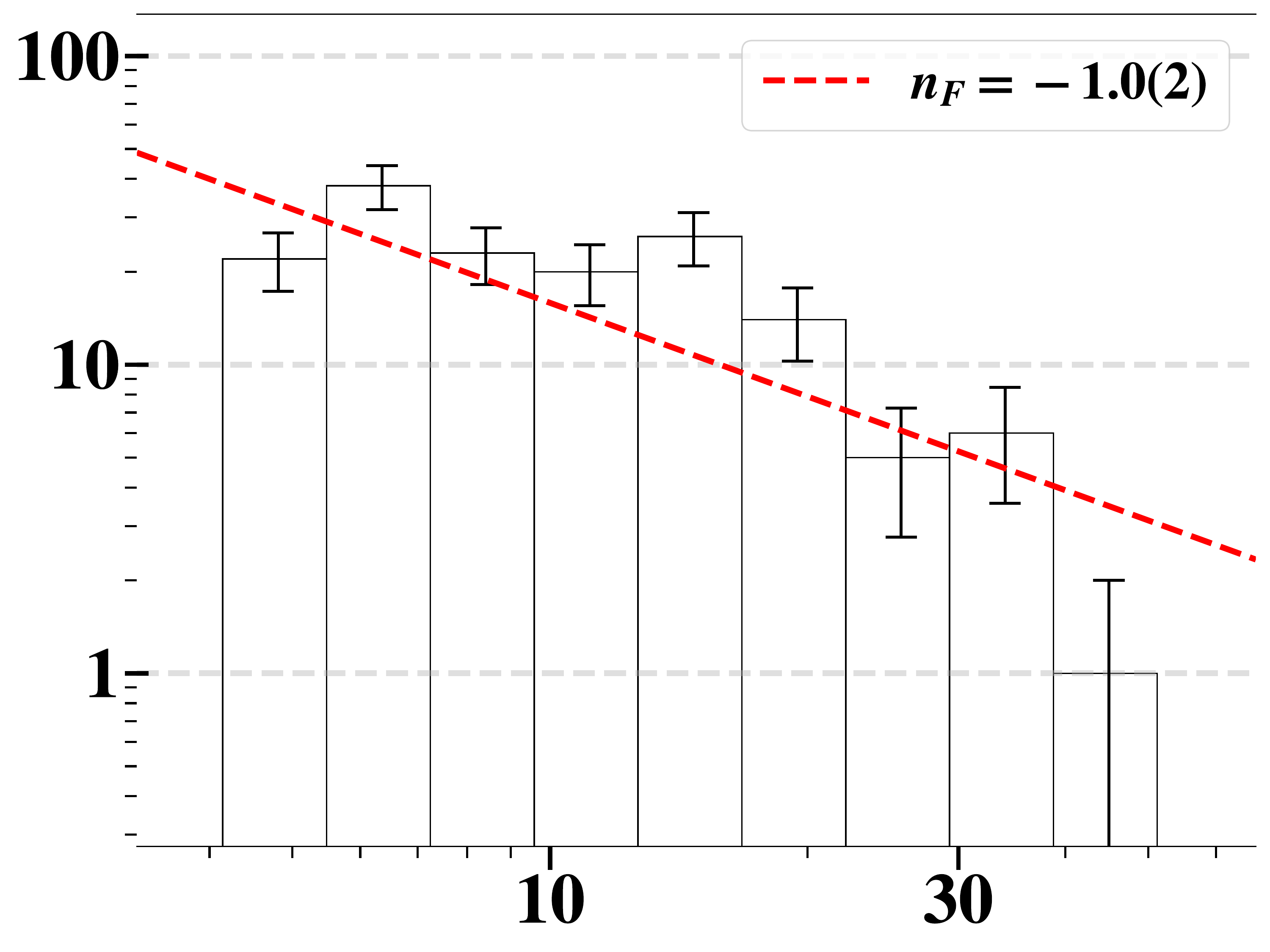} & \includegraphics[width=3.62cm]{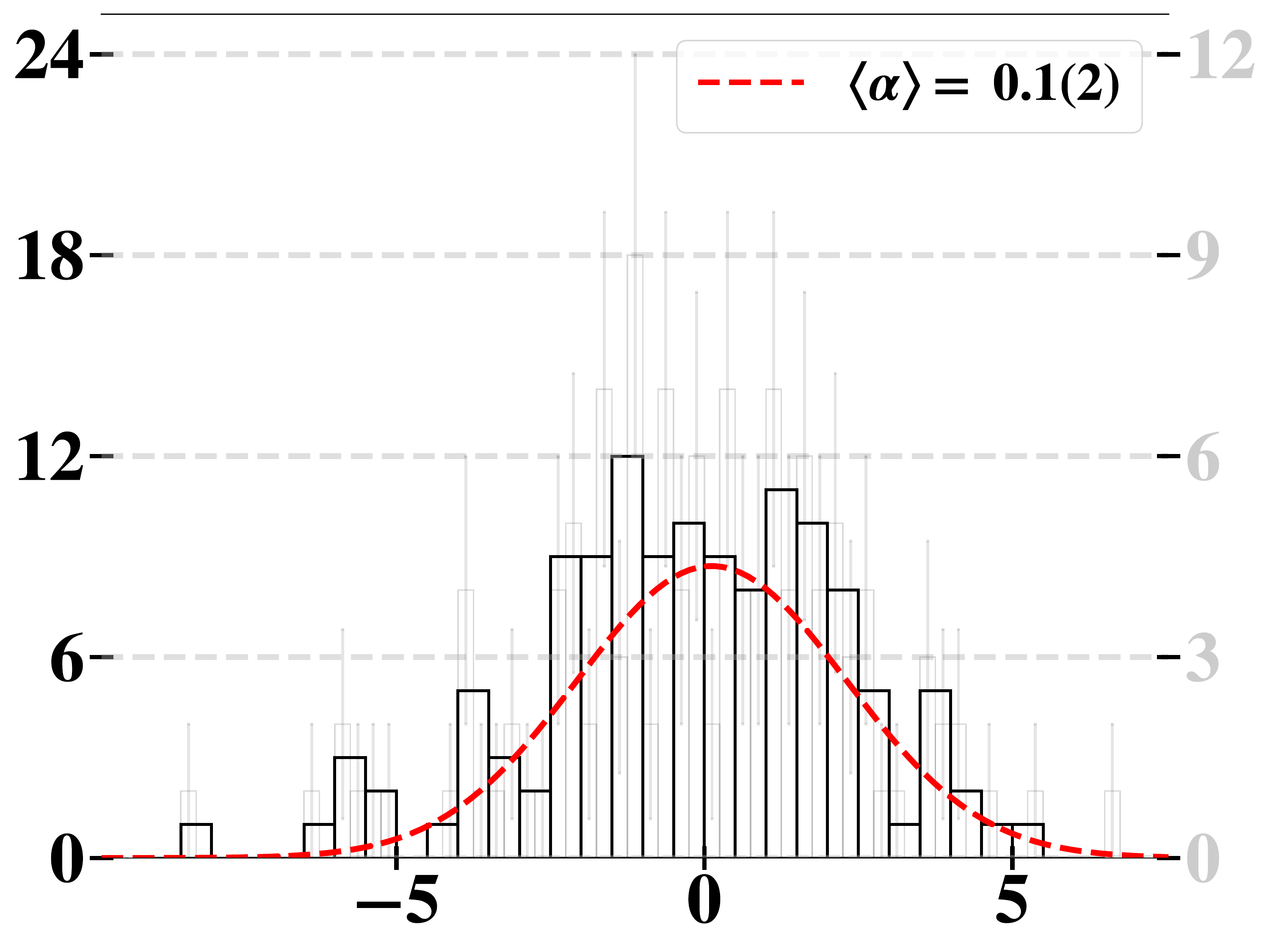}\hspace{0.1cm} \includegraphics[width=3.62cm]{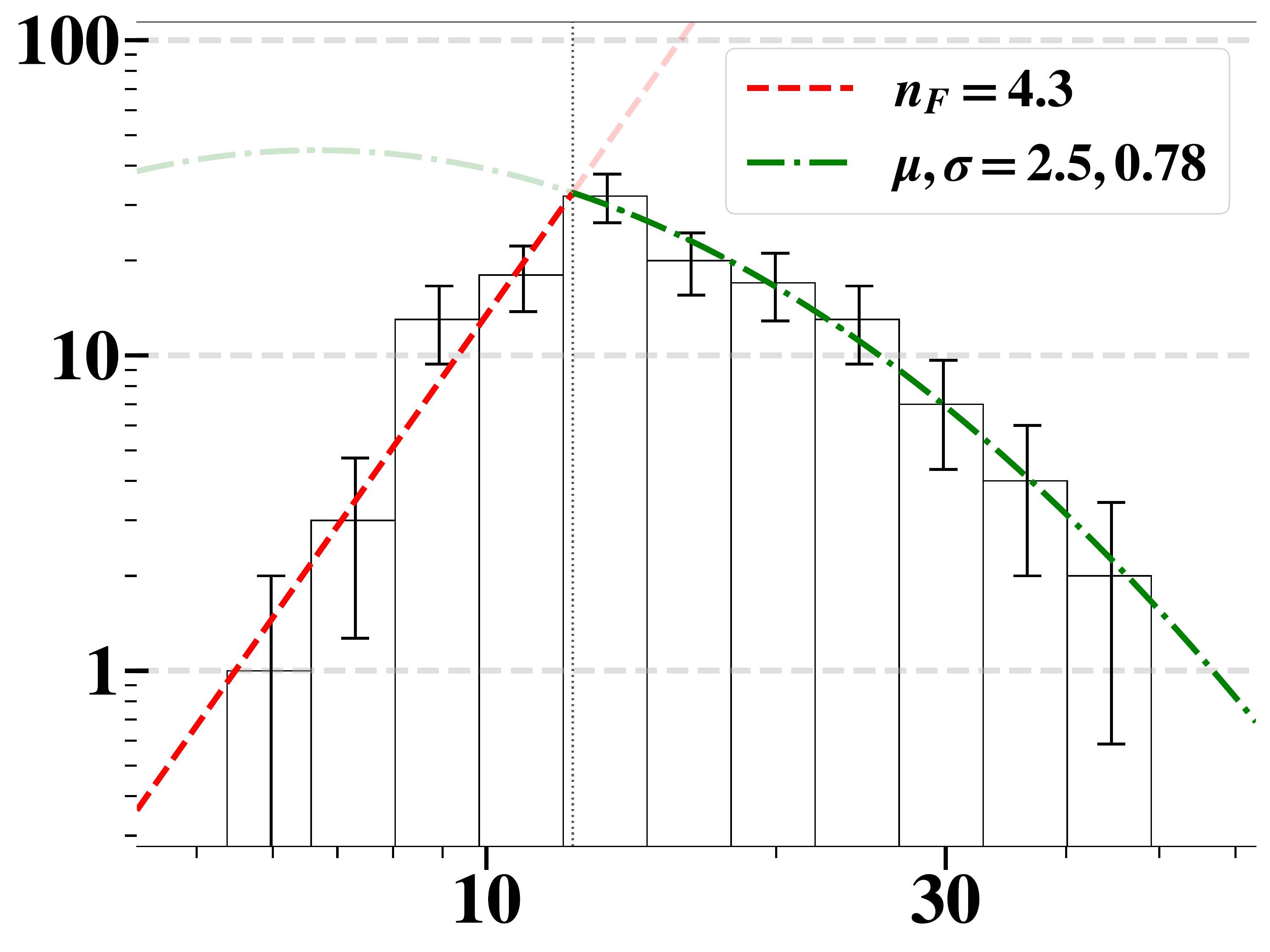}\vspace{-0.15cm}\\

(h) J1006+3015 & (i) J1336+3414 \vspace{0.13cm}\\

\includegraphics[width=3.62cm]{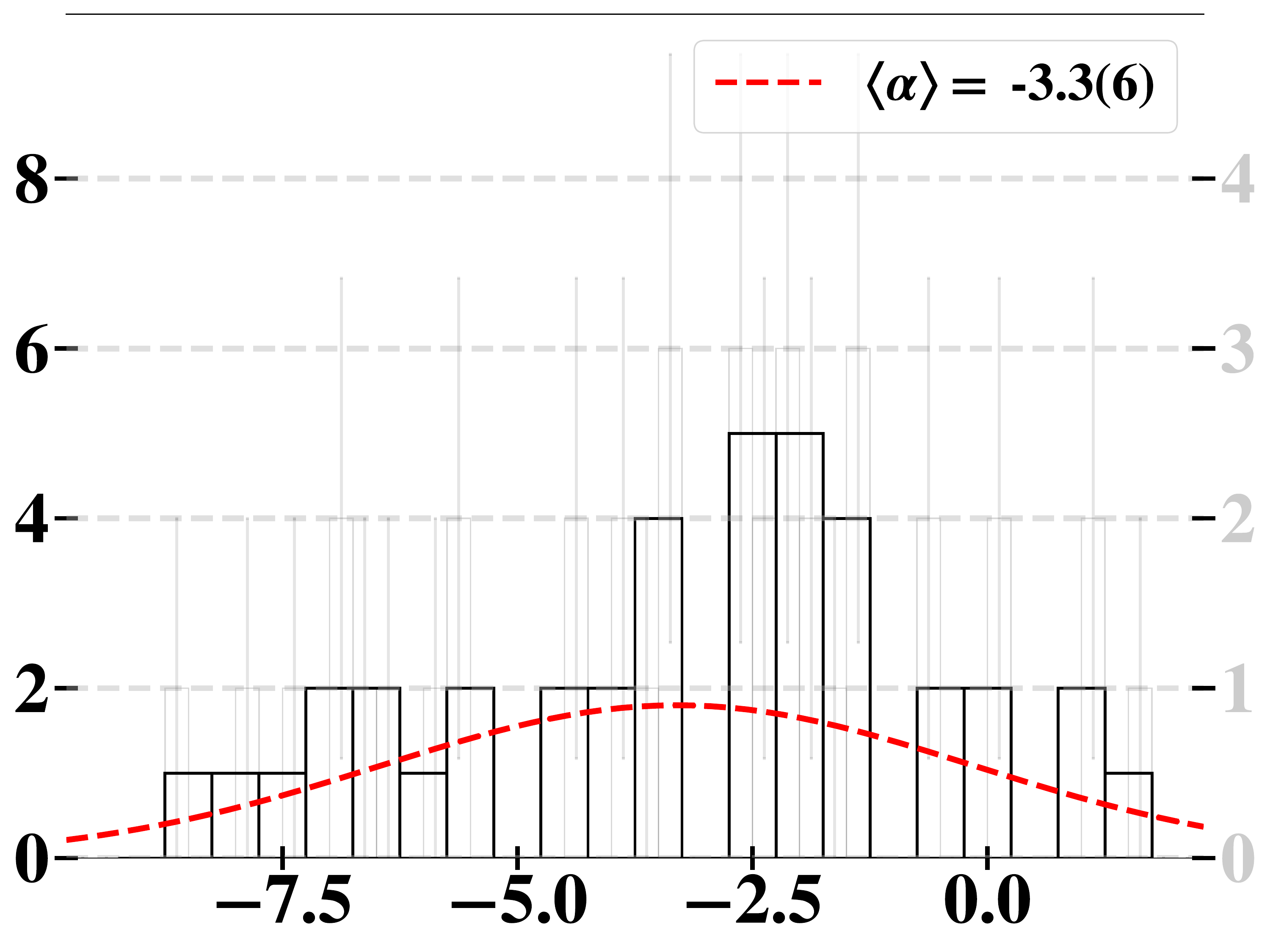}\hspace{0.1cm} \includegraphics[width=3.62cm]{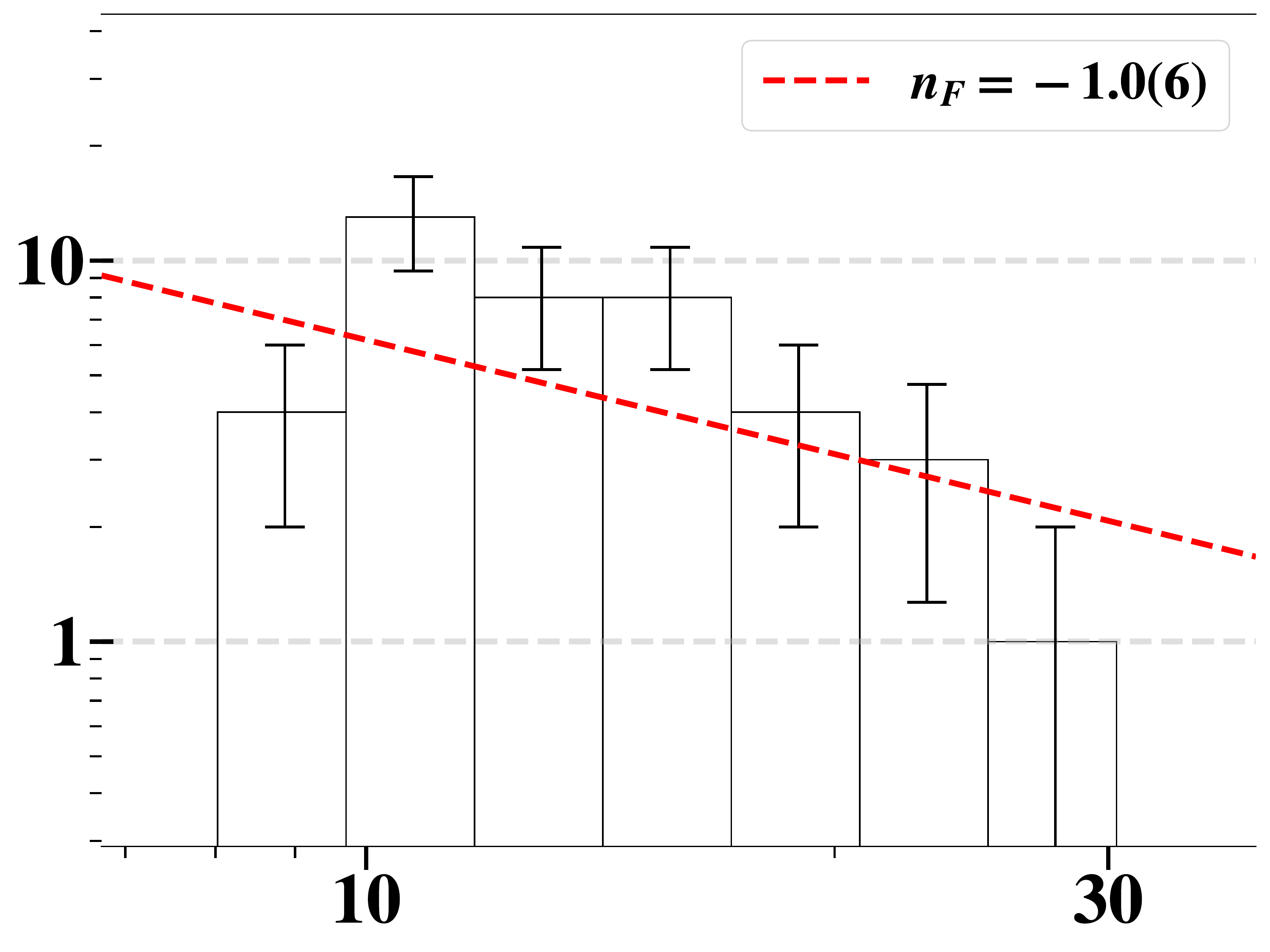} & \includegraphics[width=3.62cm]{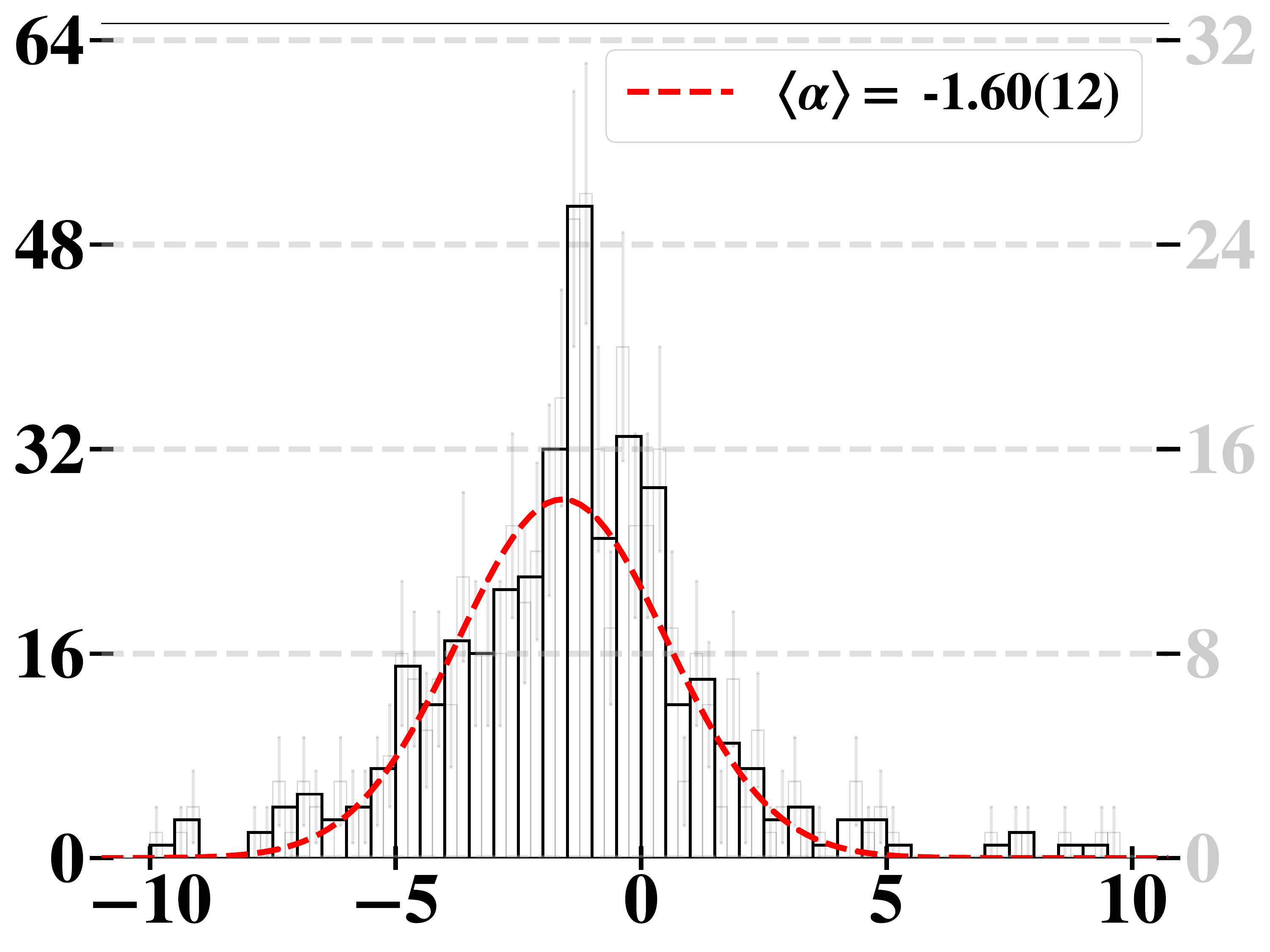}\hspace{0.1cm} \includegraphics[width=3.62cm]{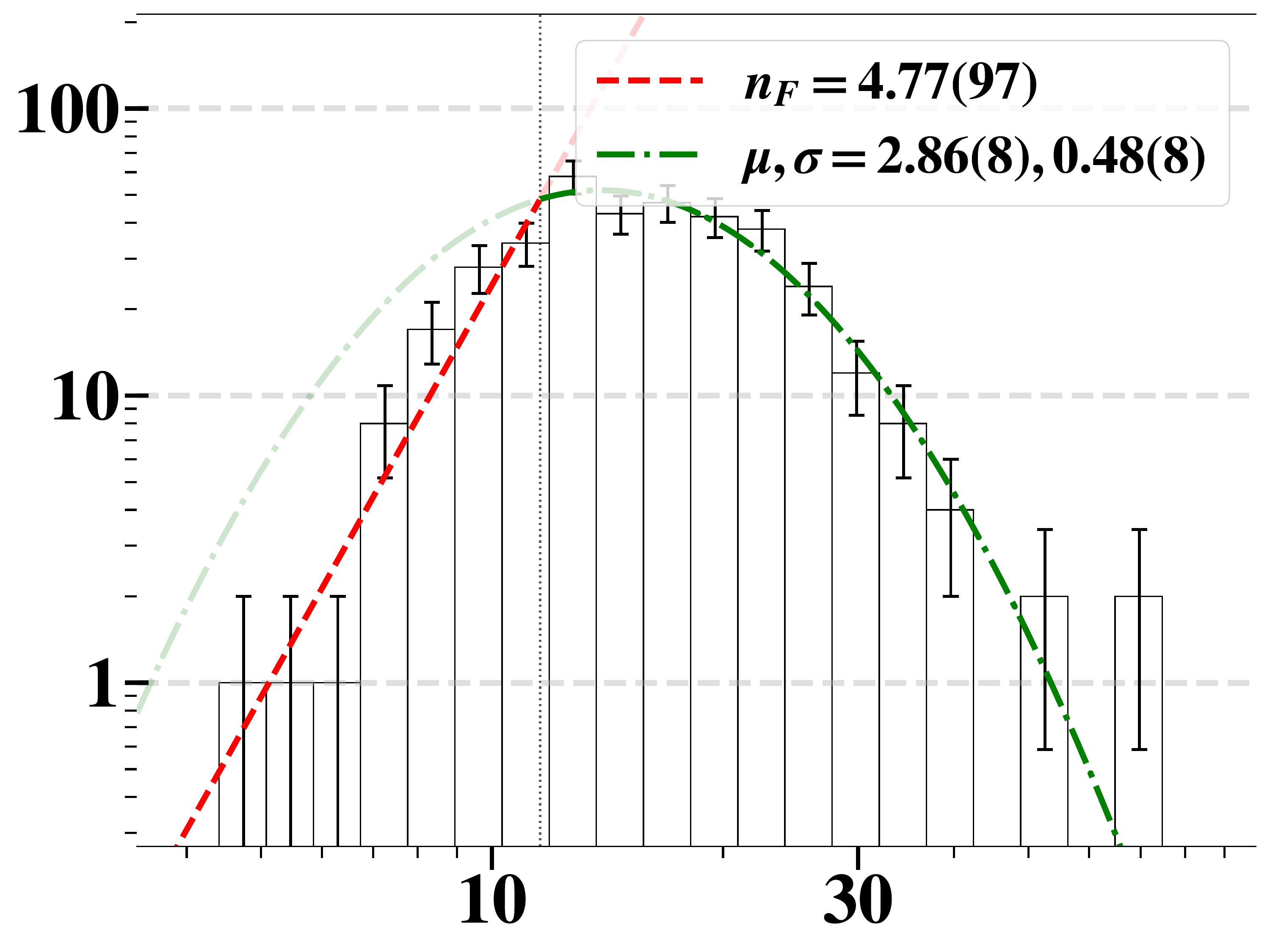}\vspace{-0.15cm}\\

(j) J1400+2125 & (k) J1538+2345 \vspace{0.13cm}\\

\includegraphics[width=3.62cm]{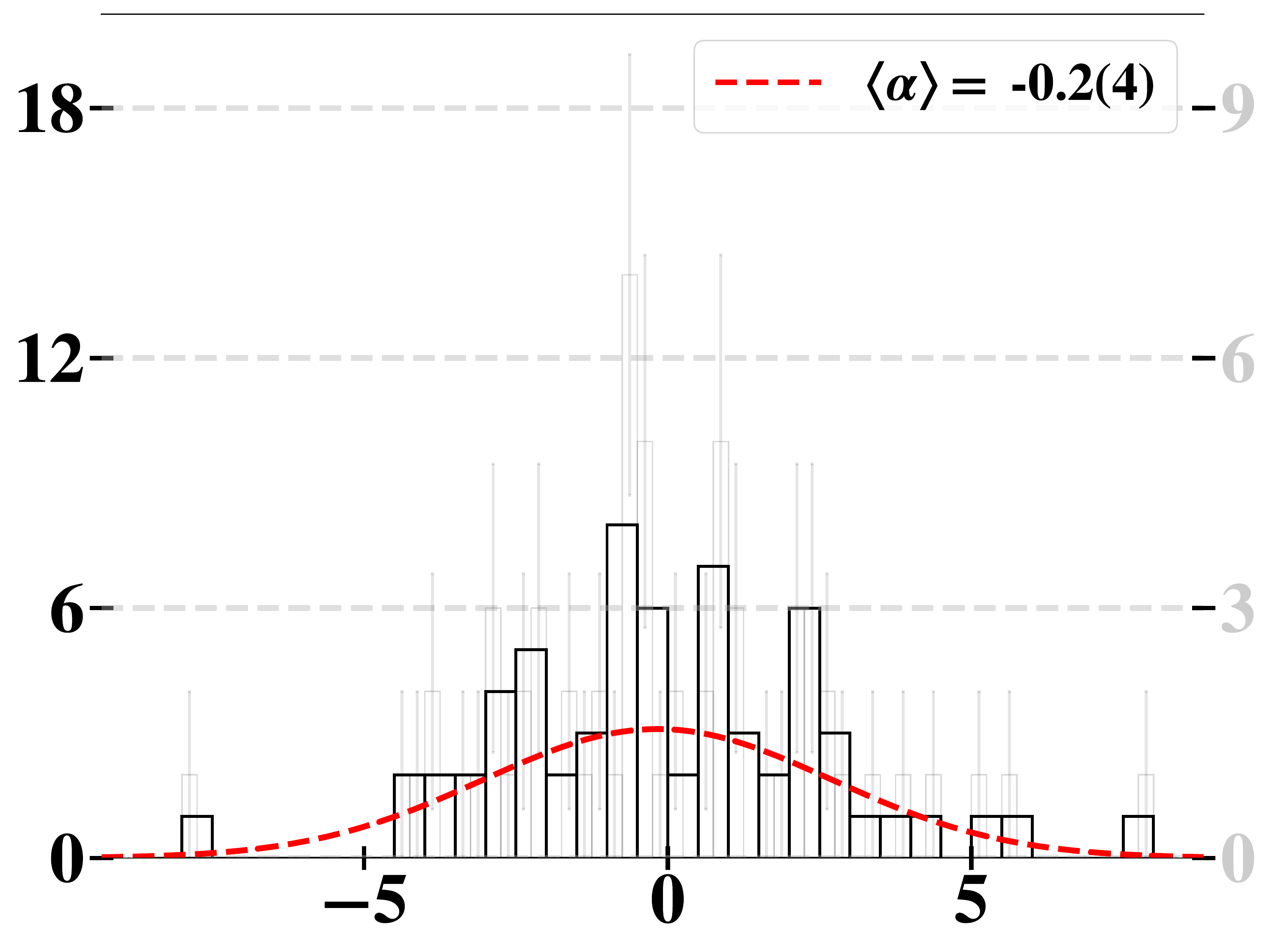}\hspace{0.1cm} \includegraphics[width=3.62cm]{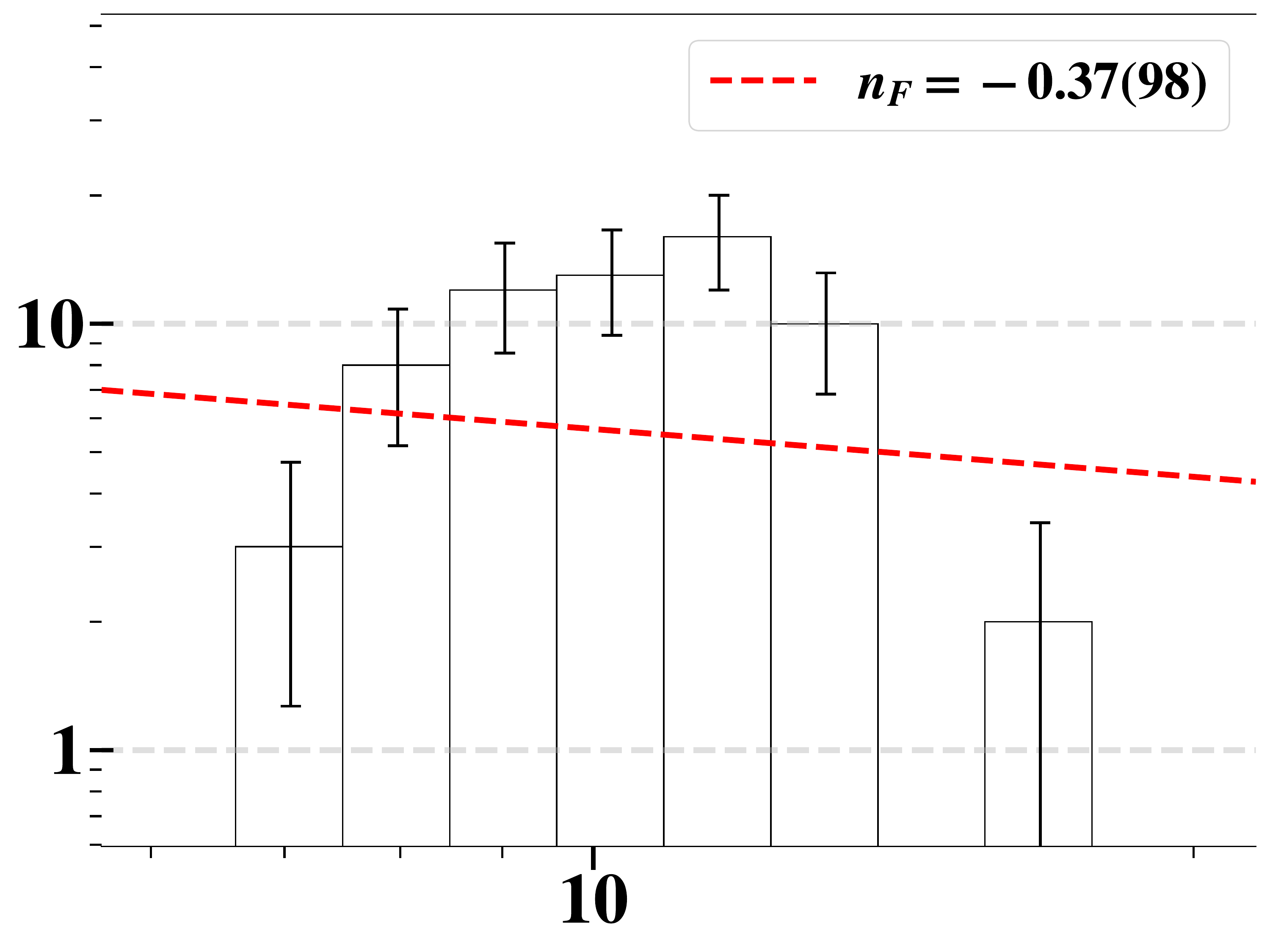} & \includegraphics[width=3.62cm]{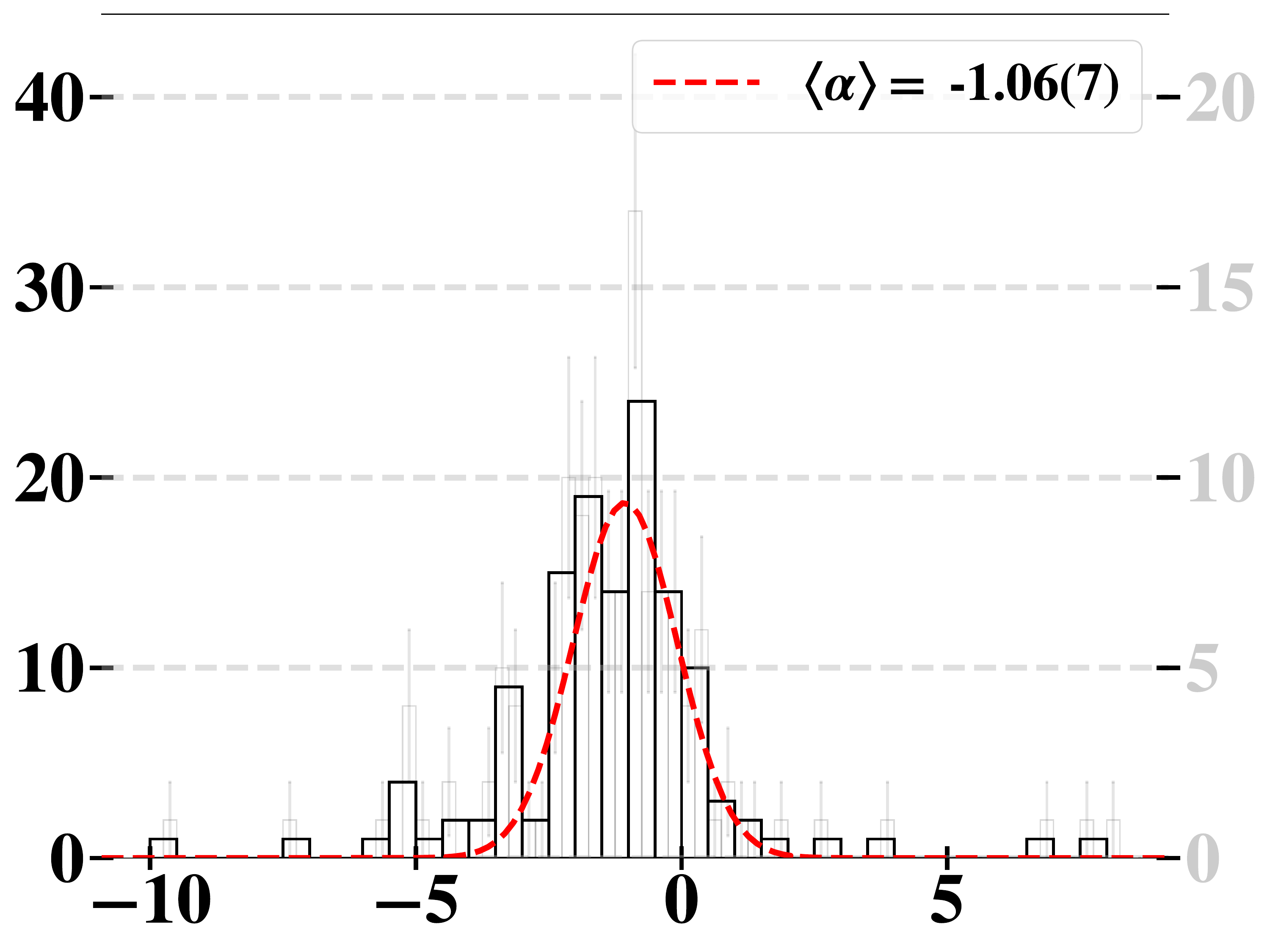}\hspace{0.1cm} \includegraphics[width=3.62cm]{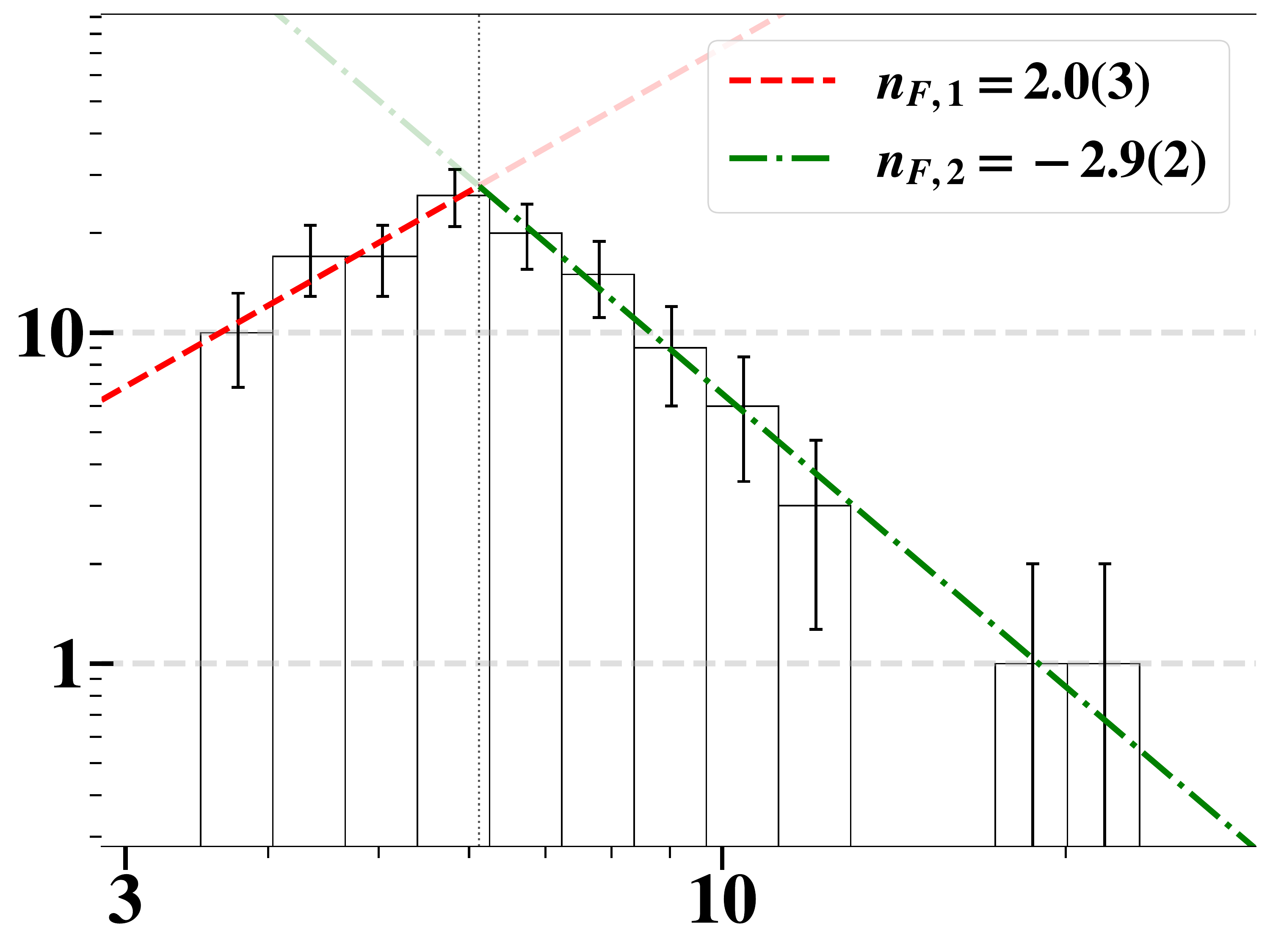}\vspace{-0.15cm}\\

(l) J1848+1516 & (m) J1931+4229 \vspace{0.13cm} \\

\includegraphics[width=3.62cm]{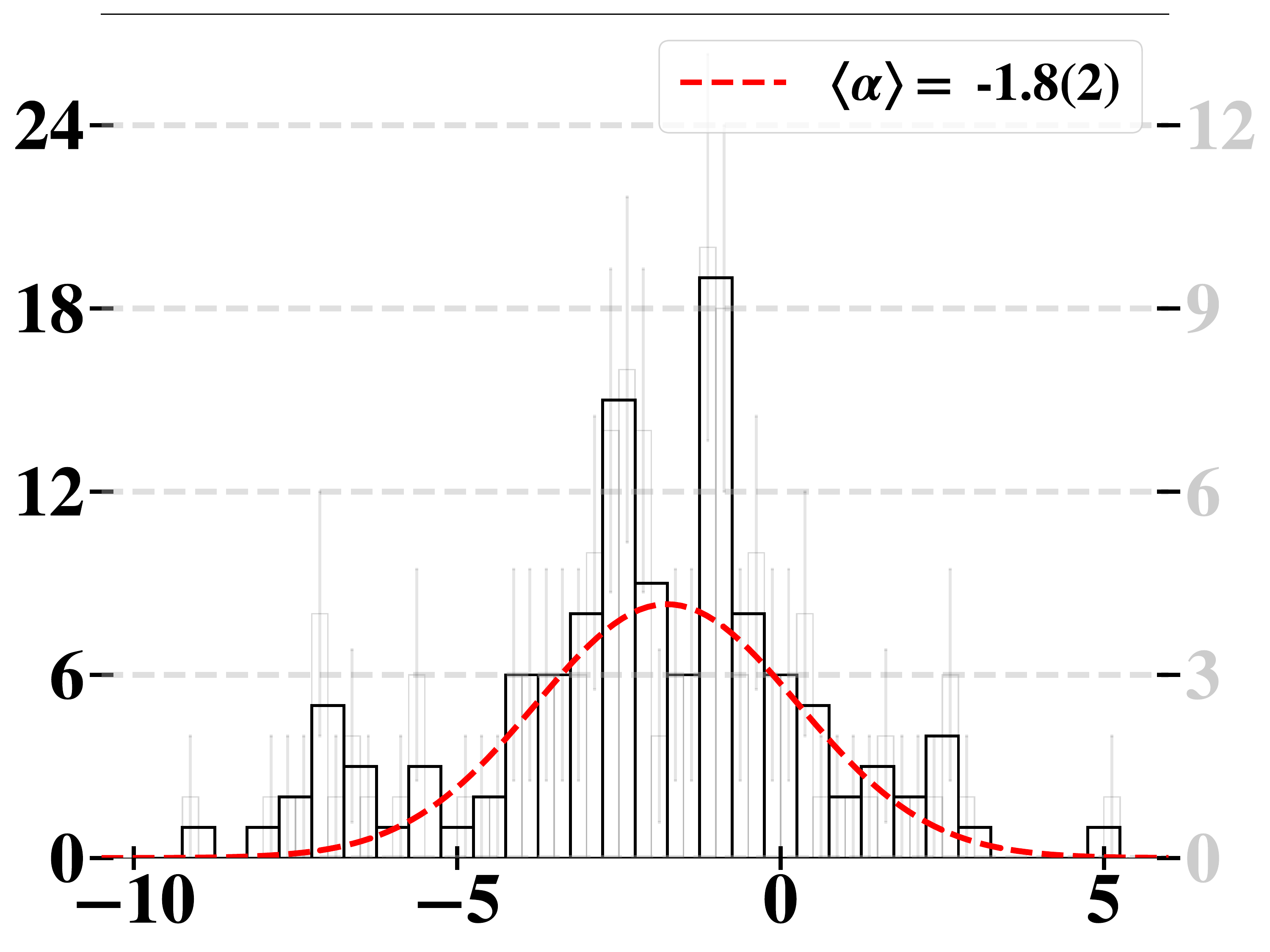}\hspace{0.1cm} \includegraphics[width=3.62cm]{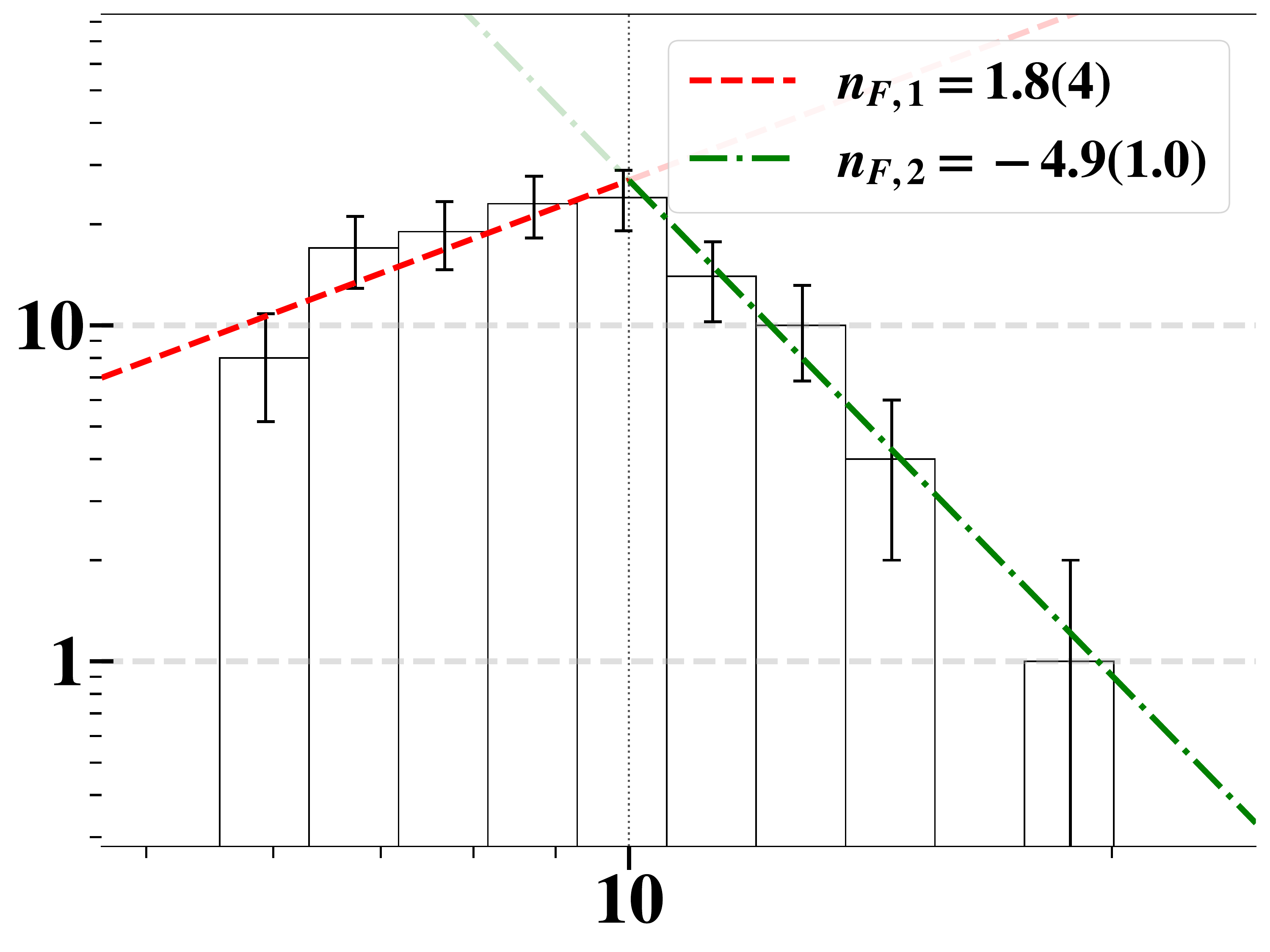} & \includegraphics[width=3.62cm]{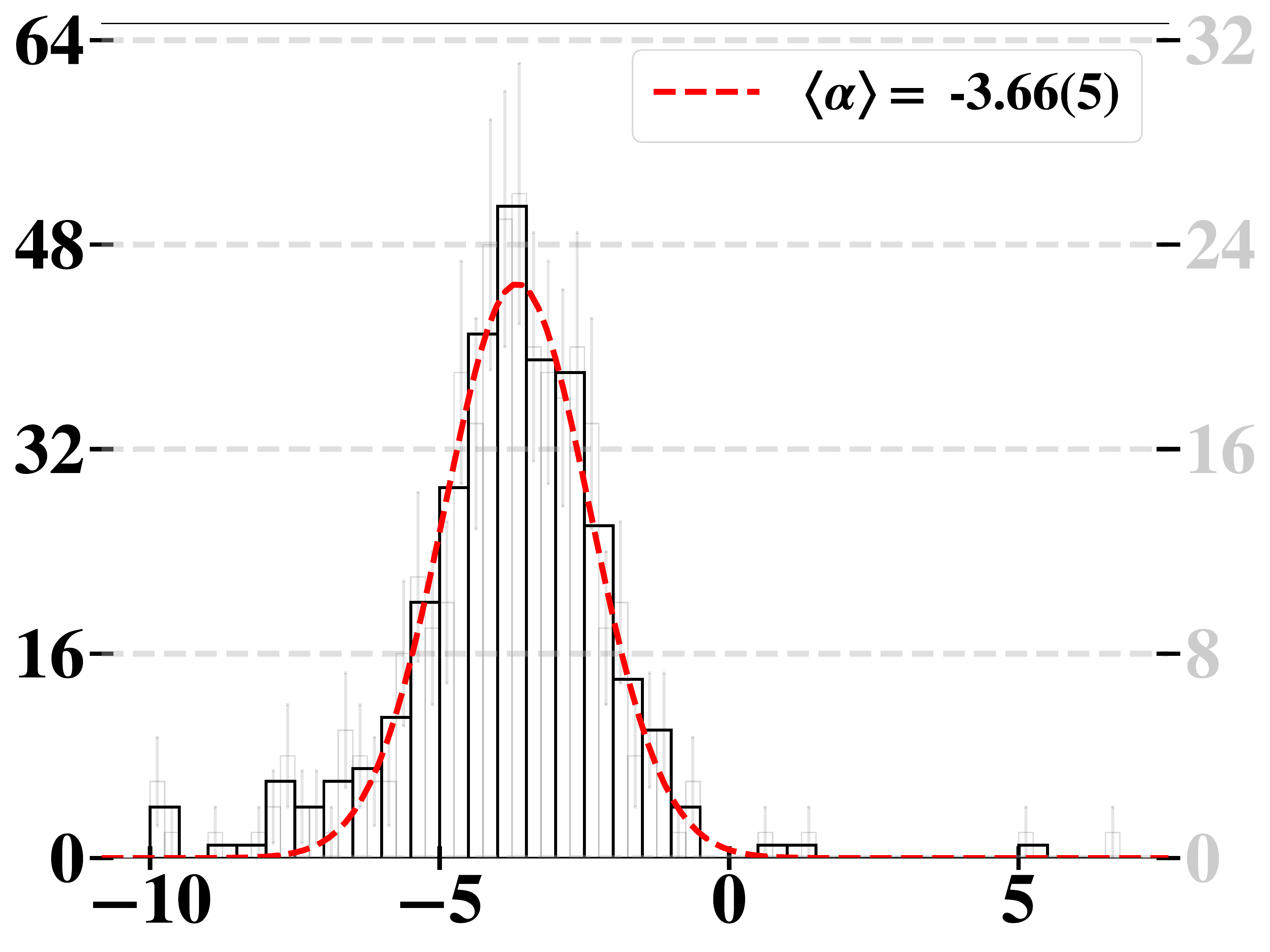}\hspace{0.1cm} \includegraphics[width=3.62cm]{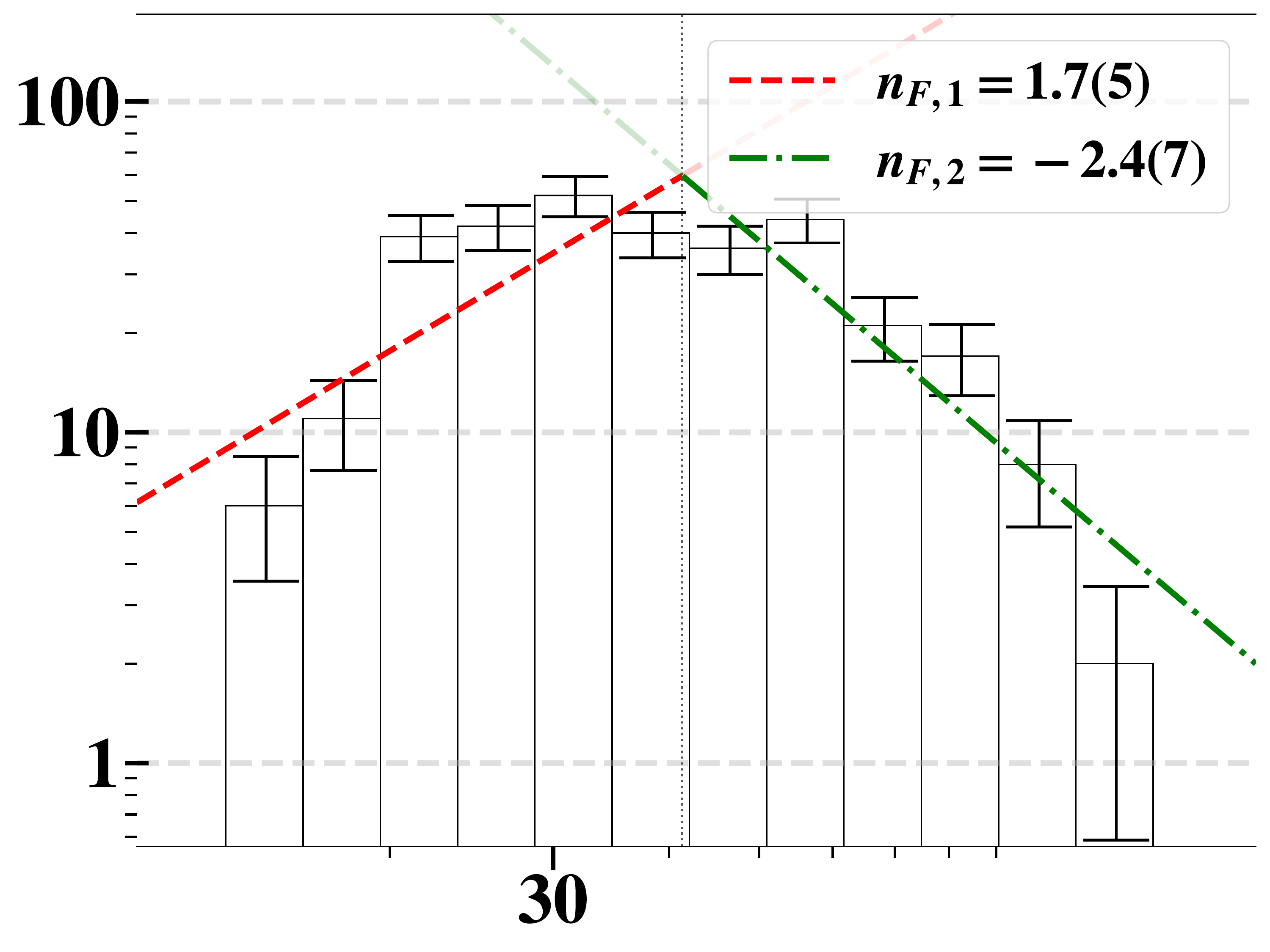}\vspace{-0.15cm} \\

\includegraphics[width=3.62cm]{Plots/spectralFits/spectralFitAxis.pdf}\hspace{0.1cm} \includegraphics[width=3.62cm]{Plots/spectralFits/fluxDensityAxis.pdf} &   \includegraphics[width=3.62cm]{Plots/spectralFits/spectralFitAxis.pdf}\hspace{0.1cm} \includegraphics[width=3.62cm]{Plots/spectralFits/fluxDensityAxis.pdf} \\
(n) J2215+4524 & (o) J2325$-$0530 \\
    \end{tabular}
    \vspace{-0.1cm}
    \caption{The (left columns) spectral flux density fit with a Gaussian function, and (right columns) burst pulse amplitude distribution fit with a function that maximised the AICc for the distribution, for 14 sources observed during the census. Each of these sources had at least 40 bursts detected in order to provide a reasonable sampling of the source properties, with PSR J2108+4524 excluded due to strong scattering preventing a good fit to the parameters (see \S~\ref{sec:discussspectral}). See the text (\S~\ref{sec:singlepulseres}) for further details on the methodology.}
    \label{fig:spectralmodfits}
\end{figure*}

\subsection{Novel Source Rotation Periods}\label{sec:newperiod}

From the $\singlepulsedetected$ sources with detectable single-pulse emission, 9 sources did not have a known period when initially observed during this census (J0348+79, J0939+45, J1006+3015, J1218+47, J1329+13, J1400+2125, J1931+4229, J2105+19 and J2138+69). Three of these sources, J1006+3015, J1400+2125 and J1931+4229, were found to have a sufficient burst rate to determine a rotation period, which can be found in Tables \ref{tab:obs_summary} and \ref{tab:newephemerides}. The reported period of J2105+19 is refined from the recent report of \SI{3.5297}{\second} from \cite{tyulbashevWeakPulsar2022}, as determined from an observation with the LOFAR core in June 2021, which will be discussed in future work \citep{McKennaInPrep}.

In the cases of J1006+3015 and J1931+4229, long enough observations of the sources in a single observing session were able to generate a sufficient number of times of arrival to determine the underlying source period through the standard brute force method. J1400+2125 required a brute force on the differences in times of arrival, generated from two sets of single--pulses observed across observations on two sequential days.

The rotation period of PSR J1336+3414 was also found to differ by a factor of 2 from the published rotation period of \SI{3.013}{\second}, at \SI{1.5066}{\second}. This is discussed further in~\S\ref{sec:j1336}.

\begin{figure}
    \centering 
    \includegraphics[width = 0.45\textwidth]{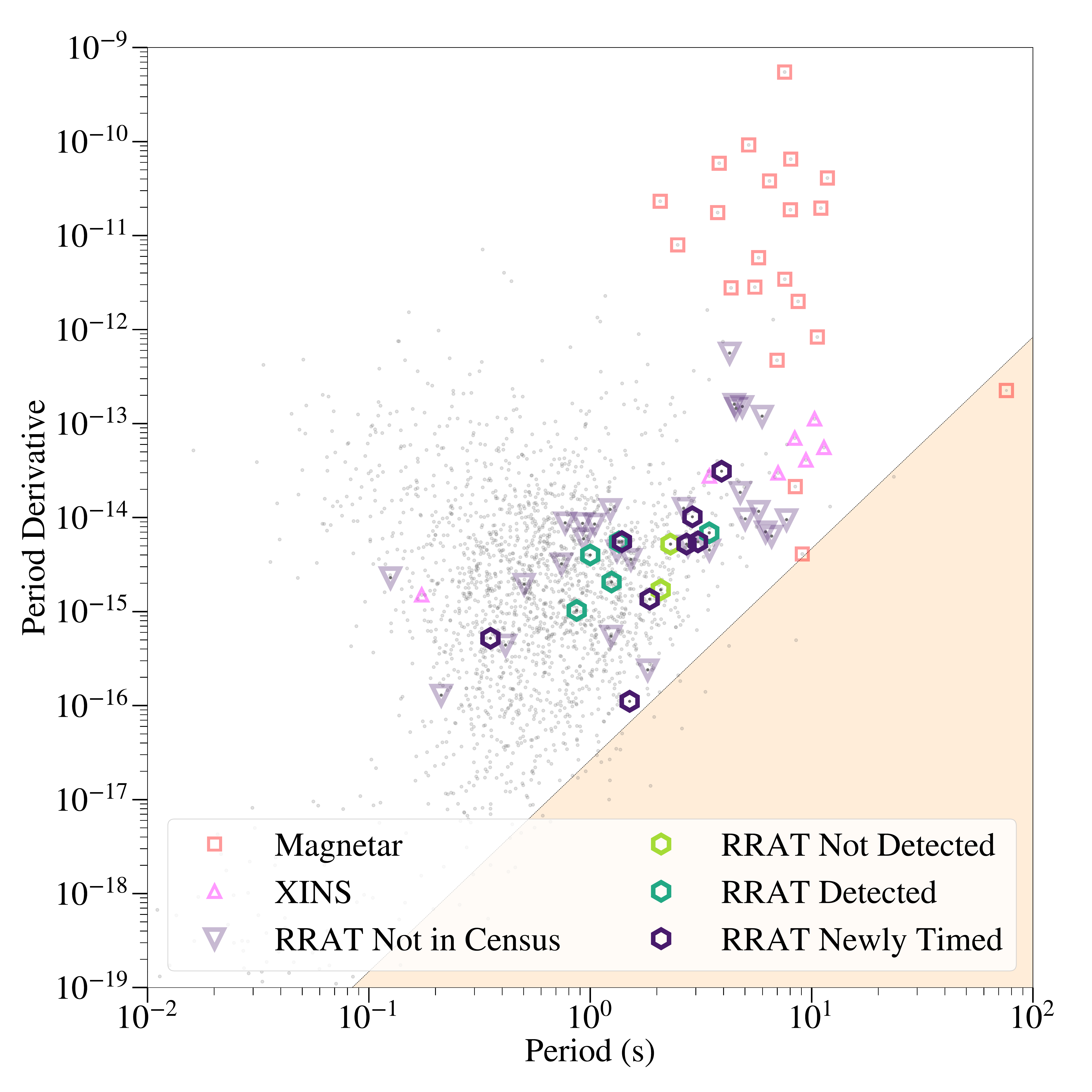}
    \caption{A period-period derivative plot containing the sources of interest in this census, modified from the work of \protect\cite{psrqpy}. The background gray dots represent the normal pulsar population, while magnetars and x-ray isolated neutron stars (XINS) have been highlighted due to their regular comparison to RRAT populations. RRATs that were (1) not observed, (2) not detected, (3) detected and (4) newly timed as a part of this work have been separately highlighted. The shaded region indicates the portion of the phase space that falls below the radio pulsar deathline that can be found as Eq.~4 in ~\protect\cite{bingDeathLine2000}. See~\S\ref{sec:discussproperties} for further details.}
    \label{fig:rrat_ppdot}
\end{figure}

\subsection{Timing}\label{sec:timnigres}
From the $\detected$ detected sources, $\timeable$ were detected at a high enough burst rate or periodic brightness to allow for pulsar timing with I--LOFAR\footnote{We consider an average burst rate of \SI{0.5}{\per\hour} for a single-pulse source, or a SNR greater than 10 across a typical \SI{2}{\hour} observation for a more standard pulsar, to be the minimum criteria for a source to be timed.}. From these $\timeable$, $\wecouldtime$ sources either did not have a full set of pulsar timing parameters available in \texttt{psrcat}.

To date, $\wetimed$ of these sources have had sufficient follow-up to produce stable timing ephemerides, which are presented in Table~\ref{tab:newephemerides}. Each of these sources can be considered to be RRATs. 

While J2215+4524 has shown periodic emission similar to that of an intermittent pulsar~(a pulsar which performs near instantaneous changes between strong and weak emission modes on timescales of weeks, months or even years \citep{Lyne2009}), however a mode change has not been observed for the source as a part of this work. While single-pulse emission is consistent, periodic emission is frequently not detectable with I--LOFAR, and consequently is not considered to be a reliable method for timing the source. 

These sources cover the length and breath of standard RRAT properties, covering an order of magnitude in rotation period, from \SIrange{0.35}{3.92}{\second}, period derivatives across three orders of magnitude, \SIrange{1.1e-16}{3.1e-14}{\second\per\second} and dispersion measures from \SIrange{8}{55}{\parsec\per\cubic\centi\metre}. One further source, J0317+1328, has been timed using the LOFAR core (results unpublished), while further observations are ongoing to time the remaining $\tobetimed$ sources at I--LOFAR.

\subsection{Blind Pulse Search}\label{sec:blindresult}
In addition to performing narrow dispersion measure range searches for single pulses, all observations that were taken as a part of the census underwent a wide dispersion measure scan for single pulses, from \SIrange{10}{500}{\parsec\per\cubic\centi\metre}. This can be considered as an \SI{\totalhours}{\hour} directionally biased survey for fast transients (other RRATs, pulsars with giant-pulses and fast radio bursts) as a by-product of this work.

While a number of pulses have been detected at dispersion measures that differ significantly from the observed source, verifying the methodology can detect such pulses, these were easily tied to well known pulsars within the beam's field of view, such as B0301+19 or B1842+14.

As of October 2022, this work has resulted in one novel detection of a source at a dispersion measure of \SI{18.6}{\parsec\per\centi\metre\cubed} near CHIME source PSR J2119+49. No sources near this dispersion measure have been reported within \SI{10}{\deg} of the original beam pointing. The source was only detectable in the bottom \SI{30}{\mega\hertz} of the bandwidth, indicating that it was either off-axis of the beam pointing (up to \SI{3}{\deg}) or spectrally steep. To attempt to detect and localise this source, \SI{6}{\hour} of follow-up observations were performed using four \SI{24}{\mega\hertz} beams, one centred on the original pointing and a further three separated by \SI{2}{\deg} from the original pointing, rotated by \SI{30}{\degree} each hour, did not result in a re-detection of this source.

\begin{figure}
    \centering 
    \includegraphics[width = 0.48\textwidth]{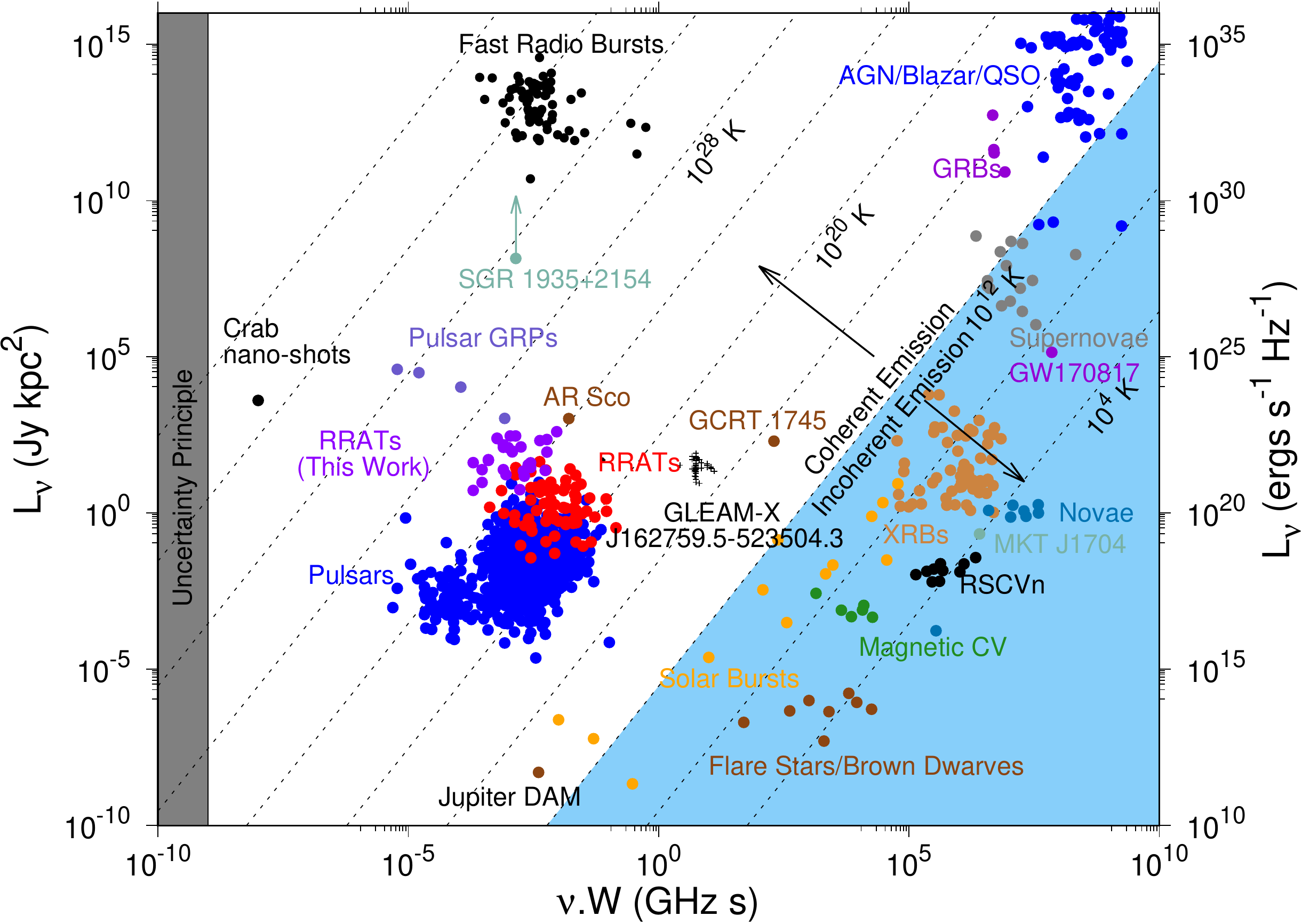}
    \caption{A phase space plot of astronomical transients signals (spectral duration against specific luminosity), modified from the work of \protect\cite{pietkaPhase2015} and \protect\cite{hurleyWalkerTransient2022} to include the brightest pulses from each of the sources detected as a part of this work. This plot includes a fix to correctly consider the frequency of data in the RRATalog, where previous plots assumed all data was taken at L--band.}
    \label{fig:phase_space}
\end{figure}

\section{Discussion}\label{sec:discuss}

\subsection{Source Properties and Detectability}\label{sec:discussproperties}

\subsubsection{Source Dispersion Measures}\label{sec:discussdm}

As seen in Fig.~\ref{fig:rrat_dms}, both the RRAT population as a whole, and the subgroup detected by the census is heavily biased towards lower dispersion measures, with only 6 of the detected source having a dispersion measure above \SI{40}{\parsec\per\centi\metre\cubed}. 
 
This is likely a result of the scattering caused by inhomogeneities in the medium along the line of sight between the observing telescope and the neutron star, causing the pulse emission to be dispersed in time (reducing the pulse SNR proportionally by this timescale, $\sqrt{t}$) and fall below the noise floor of the instrument as a result. 

Scattering effects are particularly strong at the lower frequencies used by many of the telescopes behind the catalogues used for this census. As described in \cite{cordesPropgation2002}, the scattering timescale of a radio pulsar is dependent on the dispersion measure ($\log\tau_\text{sc}\sim\log\text{DM} + \left(\log\text{DM}\right)^2$) and observing frequency ($\tau\sim\nu^{-4}$). While lower dispersion measure sources are still detectable, high dispersion measures cause the pulses to be broadened by an order of magnitude or even more in time. Using Eq.~10 of \citeauthor{cordesPropgation2002}, we can estimate the scatter broadening of a RRAT with a dispersion measure of 10, 30, 100 and \SI{300}{\parsec\per\centi\metre\cubed} observed at \SI{150}{\mega\hertz} to be 0.02, 0.29, 24 and \SI{4100}{\milli\second}. With the described frequency scaling, these are expected to be more than three orders of magnitude larger at \SI{150}{\mega\hertz} as compared to L-band (\SI{1.5}{\giga\hertz}), and nearly 30 times larger than at P-band (\SI{400}{\mega\hertz}).

Given the typical brightness of these sources, the sensitivity of I--LOFAR, and typical observed pulse widths from these sources at higher frequencies, the potential to detect a RRAT with a dispersion measure beyond \SI{100}{\parsec\per\centi\metre\cubed} is highly unlikely.

Similarly, these high dispersion measure sources must be close to the Galactic plane due to the sharp falloff of free electrons away from the Galactic disk. As previously described, the disk also resulted in significantly higher $T_{\text{sky}}$ measurements, which can severely hamper the instantaneous sensitivity of I--LOFAR towards these sources, which is a further compounding effect that reduces the detectability of these sources.

\subsubsection{Source Rotation Periods}

As seen in Fig.~\ref{fig:rrat_periods}, the majority of the sources observed but undetected in this census do not have a known rotation period. This is to be expected as in order for the period of a RRAT to be determined it must produce pulses that are sufficiently bright to meet the detectability criteria of the observing telescope, and have a sufficient burst rate at that sensitivity for multiple time-of-arrival measurements in a short window of time to allow for a rotation period to be brute-forced. Observations with I--LOFAR over a 5-hour window may not have been sufficient time to detect a single pulse in the case that the pulse-amplitude distribution has a sufficiently long, but low-probability tail to produce pulses within our sensitivity range.

While longer period sources may have been able to offset the effects of scattering discussed in~\S\ref{sec:discussdm} due to wider intrinsic pulse widths at typical duty cycles, fewer rotations of the pulsar are observed, reducing the number of samples taken from the tail of the pulse amplitude distribution.

\subsubsection{Pulse Amplitude Distributions}
The pulse amplitude distributions are visualised in Fig.~\ref{fig:spectralmodfits}, with the underlying fitted data in Table~\ref{tab:pulseamplitudetab}. The majority of these distributions are well fit by a combination of powerlaw or log-normal distribution, with the exceptions J1848+1516 and J0746+5514.

In the case of J1848+1516, this is expected to be a result of the detected pulse brightness being extremely close to the sensitivity limits of the I--LOFAR instrument. From the 68 detected pulses, only 40 per cent were above a SNR of 8, while only 13 per cent were above a SNR of 9. This low sample size, taken from the tail of the pulse amplitude distribution, results in a skewed distribution which should not be considered to be an accurate model of the underlying distribution (which should be visible with higher sensitivity instruments).

J0746+5514 appears to have a flat pulse amplitude distribution. While this may be in part due to the first bin causing a positive shift on the fitted function, the raw data for this source is extremely erratic. The observed pulses are at an abnormally high significance, with only 16 per cent of the 48 bursts below a SNR of 9 and more than half of the observed pulses have a SNR above 13, making the pulses from this source extremely bright in the data. As presented in Table~\ref{tab:obs_summary}, the standard deviation of the pulse width is more than 60 per cent of the mean pulse width (the observed pulses have widths varying from \SIrange{3}{77}{\milli\second}). The flat distribution is likely a consequence of the relatively low sample of pulses from such a diverse population, and will require further follow-up to accurately characterise the pulse amplitude distribution.

\subsubsection{Spectral Indices}\label{sec:discussspectral}
While spectral indices of RRATs have been shown to be more extreme than the normal pulsar population \citep{shapiro-albertRadioPropertiesRotating2018}, it is expected that the single-pulse spectral indices presented in Table~\ref{tab:obs_summary} and Fig.~\ref{fig:spectralmodfits} are biased towards being steeper than the actual emission properties. Multiple different factors contribute to this, including (a) the position uncertainty on a number of the sources (discussed in~\S\ref{sec:sources}) as the beam sensitivity is reduced for off-axis sources as the beam narrows at higher frequencies, and (b) reduced SNRs after splitting the bandwidth into \SI{10}{\mega\hertz} segment for low significance pulses in order to better sample the spectral behaviour of the sources. (b) will have a particularly strong effect on sources with more extreme spectral indices (both positive and negative), as the gain of I--LOFAR is relatively flat across the observed bandwidth, potentially causing the dimmer emission to fall below the noise floor near the edges of the bandwidth.

We can identify sources that have these biases when they have been detected through both single-pulse and periodic, as after folding the data to produce a folded archive there will likely be sufficient data to have a strong SNR across the entire sampled bandwidth. Consequently, the effects of (a) can be seen for the untimed source J0317+1328 ($\alpha_{\text{sp}}=-5.3(8)$ and $\alpha_{\text{f}}=-1.8(4)$), while (b) is likely contributing to the spectral index disparity of J1538+2345 ($\alpha_{\text{sp}}=-1.6(12)$ and $\alpha_{\text{f}}=-0.51(17)$) and, despite the scattering causing a strong positive bias in the single pulse data, J1848+1516 ($\alpha_{\text{sp}}=-0.2(4)$ and $\alpha_{\text{f}}=0.52(14)$).

J2108+4516, a binary pulsar, is a case where the spectral index is significantly more positive in the single-pulse data as opposed to the periodic emission ($\alpha_{\text{sp}}=1.91(14)$ and $\alpha_{\text{f}}=-0.4(4)$). This is likely as a result of the source showing strong signs of scattering (in the folded profile visible in Fig.~\ref{fig:periodicprofiles}(k)), with the resulting diminished SNRs at lower frequencies causing the source to appear to have a positive spectral index in the single-pulse data.

\subsubsection{Further Limits on Detectability}

Some sources in the PRAO catalogue may have not been detected due to limitations in the reported positions of sources. Due to the fixed declination and wide full-width-half-max (FWHM) of their beams, position uncertainty may be present in their initial reported positions which can only be reduced with the detection of further pulses (to report a better mean right ascension, and more accurate declination). 

This effect is noted specifically for PSR J1336+3414~(\S\ref{sec:j1336}), where the reported position and timed position differed by \SI{0.48}{\degree}, (between a third and a fifth of the beam FWHM depending on the observing frequency). Correcting for the position of this source is also correlated with the detection of off-pulse emission that leads the main pulse train, demonstrating even these small changes can have an effect on detectability of pulses from these sources. Weaker sources on the edge of detectability would be unlikely to be detected with such a large offset between the true sky position and the observed sky position.

\subsubsection{Phase Space}

The transient phase space plot in Fig.~\ref{fig:phase_space} shows that the pulses detected as a part of this census have been brighter but have shorter spectral duration than emission previous reported for RRATs.

In terms of luminosity, the baseline data for RRATs is generated from entries in the RRATalog, which has been described in~\S\ref{sec:rratalog}, where sources are primarily described at either L-band or P-band observations. With the expected brightening of these sources at lower frequencies, combined with the selection effects from the lower relative sensitivity of I--LOFAR compared to other instruments at their respective frequencies, it is not a surprise that the results from this work tend to be at a higher brightness than the previous RRAT population on this plot.

While the multiple surveys cited as the data sources for the RRATalog data provide the peak pulse brightness for each source, it is unclear what form of sampling is used to determine the width values, with different works using a combination of median, mean and brightest pulse sampling to provide the pulse width. For this work, the width of the brightest pulse for each source was used to generate the data for the plot.

Regardless, a clear separation is visible between the sources detected as a part of this work and the general RRAT population. This is likely due to the selection effects previously mentioned that result in the telescope detecting strong, minimally scattered pulses with similar widths to those presented in the RRATalog, resulting in an order of magnitude increase in the luminosity, and order of magnitude reduction in the frequency-duration samples from the lower observing frequency.

\subsubsection{Comparisons to Previous LOFAR Work}

During the work of \citetalias{karako-argamanDiscoveryFollowupRotating2015}, 5 of the sources characterised as a part of this work were detected using 20--22 stations of the LOFAR core, while our reprocessing of the LOTAAS survey (\S\ref{sec:archivalmining}, \citetalias{michilliSinglepulseClassifierLOFAR2018}, \citetalias{sanidasLOFARTiedArrayAllSky2019b}) in Table~\ref{tab:lotaas_summary} led to 12 detected single-pulse sources, with 2 overlapping between the works. 

Comparing the observed burst rates of these 15 detected sources, there are two distinct categories of sources: those with similar observed burst rates at both the 8-stations of the LOFAR core and I--LOFAR, and those with significant increases in observed burst rate as compared to this work.

The former group of 8 sources (J0054+6650, J0102+5356, J0139+3336, J0746+5514, J1006+3015, J1336+3414, J1400+2125, J2325$-$0530) can be considered a group of sources with the potential to be monitored with an international LOFAR station, with any observing time offering a relatively high completeness as compared to significantly more sensitive configurations of the LOFAR core. These sources may have a discontinuity in their pulse amplitude distributions between the emission during the majority of rotations and the observable giant-pulses detected during this work.

It is notable that out of the 7 sources that fall in the latter group (J0201+7005, J0332+79, J1404+1159, J1538+2345, J1848+1510, J2202+2134, J2215+4524), all but one source, J0332+79, were found to produce detectable periodic emission in at least one observation during the census and further follow-up observations. While none of these sources were consistently detectable at I--LOFAR, J1538+2345 produced consistent periodic emission when observed with FAST~\citep{luStudyThreeRotating2019a}, and J1848+1510 was found to produce persistent periodic emission during observations present in the LOFAR LTA. Consequently, there is an argument to re-classify these sources as highly variable pulsars rather than RRATs, given an increase in sensitivity allows for consistent detection of periodic emission at some sites.

Except for J1404+1159 and J0854+5449, the sources with periodic emission detected during the LOTAAS survey and presented in Table~\ref{tab:lotaas_summary} have shown significant flux variability between observations at I--LOFAR. Consequently, a comparison of the emission properties between these epochs is unlikely to offer any further insight beyond what is presented elsewhere within this work.

\subsection{Sources of Interest}

\subsubsection{PSR J0209+5759}

J0209+5759 is a nulling pulsar reported by the CHIME/FRB collaboration, and its properties at \SI{600}{\mega\hertz} have been discussed in \cite{goodFirstDiscoveryNew2020}. It has a high level of nulling, with \citeauthor{goodFirstDiscoveryNew2020} reporting a lower limit on the nulling fraction of 21 per cent, though while active, they typically can observe \SI{21.4(5)}{pulses \per\hour}. They reported two estimates of the dispersion measure of the source, \SI{55.282}{\parsec\per\centi\metre\cubed}, though this was calculated through PRESTO's \texttt{prepfold} rather than through analysis of single pulses (but was used to generate the reported timing ephemeris), and \SI{55.3(6)}{\parsec\per\centi\metre\cubed} from the CHIME/FRB system metadata.

At I--LOFAR, this source has been detected through both single-pulse and periodic emission. Analysis of the nulling behaviour across \SI{26.6}{\hour} of observations,  indicates an upper bound for the nulling fraction of 95.3(4) per cent for
a 3-sigma detection threshold of emission during \SI{30}{\second} integrations, though given the source likely has a lower SNR at I--LOFAR, and the length of the integrations of the folded data, this could be an overestimate of the true nulling fraction.

The source has a significantly lower single-burst rate at I--LOFAR as compared to CHIME, with only 1.6(2) bursts observed per hour on average, though this peaked at \SI{7(1)}{\per\hour} during a \SI{40}{\minute} observation in November 2021.

The dispersion measure of the source at I--LOFAR was measured as \SI{55.855(4)}{\parsec\per\centi\metre\cubed}, within the confidence interval of the CHIME/FRB measurement, but offset from the \texttt{prepfold} estimates. At \SI{151}{\mega\hertz}, utilising their reported \texttt{prepfold} dispersion measure measurement results in a distinct dispersive delay and appreciable SNR loss when used on the brightest observed pulses and the active mode periodic emission.

Inspecting the LOTAAS archival pointing of the source indicates that the nulling behaviour was detectable during the survey (using the same methodology as I--LOFAR indicated a nulling fraction of 95(2) per cent during the hour-long observation), however no single pulses were detected during the observation. This may be a consequence of the relatively flat spectral index combined with the lower observing frequencies of the LOTAAS survey (typically \SIrange{119}{151}{\mega\hertz}) causing pulses to be below the sensitivity limit of the instrument, or the source may not have produced pulses bright enough to be detected by either I--LOFAR or the 8-core stations used during the LOTAAS survey (41 per cent of observations at I--LOFAR did not result in a detection of J0209+5759).

\subsubsection{PSR J0226+3356}
While initially announced as a RRAT in early releases of the CHIME catalogue, the recent work of \citetalias{dongCHIME2022} reclassifies this source as a pulsar due to persistent periodic emission that has been detected with the CHIME telescope.

Observations at I--LOFAR have yet to detect a burst from this source, however periodic emission has been detected in all but one observation taken to date. 

\subsubsection{PSR J1006+3015}
J1006+3015 was reported by the PRAO catalogue, and was found to have a period of \SI{3.066}{\second} and further timed as a part of this work. While an uncertainty was not provided, this result is similar to the multi-epoch period estimate of \SI{3.069}{\second} presented in \cite{smirnovRRATVariability2022}.

While relatively infrequent as compared to J2325$-$0530, this source has consistently produced extremely high significance pulses observed during this work, with 8 of the 10 highest SNR bursts during the census observed during follow-up observations of this source, and an overall brightness ratio between the brightest and dimmest pulse of 16. The brightest pulses have demonstrated interesting morphologies, such as significant temporal variance in amplitude during otherwise Gaussian-shaped pulses (as visible in the best-pulse plot, Fig.~\ref{fig:singleprofiles}(l)) or multi-peaked pulses. Further follow-up with a more sensitive instrument may help determine if these are unique traits to these bright pulses, or an underlying variability that falls below the noise floor for observations with I--LOFAR.

\subsubsection{PSR J1132+2513}
J1132+2513 is reported by the PRAO catalogue, and, as a source of RRAT--like bursts, though was previously only reported as a periodic source during the GBT350 Drift Scan Survey~\footnote{While reported on the \href{https://web.archive.org/web/20220626134940/http://astro.phys.wvu.edu/GBTdrift350/}{survey webpage}, the source is never discussed in any publications regarding the survey, has an unknown discovery date and discovery plot with half the current period.} (J1132+25) and as a known source by LOTAAS (J1134+25). During this work, no single pulses were detected, though the periodic emission reported during the LOTAAS survey was detected.

This behaviour appears to align with the previous report from \cite{smirnovRRATVariability2022} which indicates the source is an intermittent source of single-pulse emission, at a low single-to-noise ratio. Given this pulsar currently does not have a full timing ephemeris in the pulsar catalogue (entry as J1132+25), future observations performed at I--LOFAR with the goal of timing the source through the detectable periodic emission may result in the detection of single pulse emission from the source.

\subsubsection{PSR J1329+13}\label{sec:J1329}
J1329+13 is a source from the PRAO catalogue with an extremely low observed burst rate \citep{tyulbashevDetection25New2018}, with the current online version of the PRAO catalogue only reporting a single pulse being detected from the source with the BSA LPI. 

From \SI{50}{\hour} of observations with I--LOFAR, only 5 pulses have been detected from the source. The first two pulses were observed within the space of a week in February 2021, and it was not detected again until two further pulses were detected in November and early December 2021. Assuming there is no bias on the temporal distribution of pulses (due to nulling, presence in binary system or intermittency), this would give the source a burst rate of \SI{0.10(5)}{\per\hour} for an ILT station, and \SI{0.008(8)}{\per\hour} at the BSA LPI. 

If the time between pulses were to form a standard Poisson distribution, unless the source has a strongly positive spectral index (which is unlikely given the properties of the few observed pulses), the BSA LPI burst rate should be expected to be closer to that of the I--LOFAR measurement, if not higher due to the increased sensitivity of the instrument. As a result, it can be inferred that this source likely acts similarly to an intermittent neutron star.

A blind periodicity search across the 5 observations with any observed pulses using riptide \citep{morelloOptimalPeriodicitySearching2020b} did not result in a credible detection of periodic emission from the source, with 7$\sigma$ upper limits of between \SI{1.6}{\milli\jansky} and \SI{1.0}{\milli\jansky} assuming a 1 per cent duty cycle and that the source was active for the entirety of the observations.

\subsubsection{PSR J1336+3414}\label{sec:j1336}
The reported timing ephemeris for BSA PSR J1336+3414 differs significantly from the original parameters published in \cite{tyulbashevDetection25New2018} and subsequent papers. Pulsar timing indicates it is offset by \SI{0.48}{\deg} from the reported position, and the rotation period was found to be half of the originally reported \SI{3.013}{\second}. 

The confusion in the reported source period may have been the result of an infrequent off-pulse component visible in the pulsar timing residuals. After updating the source pointing to match the fit position from pulsar timing, 11 pulses (10 per cent) have been observed to have a 0.044 phase (roughly \SI{66}{\milli\second}) offset from the majority of observed pulses. Off-pulse components can cause issues for most algorithms used to brute-force RRAT periods and may have been biased to a harmonic rotation frequency, which was then reported.

J1336+3414 has both the slowest spin down rate and the oldest characteristic age of any RRAT, with \SI{1.11(3)e-16}{\second\per\second} and \SI{216}{\mega\year}, replacing J1647$-$3607 (\SI{1.29(2)e-16}{\second\per\second}) and J1739$-$2521 (\SI{120}{\mega\year}). Its rotation parameters result in it sitting extremely close to the standard radio pulsar deathline~\citep[equation 4]{bingDeathLine2000}, making it an interesting source for modelling the emission properties of ageing pulsars.

\subsubsection{PSR J1538+2345}
J1538+2345 is described as initially being reported as part of the GBT \SI{350}{\mega\hertz} drift-scan survey, though did not appear in any publications until it received follow-up with both the GBT and LOFAR core to characterise and time it prior to this work, with the results discussed in \citetalias{karako-argamanDiscoveryFollowupRotating2015}.

On several occasions it has been detected through periodic folding at I--LOFAR, with the folded profile in Fig.~\ref{fig:periodicprofiles} showing a dual peaked source, with weaker emission from the leading peak compared to the lagging peak, with weak emission between the two peaks. Single-pulses are detectable from both peaks, and form two distinct pulse trains with a separation of \SI{0.024}{} rotations when times of arrival are analysed. 30 per cent of the pulses align with the first peak, while the remainder align with the second. No pulses at I--LOFAR have been detected offset from these two groups. % 30.2per cent of pulses

\subsubsection{PSR J1848+1516}
J1848+1516 is a source reported by several surveys. The RRATalog cites the discovery to the GBNCC survey as J1849+15 (no publications discuss this, survey pages do not mention the source), it was detected as a part of the LOTAAS survey in \citetalias{michilliSinglepulseClassifierLOFAR2018}, and by the PRAO as J1848+1518 in \cite{tyulbashevDetection25New2018}.

\cite{michilliLOFARTiedarrayAllsky2020} discuss the source properties as seen at a higher cumulative signal-to-noise ratio as compared to the observations taken during this work and note the source appear to show both strong pulse-to-pulse variability, and some nulling behaviour, with emission turning on and off on the order of tens of rotations. They also note that the three peaks visible in a folded profile of the source (as seen from I--LOFAR in Fig.~\ref{fig:periodicprofiles}(i)) are extremely time variable, with the primary (brightest) peak being both lagged and led by secondary and tertiary components intermittently. The data taken with I--LOFAR does not have the sensitivity required to allow us to further comment on this behaviour.

When comparing the single-pulse and folded emission TOAs collected during this work, single-pulse TOAs were found to fall within a \SI{20}{\milli\second} window centred on the main peak of the folded profile, which is less than half the width of the primary periodic profile emission peak, provided in Table \ref{tab:obs_summary}.

\subsubsection{PSR J1931+4229}
J1931+4229 was one of the earliest source detected by CHIME/FRB and announced in the CHIME/FRB Galactic Source catalogue. It was initially observed to have a high burst rate at low frequencies, and as a result a rotation period of \SI{3.921}{\second} was determined in December 2020 through the detection of 13 pulses within an observation. This result was also confirmed as a result of 9 and 11 pulses during single observations in April and June 2021.

This was the slowest source detected (and timed) as a part of this census, and the resulting ephemeris indicates it falls into a region period-period derivative phase space (see Fig.~\ref{fig:rrat_ppdot}) where sources have properties that border those of x-ray isolated neutron stars (XINS) and magnetars. It is unknown whether these sources represent a sub-set of neutron stars separate from the remainder of the RRAT population, only associated by emission properties, or are just a more extreme group of RRATs.

This source was also detected during an expanded search using the BSA LPI. In \cite{samodurovDetectionStatisticsPulse2022}, the authors discuss it and report that it has a rotation period of \SI{3.6375}{\second}. Our data has not been able to reproduce this result, and attempting to process times of arrival at I--LOFAR using this result as a trial period does not result in a coherent timing solution.

This source was not detected during re-processing of the LOTAAS archive, which was unexpected given the burst rate of \SI{2.5(2)}{\per\hour} at I--LOFAR, which indicates it was likely to be detected within an hour-long pointing, and the negative spectral index, of $-1.06(7)$, that would cause it to have a higher SNR when observed with the 8-core station in the LOTAAS survey's configuration.

\subsubsection{PSR J2215+4524}
PSR J2215+4524 was an early source announced by the CHIME/FRB collaboration. An ephemeris and overall source characteristics at~\SI{600}{\mega\hertz} are described in the recent work of \citetalias{dongCHIME2022}. We note that the position and period parameters in the ephemeris provided in Table~\ref{tab:newephemerides} have lower uncertainties than provided in their work.

\subsubsection{PSR J2355+1523}
PSR J2355+1523 was announced by the CHIME/FRB collaboration, and a timing ephemeris has been published in the recent work of \citetalias{dongCHIME2022}. 

We are highlighting the source as while it was detected on a regular basis at I--LOFAR prior to July 2021, with a burst rate of \SI{0.49(16)}{\per\hour}, since then it was only been detected 5 times at a rate of \SI{0.12(6)}{\per\hour}. Prior to July 2021, the source was detectable at a reasonable significance, with 8 of the 10 detected pulses having SNRs above 8, and two being above 11. However, none of the 5 pulses observed since July 2021 have been above a SNR of 8. 

The CHIME/FRB collaboration provides a plot of detected epochs for a number of sources, including this source\footnote{\href{https://storage.googleapis.com/chimefrb-dev.appspot.com/J2355+1523/J2355+1523_prepfold.png}{Direct link for J2355+1523}, and an \href{http://web.archive.org/web/20230213164601/https://storage.googleapis.com/chimefrb-dev.appspot.com/J2355+1523/J2355+1523_prepfold.png}{archival copy}}, with their catalogue. Inspecting the plot indicating this source might show signs of intermittency, but on shorter time scales than the delay between detections at I--LOFAR. Further coordinated observations may be required to determine the behaviour of this source, and if it is variable depending on observing frequency.

\subsection{The Potential for Further Novel Period Solutions}
While 5 of the sources detected as a part of this work have lacked a rotation period and do not meet the burst rate requirements for the periods to be solved by standard methods, there is potential for timing models to be derived from an extended monitoring campaign. The in-development, sparse-sampled period solver \texttt{altris}~\citep{morelloPrivateComms} has been demonstrated to be capable of solving RRAT periods with a limited set of TOA measurements taken from this work, and we hope that after further detections it may be possible to solve rotation periods for a number of these sources.

We highlight the near-term potential for Pushchino sources PSR J0939+45 and PSR J1218+47, and the CHIME source PSR J2138+69 to be solved by this tool. These sources have been detected on a number of occasions by I--LOFAR, despite their lack of known period. While these sources have low burst rates compared to other sources in this work (0.16(9), 0.37(10) and \SI{0.292(97)}{\per\hour}), they have been detected on a regular basis due to a number of extended observations over recent months. While these burst rates are low, with sufficient observing time allocated there should be sufficient information for the \texttt{altris} tool to solve the rotation models of these sources.

\subsection{Viability of Single LOFAR stations for RRAT Monitoring}
This work has shown that nearly a third of known Northern Hemisphere RRATs can be detected with an international LOFAR station working as a standalone instrument. Given the lack of follow-up and systemic analysis of a large amount of these sources, ongoing monitoring with the 14 international stations could allow for more details of these elusive neutron stars to be unearthed, especially in the case of sources with significantly low burst rates that can't be easily monitored.

The planned scheduling improvements being rolled out for LOFAR 2.0 will give rise to the LOFAR Mega Mode\footnote{\href{https://web.archive.org/web/20220703090811/https://www.astron.nl/what-we-look-forward-to-in-lofar-2-0-a-brain-transplant-for-lofar/}{ASTRON.nl: What we look forward to in LOFAR 2.0: A brain transplant for LOFAR}}. This automated scheduler allows for the international stations to be scheduled to be controlled independent of the core and other station while the telescope operates in `international mode', and is expected to provide a wealth of new data for pulsar timing and other monitoring campaigns. 

This work shows that RRATs (and other similarly bright radio transients) could be another pool of sources that can be sampled for monitoring efforts with the new scheduler. Several sources have been shown to have low-burst rates (J0348+79, J2355+1523), or appear extremely intermittent (J1329+13), when observed with a single station, such that they are detected too infrequently to be reliably monitored alongside other observation campaigns at a single station during `local mode' observations. Mega Mode would allow for these sources to be monitored during what would otherwise be downtime for international stations while observations are taken with the core, and help further investigate timing solutions and improve the statistical uncertainty on the observable parameters of these undersampled sources.

\subsection{Fast Radio Burst Search}\label{sec:frbdiscuss}

As mentioned in~\S\ref{sec:processing}, each observation taken as a part of this census was additionally searched for single-pulse candidates between \SIrange{10}{500}{\parsec\per\centi\metre\cubed} to contribute to a directionally biased search for FRBs as a part of this census. However, no credible candidates (apart from the Galactic sources discussed in \ref{sec:blindresult}) were detected as a part of this work.

% Given the burst amplitude distribution for FRB 20180916B provided by \cite{pastorMarazuela2021}, an estimate can be made as to the burst rate of such a source with an international LOFAR station. Across \SI{57.6}{\hour} of observations with the LOFAR core, the brightest burst was detected at a SNR of 35.1. Given the 24-core stations have around 5 times the sensitivity of an international station (six times the collecting area, 0.85 coherent beam-forming correction factor as discussed in \cite{kondratievLOFARCensusMillisecond2016a}), such a pulse would have a SNR of around 7 at an international station. With a 30 per cent increase in SNR, a pulse from such a FRB could be definitely detected at an international station. For their fit burst amplitude-frequency power law with $\Gamma = -1.5(2)$, this would happen once every 84 hours during the active mode of R3.

A blind FRB search at \SI{145}{\mega\hertz} has been performed using the ARTEMIS backend at the Rawlings (UK) and Nançay (FR) LOFAR stations \citep{karastergioArtemis2015}, using $8\times$\SI{ 6}{\mega\hertz} bandwidth beams to increase the observed sky area from a normal HBA beam field of view of \SI{3.6}{\square\deg} to \SI{28.8}{\square\deg} for a similar amount of time to this census (\SI{1446}{\hour}). From this, they determined an upper limit on the fast radio burst rate of \SI{29}{\per\sky\per\day} in the \SIrange{143}{149}{MHz} range, and a dispersion measure search for pulses at up to \SI{300}{\parsec\per\centi\metre\cubed}.

This work consisted of a similar amount of observing time, but this work has used a wider bandwidth, with an upper limit of \SI{89}{\mega\hertz} after RFI flagging, rather than the wider field of view used in  the work of \citeauthor{karastergioArtemis2015}. Considering the overall volume of sky probed, the sky sampled as a part of this work was probed to a depth 7.6 times greater than their work, however as a consequence of the reduced field of view, we only reach an upper limit on the low--frequency fast radio burst rate that is 92 per cent ($\sim$\SI{29}{\per\sky\per\day}) of that presented by \citeauthor{karastergioArtemis2015}. % Rate comparison = Tobs ratio*Volume ratio - (1408/1446)*(((sqrt(89/6))^(1.5))/8)
It should additionally be noted that a fraction of time spent observing was near the Galactic plane, which can be considered occulting due to the high sky temperatures severely decreasing the telescope's sensitivity (\S\ref{sec:sensitivity}).

%Multiple orders of magnitude more observing time would be necessary in order to approach the upper limit set by \citeauthor{karastergioArtemis2015} in the standard observing mode used for this work.

It is worth highlighting that the \texttt{heimdall} search software used for the census is optimised to detect broadband, spectrally flat sources, which doesn't describe the burst structures of many types of fast radio transients. Such behaviour has been clearly documented for FRB 20180916B at the LOFAR core, see~\cite[Fig.~2 and Table~1 of][]{pleunisLOFARR32021}. As a result, reprocessing the data across subsections of the bandwidth is planned for the future, and may result in previously missed candidates being detected.

\section{Conclusions and Future Work}\label{sec:conclusions}
We have shown that a reasonable number of Northern Hemisphere RRATs can be both detected and monitored with a single international LOFAR station. From the 29 detected sources, 4 rotation periods have been updated and 8 new phase-coherent pulsar timing ephemerides have been presented. 15 sources had sufficient detected bursts that their emission properties could be studied.

Observations are ongoing at I--LOFAR to continue the census for newly announced sources and slowly observe the sources that were not initially observed from the RRATalog (as described in~\S\ref{sec:rratalog}), while also monitoring and timing a number of the sources mentioned in this work. Plans are in place to start extracting full Stokes $IQUV$ data to characterise the polarisation properties of pulses from these sources in the near future.

Plans were already made during the census to inspect some sources detected in early 2021 and 2022 at other frequencies and higher sensitivities at \SI{150}{\mega\hertz}. As a result, observations and initial analysis have been completed for a sub-set of the discussed sources at some of the most sensitive instrument at their respective observing bandwidths. These include observations at FAST (L-band), the LOFAR core (24 stations, \SI{150}{\mega\hertz}) and NenuFAR (72 mini-arrays, \SI{50}{\mega\hertz}).

The RRAT census dataset itself also has a high legacy value, which is likely to yield more scientific results in the future. For instance, blind periodicity searches could yet be performed on the census data; these were computationally infeasible at the time of data collection, but this is becoming ever more tractable with time. To aid in such investigations, a publicly accessible I–LOFAR data archive is currently being created, the details of which will be reported when it is complete. Such an archive will facilitate archival analysis as new ideas arise and/or when improved/faster algorithms allow for deeper studies of extant data.

\section*{Acknowledgements}\label{sec:acknowledgements}
The Irish LOFAR Consortium consists of Trinity College Dublin, Armagh Observatory and Planetarium, University College Dublin, Dublin City University, University College Cork, University of Galway, Dublin Institute for Advance Studies and Technological University of the Shannon Athlone. It receives generous funding from Science Foundation Ireland and the Department of Further and Higher Education, Research, Innovation and Science.

The REALTA compute cluster was funded by Science Foundation Ireland.

We acknowledge use of the BSA-Analytics project catalogues, provided at \url{https://bsa-analytics.prao.ru/en/}, the CHIME/FRB Public Database, provided at \url{https://www.chime-frb.ca/} by the CHIME/FRB Collaboration and the RRATalog, provided at \url{http://astro.phys.wvu.edu/rratalog/} by Cui and McLaughlin.

D.J.McK is receiving funding under the Government of Ireland Postgraduate Scholarship (GOIPG/2019/2798) administered by the Irish Research Council (IRC).

%%%%%%%%%%%%%%%%%%%%%%%%%%%%%%%%%%%%%%%%%%%%%%%%%%
\section*{Data Availability}\label{sec:data}
The data of single pulses, pulse profiles and derived data, can be found on Zenodo~\citep{dmckennaCensusData2023}. The code used to generate the final data products can be found on GitHub\footnote{\url{https://github.com/David-McKenna/RRATCensusScripts}}~\citep{dmckennaCensusScripts2023}. The timing ephemerides found in Table~\ref{tab:newephemerides} can be found on Zenodo~\citep{dmckennaTimingData2023}.

Access to the archived datasets described in~\S\ref{sec:processing} can be made available upon request to the authors.



\bibliographystyle{mnras}
\bibliography{rratCensus} % if your bibtex file is called example.bib

%\begin{thebibliography}{99}
%\end{thebibliography}

%%%%%%%%%%%%%%%%%%%%%%%%%%%%%%%%%%%%%%%%%%%%%%%%%%

%%%%%%%%%%%%%%%%% APPENDICES %%%%%%%%%%%%%%%%%%%%%
\newpage
\appendix

\section{Observed sources}
Table~\ref{tab:sourcemappings} contains the name pairs for sources that were discussed in this work, but differ in name from their original announcement from a catalogue mentioned in \ref{sec:sources}. This table prioritises timing solutions from this paper or other recent works, or the name associated with the source in v1.67 of \texttt{psrcat}. The resolution of declination parameters has reduced to only include the degree digits in the case that a source has not been timed, nor observed with a telescope that has a sufficient resolution to determine the sky location with a sub-minute resolution.

Table~\ref{tab:sourcetobs} contains a list of all sources observed as a part of this census that met the minimum \SI{5}{\hour} observing time criteria. The provided list of observing time covers any observations that were performed, searched and archived between August 2020 and October 2022.

Non-detected sources can be considered to have an upper bound on their burst rate of at least \SI{0.2(2)}{\per\hour} at the sensitivity of a single international LOFAR station at \SI{151}{\mega\hertz} as a result of this work.

\begin{table*}
    \centering
    \caption{A table noting the differences in detected source names between the original catalogue names, and the current names used in this work from pulsar timing results, other publications, or \texttt{psrcat}.}
    \label{tab:sourcemappings}
    
    \begin{minipage}{0.44\columnwidth}
\centering
    \begin{tabular}{ll}
    \hline\hline
    Cat. Name & Source Name \\
    \hline
J0054+66 & J0054+6650 \\
J0103+54 & J0102+5356 \\
J0139+3310 & J0139+3336 \\
J0209+58 & J0209+5759 \\
J0227+33 & J0226+3356 \\
J0318+1341 & J0317+1328 \\
\hline\hline
    \end{tabular}
\end{minipage}\hspace{0.02\columnwidth}
\begin{minipage}{0.44\columnwidth}
\centering
\begin{tabular}{ll}
\hline\hline

Cat. Name & Source Name \\
\hline
\multirow{2}{*}{J0332+7910} & J0332+79 \\
& J0337+79\textsuperscript{\textdagger}\\
J0746+55 & J0746+5514 \\
J0854+54 & J0854+5449 \\
J1005+3015 & J1006+3015 \\
J1132+2515 & J1132+2513 \\
\hline\hline
    \end{tabular}
\end{minipage}\hspace{0.02\columnwidth}
\begin{minipage}{0.44\columnwidth}
\centering
\begin{tabular}{ll}
\hline\hline

Cat. Name & Source Name \\
\hline
J1329+1349 & J1329+13 \\
J1336+3346 & J1336+3414 \\
J1400+2127 & J1400+2125 \\
J1404+1210 & J1404+1159 \\
J1849+15 & J1848+1516 \\
J2105+1917 & J2105+19 \\
\hline\hline
    \end{tabular}
\end{minipage}\hspace{0.02\columnwidth}
\begin{minipage}{0.44\columnwidth}
\centering
\begin{tabular}{ll}
\hline\hline

Cat. Name & Source Name \\
\hline
J2108+45 & J2108+4516 \\
J2202+2147 & J2202+2134 \\
J2215+45 & J2215+4524 \\
J2355+15 & J2355+1523 \\
\\
\\
\hline\hline
    \end{tabular}
    \end{minipage}
    
    \bigskip
    {\footnotesize \textsuperscript{\textdagger} While the former entry is used in this paper, \texttt{psrcat} appears to have an updated source name and position for this source, however the cited source does not refer to the original or new source name, or any source with the associated parameters.}
\end{table*}

\begin{table*}
\centering
\caption{A list of all 114 pointings inspected for the $\observed$ observed sources, based on the source name initially provided by the catalogue prior to the first observation of the source. Sources denoted with dagger suffix are referred to by an updated name in previous sections (see mappings in Table~\ref{tab:sourcemappings}). Sources with a star suffix have had multiple pointings observed, see~\S\ref{sec:overlap} for further details on these sources.}
\label{tab:sourcetobs}
\begin{minipage}{0.3\columnwidth}
\centering
\begin{tabular}{ll}
\hline\hline
Source & Time  \\
& \SI{}{\hour} \\
\hline
J0012+54   & 9.40  \\
J0013+29   & 5.17  \\
J0027+84   & 5.20  \\
J0034+27   & 6.75  \\
J0054+66\textsuperscript{\textdagger}   & 31.50 \\
J0054+69   & 5.10  \\
J0103+54\textsuperscript{\textdagger}   & 27.28 \\
J0121+5329 & 5.15  \\
J0139+3310\textsuperscript{\textdagger} & 16.63 \\
J0201+7005 & 10.90 \\
J0209+58\textsuperscript{\textdagger}   & 26.62 \\
J0227+33\textsuperscript{\textdagger}   & 18.00 \\
J0301+20   & 5.33  \\
J0305+4001 & 5.05  \\
J0318+1341\textsuperscript{\textdagger} & 10.07 \\
J0327+09   & 5.95  \\
J0332+7910 & 15.42 \\
J0348+79   & 26.73 \\
J0452+1651 & 5.15  \\
J0503+22   & 5.62  \\
J0517+24   & 5.02  \\
J0534+3407 & 6.03  \\
J0544+20   & 5.17  \\
\hline\hline
\end{tabular}
\end{minipage}\hspace{0.1\columnwidth}
\begin{minipage}{0.3\columnwidth}
\centering
\begin{tabular}{ll}
\hline\hline
Source     &  Time  \\
& \SI{}{\hour} \\
\hline
J0609+1635 & 5.10  \\
J0625+1730 & 5.30  \\
J0627+16   & 5.35  \\
J0630+25   & 5.32  \\
J0640+0744 & 7.65  \\
J0653$-$06   & 5.37  \\
J0658+29   & 5.68  \\
J0740+17   & 5.67  \\
J0746+55\textsuperscript{\textdagger}   & 43.20 \\
J0803+3410 & 6.85  \\
J0812+8626 & 5.17  \\
J0854+54\textsuperscript{\textdagger}   & 24.20 \\
J0939+45   & 18.47 \\
J0941+1621 & 5.40  \\
J1005+3015\textsuperscript{\textdagger} & 95.62 \\
J1048+53   & 5.42  \\
J1106+02   & 5.17  \\
J1130+09\textsuperscript{*}   & 6.17  \\
J1132+0921\textsuperscript{*} & 6.15  \\
J1132+2515\textsuperscript{\textdagger} & 11.23 \\
J1218+47   & 34.68 \\
J1246+53   & 7.23  \\
J1329+1349\textsuperscript{\textdagger} & 49.62 \\
\hline\hline
\end{tabular}
\end{minipage}\hspace{0.1\columnwidth}
\begin{minipage}{0.3\columnwidth}
\centering
\begin{tabular}{ll}
\hline\hline
Source     &  Time  \\
& \SI{}{\hour} \\
\hline
J1336+3346\textsuperscript{\textdagger} & 79.00 \\
J1346+06   & 5.17  \\
J1346+0622 & 6.40  \\
J1354+24   & 5.40  \\
J1400+2127\textsuperscript{\textdagger} & 86.67 \\
J1404+1210\textsuperscript{\textdagger} & 5.73  \\
J1430+22   & 5.25  \\
J1432+09   & 5.42  \\
J1439+76   & 6.43  \\
J1502+2813 & 10.53 \\
J1538+2345 & 38.28 \\
J1541+47   & 7.03  \\
J1550+09   & 6.88  \\
J1550+0943 & 5.13  \\
J1554+18   & 5.17  \\
J1555+0108 & 8.18  \\
J1603+18   & 5.02  \\
J1639+21   & 5.05  \\
J1732+2700 & 9.35  \\
J1737+24   & 5.18  \\
J1838+5051 & 5.68  \\
J1841$-$0448 & 5.30  \\
J1849+15\textsuperscript{\textdagger}   & 32.50 \\
\hline\hline
\end{tabular}
\end{minipage}\hspace{0.1\columnwidth}
\begin{minipage}{0.3\columnwidth}
\centering
\begin{tabular}{ll}
\hline\hline
Source     &  Time  \\
& \SI{}{\hour} \\
\hline
J1850+15   & 5.37  \\
J1907+34   & 5.18  \\
J1917+1723 & 5.18  \\
J1919+17   & 5.18  \\
J1928+15   & 5.18  \\
J1928+17   & 5.10  \\
J1930+0104 & 5.17  \\
J1931+4229 & 53.72 \\
J1943+58   & 6.37  \\
J1945+61   & 5.90  \\
J1946+24   & 6.48  \\
J2002+13   & 5.08  \\
J2005+38   & 5.33  \\
J2007+20   & 5.15  \\
J2008+37   & 7.33  \\
J2047+1259 & 5.32  \\
J2047+13   & 5.18  \\
J2052+1308 & 5.08  \\
J2057+46   & 5.45  \\
J2105+1917 & 7.65  \\
J2105+6223 & 5.17  \\
J2107+26   & 5.18  \\
J2107+2606 & 5.17  \\
\hline\hline
\end{tabular}
\end{minipage}\hspace{0.1\columnwidth}
\begin{minipage}{0.3\columnwidth}
\centering
\begin{tabular}{ll}
\hline\hline
Source     &  Time  \\
& \SI{}{\hour} \\
\hline
J2108+45   & 6.23  \\
J2113+73   & 5.07  \\
J2116+37   & 5.65  \\
J2119+49   & 11.07 \\
J2135+3032 & 5.28  \\
J2138+69   & 30.83 \\
J2146+2148 & 5.18  \\
J2202+2147\textsuperscript{\textdagger} & 18.07 \\
J2205+2244 & 5.15  \\
J2208+46   & 5.22  \\
J2210+2118 & 5.07  \\
J2215+45\textsuperscript{\textdagger}   & 37.23 \\
J2221+81   & 6.95  \\
J2225+35   & 5.18  \\
J2237+28   & 5.17  \\
J2252+2451 & 5.20  \\
J2311+67   & 5.15  \\
J2316+75   & 5.17  \\
J2325$-$0530 & 7.03  \\
J2329+04   & 5.43  \\
J2355+15\textsuperscript{\textdagger}   & 51.92 \\
J2357+24   & 7.45  \\\\\hline\hline
\end{tabular}
\end{minipage}
\end{table*}

\section{Single-Pulse Profiles}
Fig.~\ref{fig:singleprofiles} provide plots of the 1-dimensional pulse profile for the brightest burst detected from each of the sources with single-pulse properties in Table~\ref{tab:obs_summary}.

These figures contain both the raw-sampled data, and a box-car averaged profile in the case of low-peak-significance pulses.

\begin{figure*}
    \centering

    \begin{tabular}{ccc}
\adjincludegraphics[height=0.14\textheight]{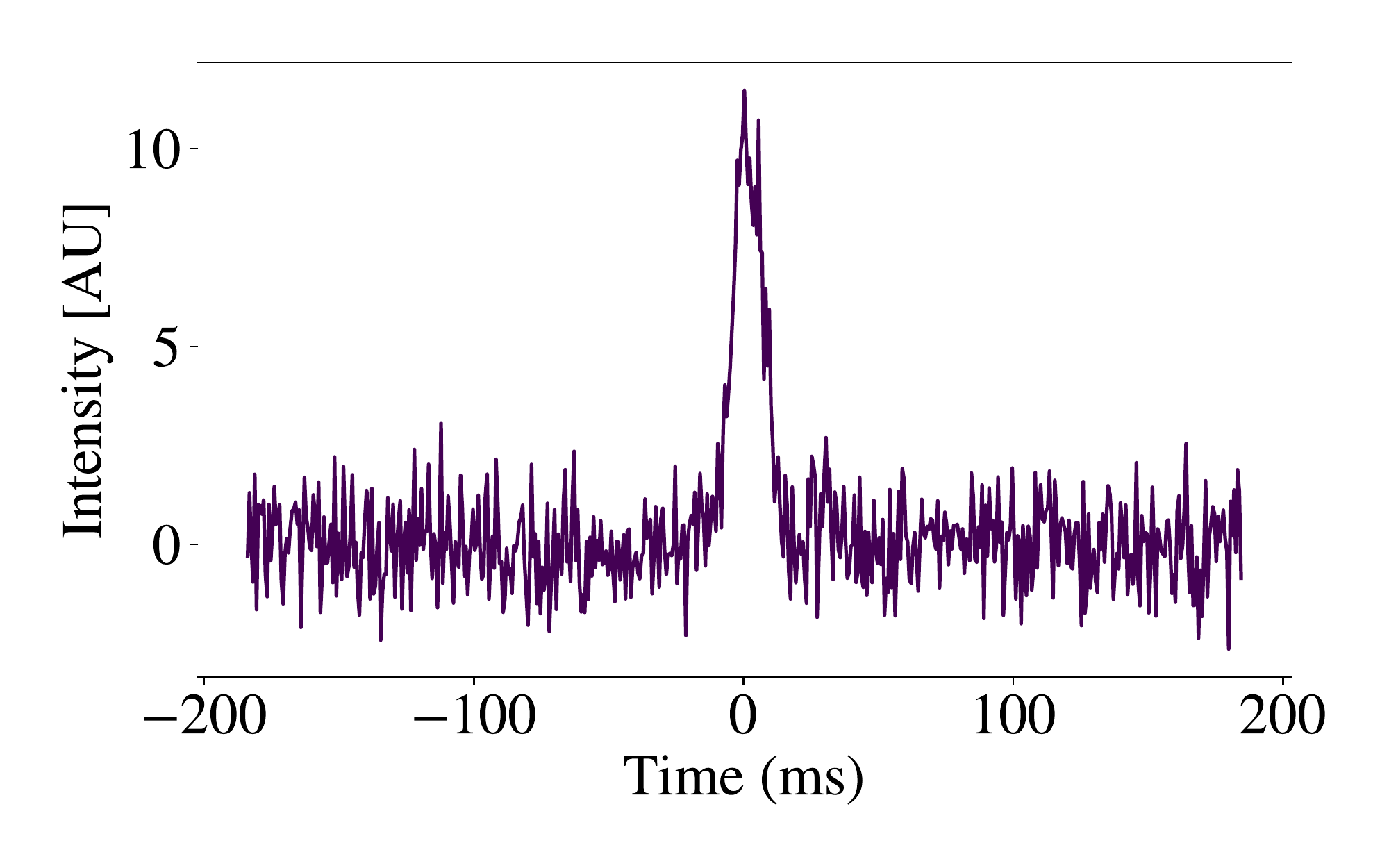} &\adjincludegraphics[height=0.14\textheight]{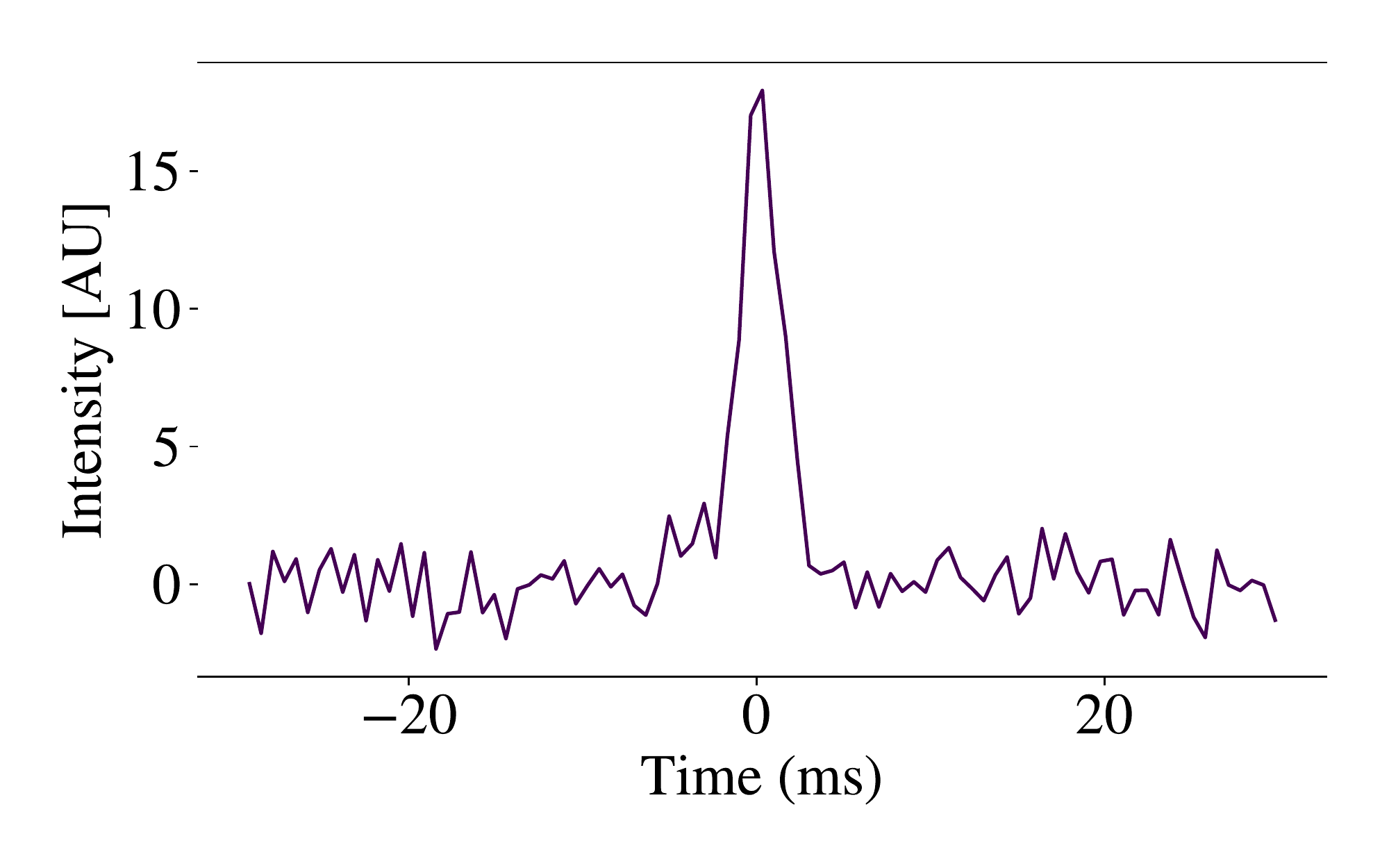} &\adjincludegraphics[height=0.14\textheight]{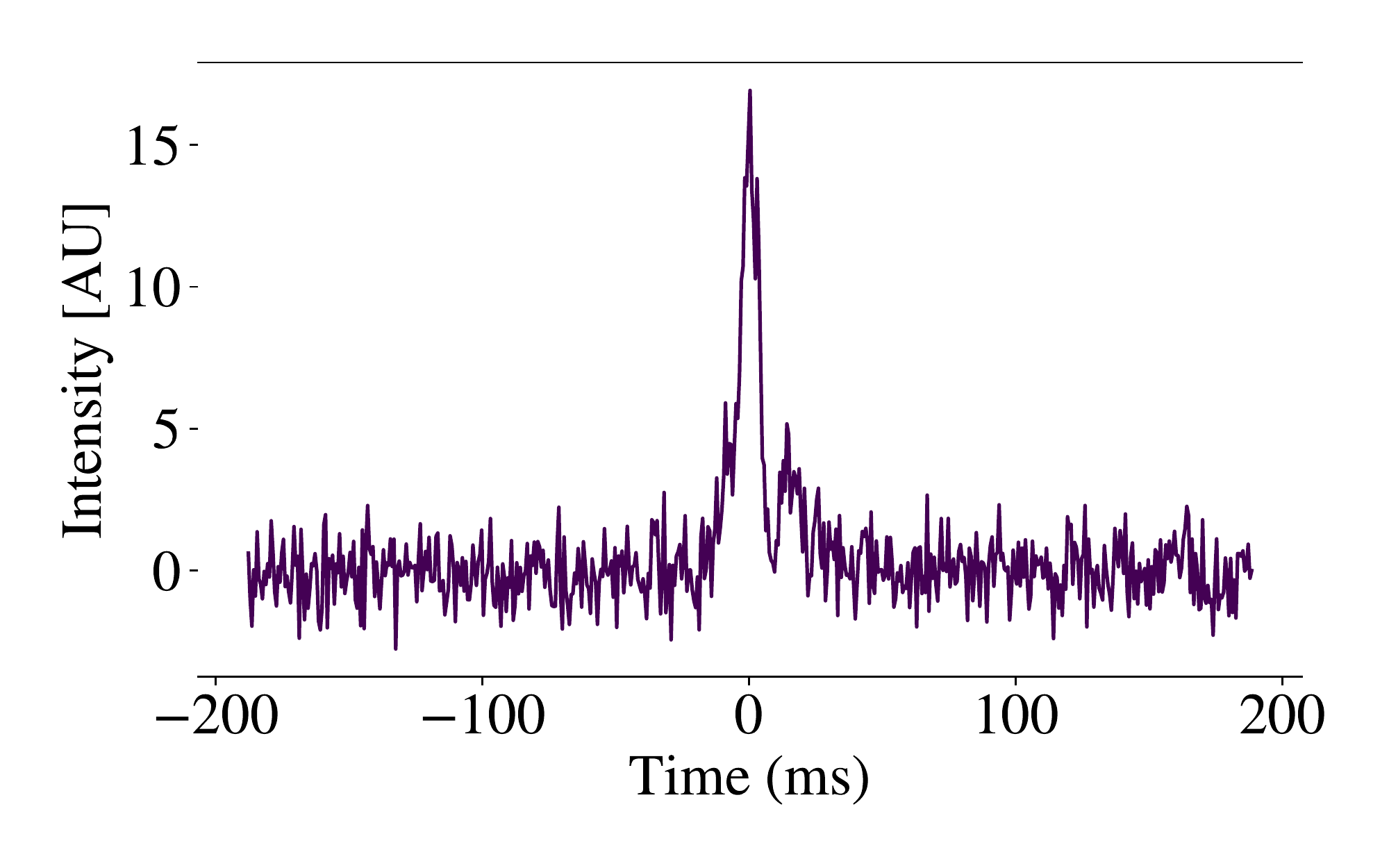} \\
\hphantom{3} (a) J0054+6650 &\hphantom{3} (b) J0102+5356 &\hphantom{3} (c) J0139+3336 \\\\
\adjincludegraphics[height=0.14\textheight]{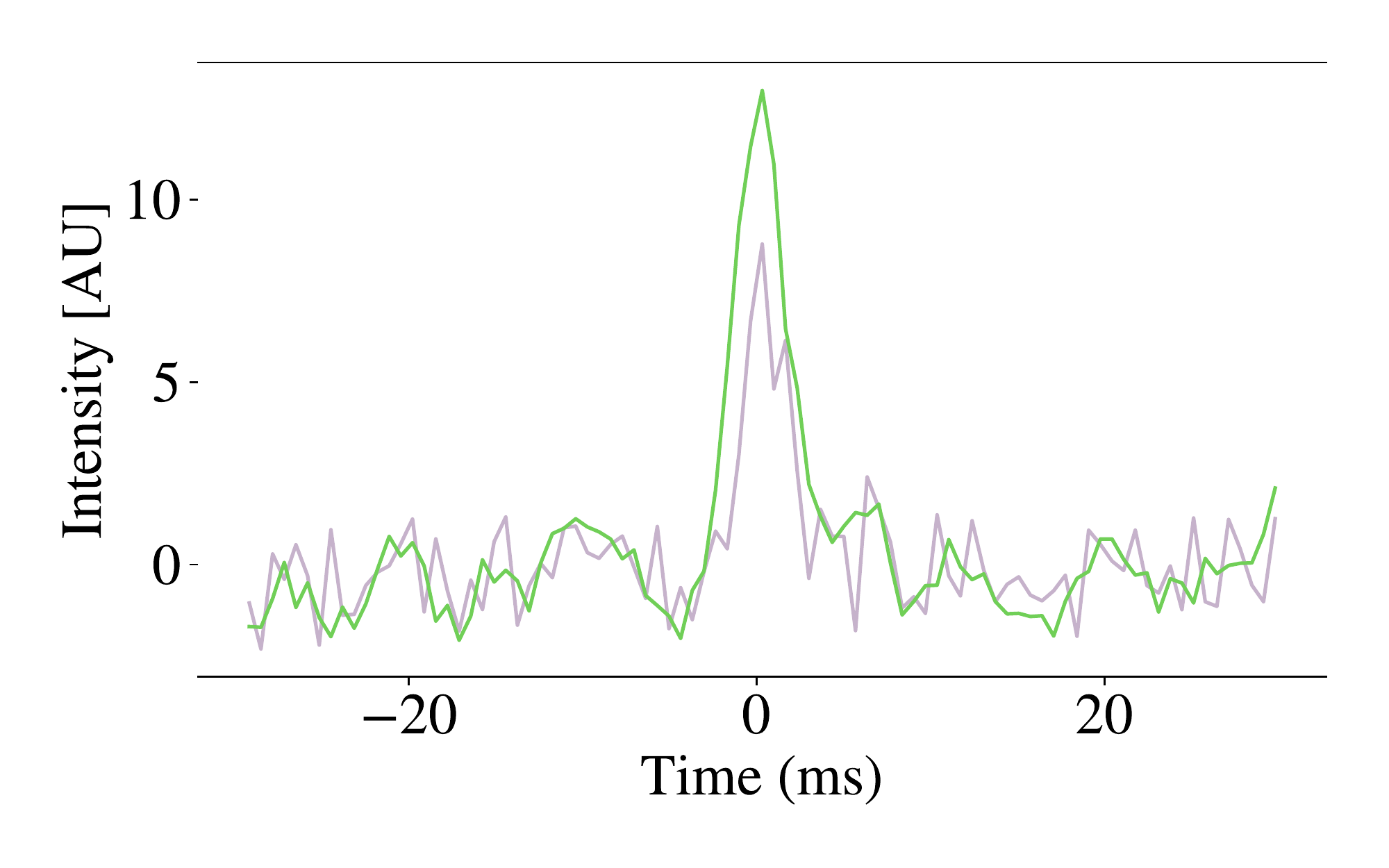} &\adjincludegraphics[height=0.14\textheight]{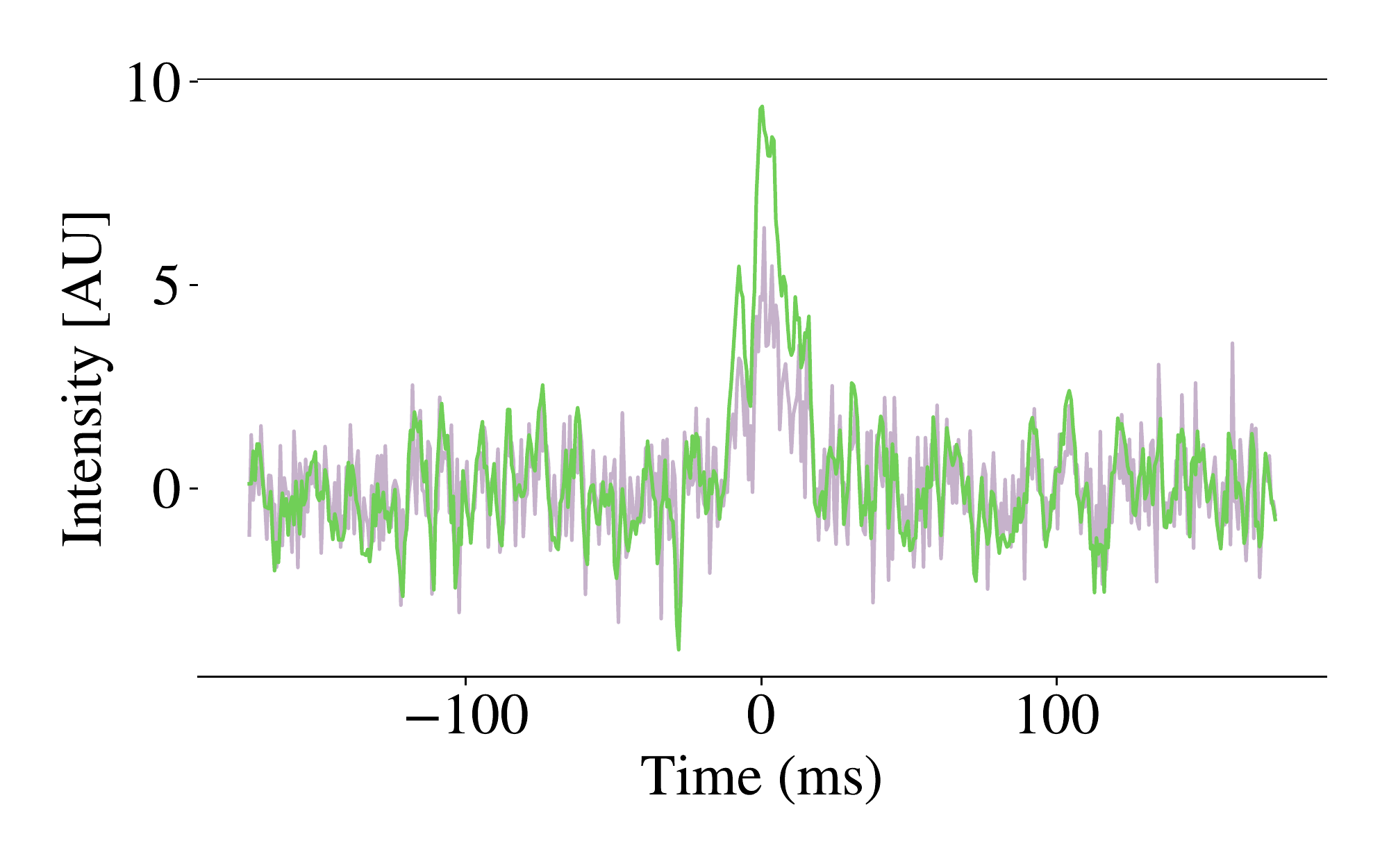} &\adjincludegraphics[height=0.14\textheight]{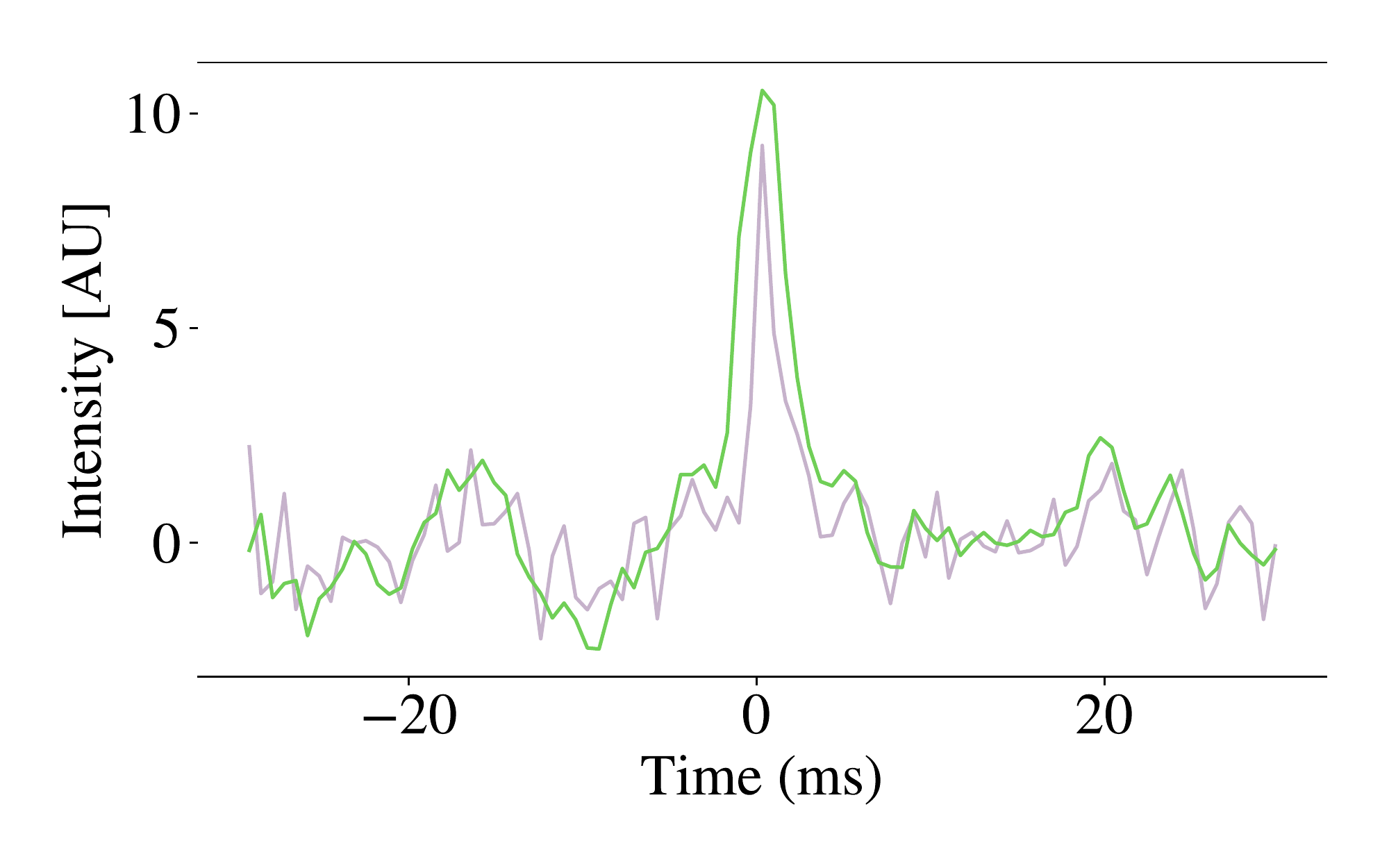} \\
\hphantom{3} (d) J0201+7005 &\hphantom{3} (e) J0209+5759 &\hphantom{3} (f) J0317+1328 \\\\
\adjincludegraphics[height=0.14\textheight]{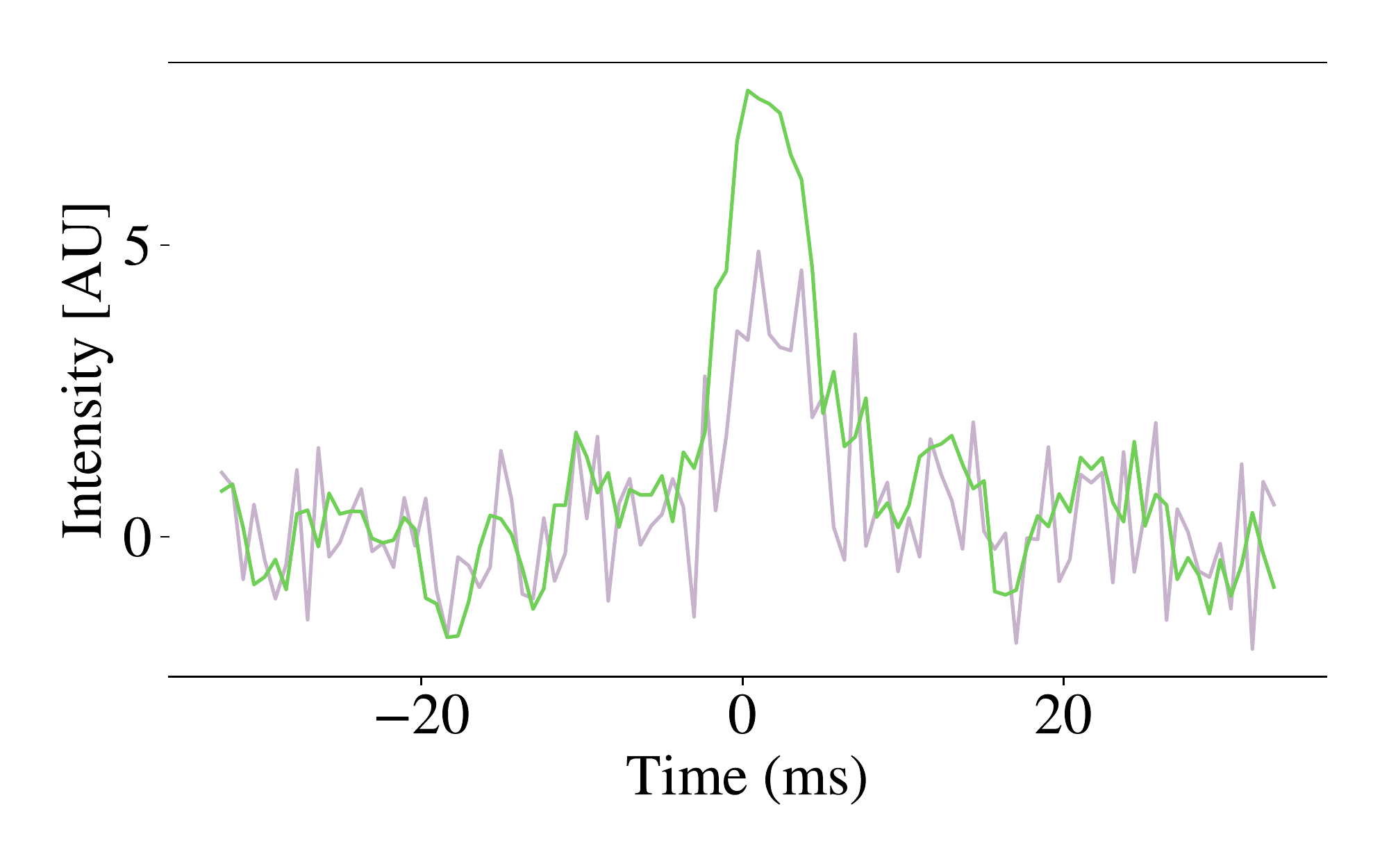} &\adjincludegraphics[height=0.14\textheight]{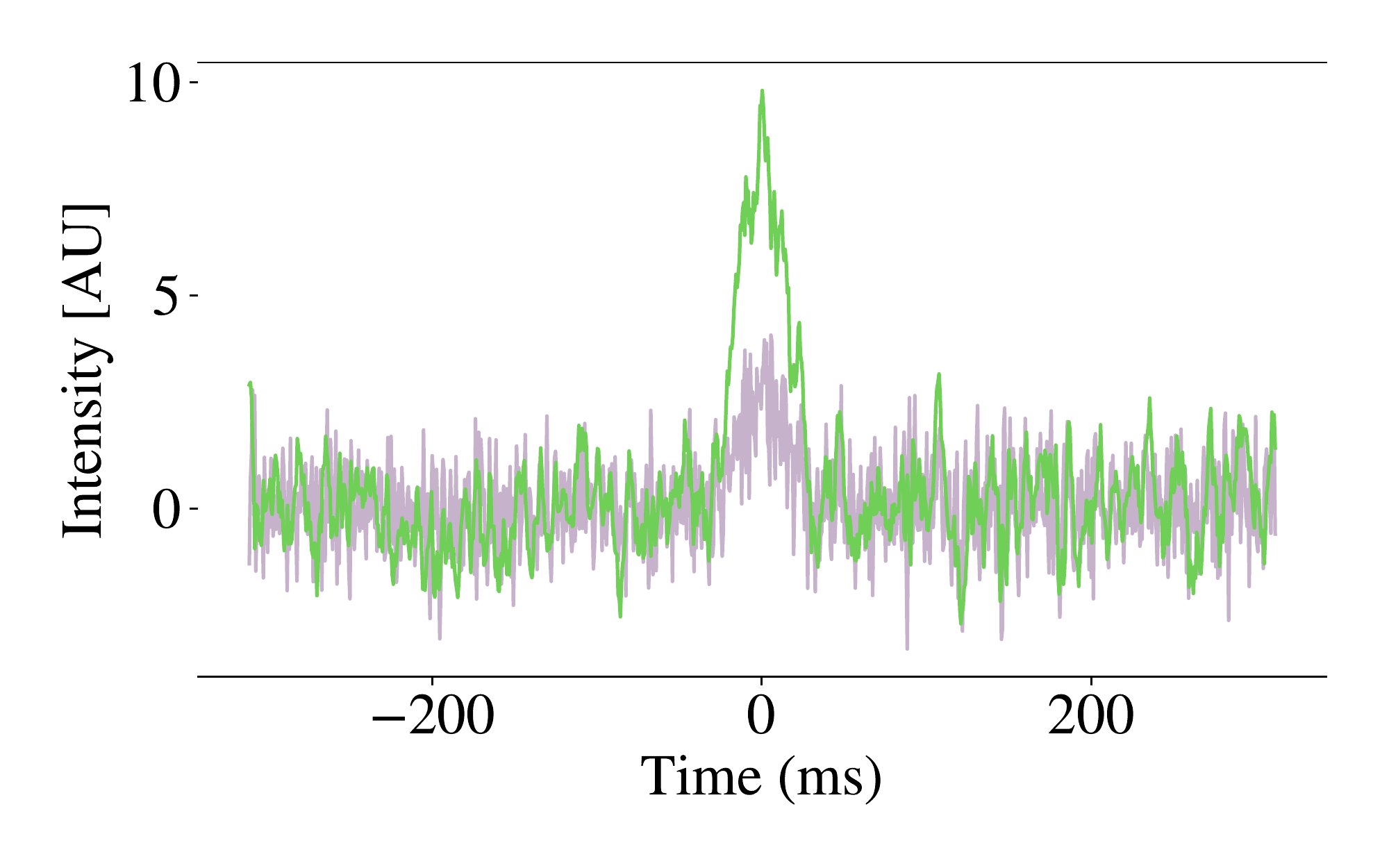} &\adjincludegraphics[height=0.14\textheight]{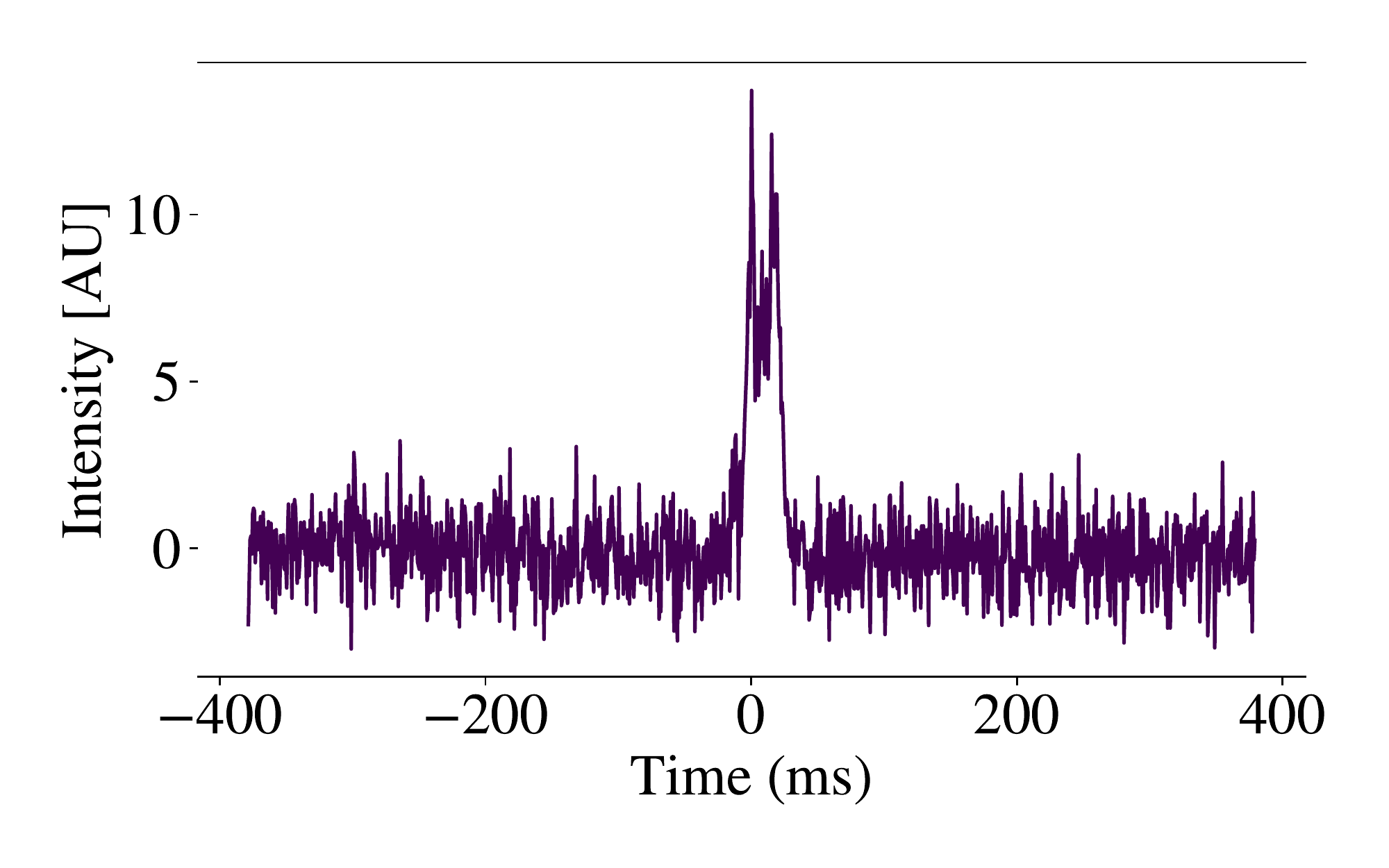} \\
\hphantom{3} (g) J0332+79 &\hphantom{5} (h) J0348+79 &\hphantom{3} (i) J0746+5514 \\\\
\adjincludegraphics[height=0.14\textheight]{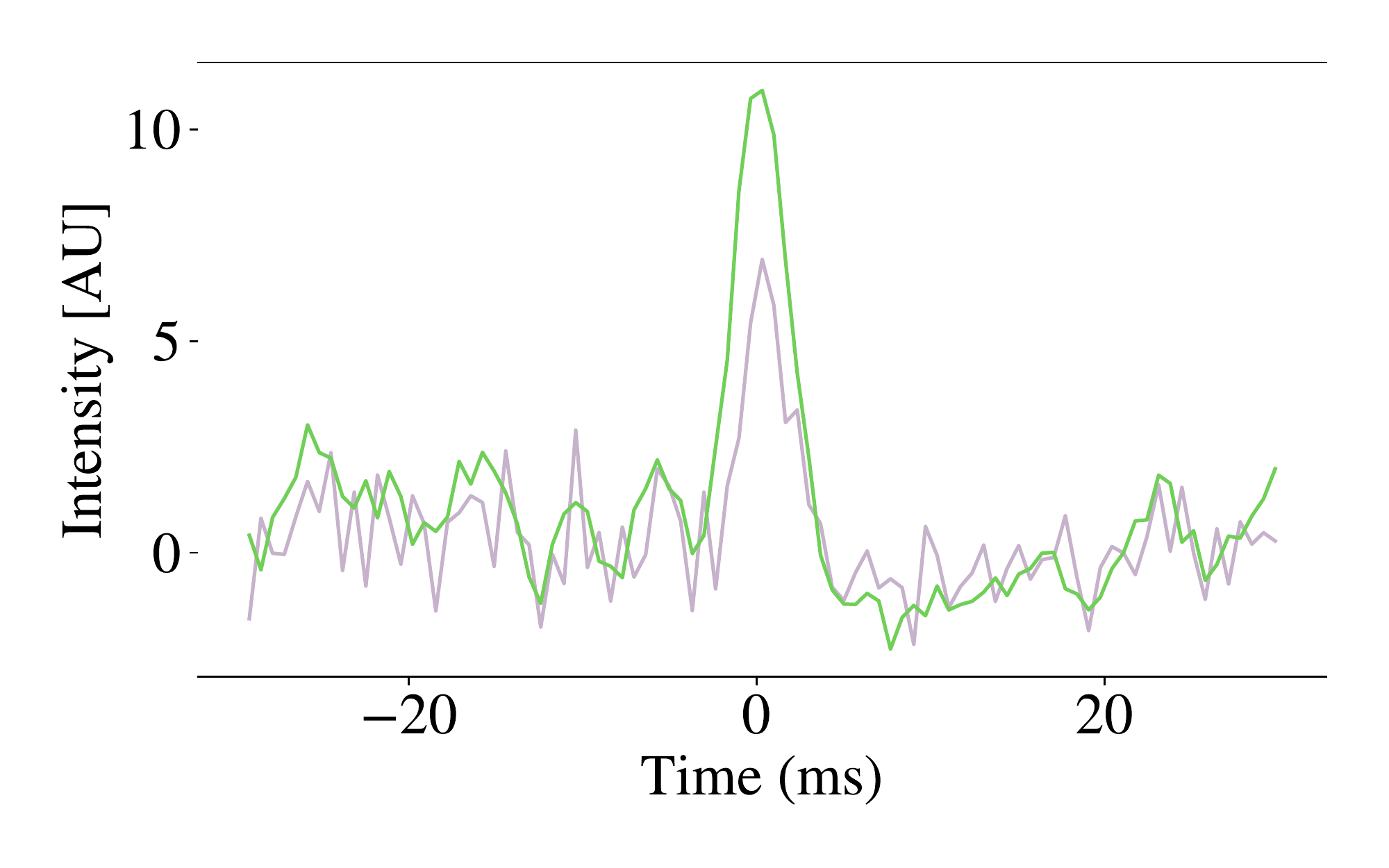} &\adjincludegraphics[height=0.14\textheight]{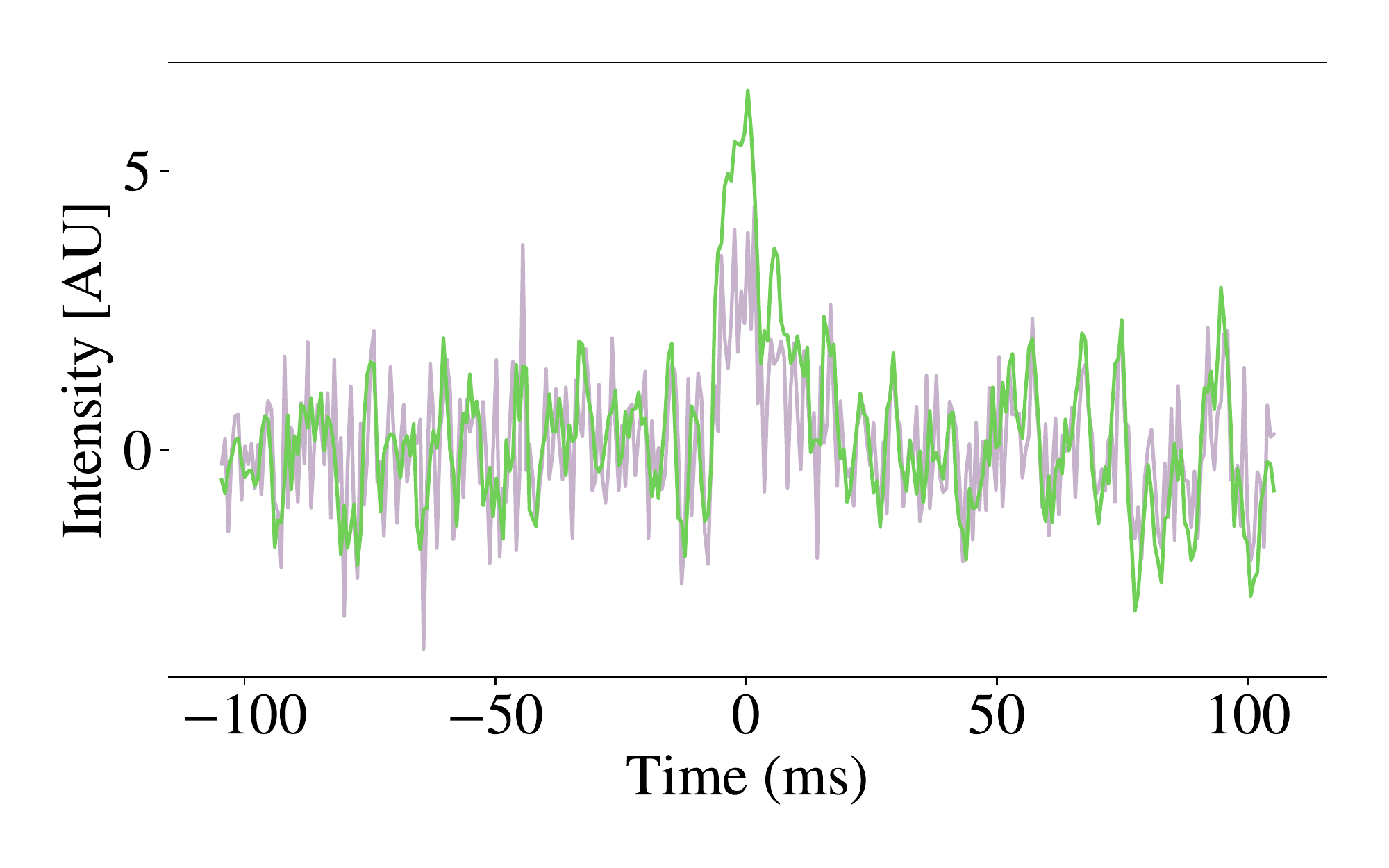} &\adjincludegraphics[height=0.14\textheight]{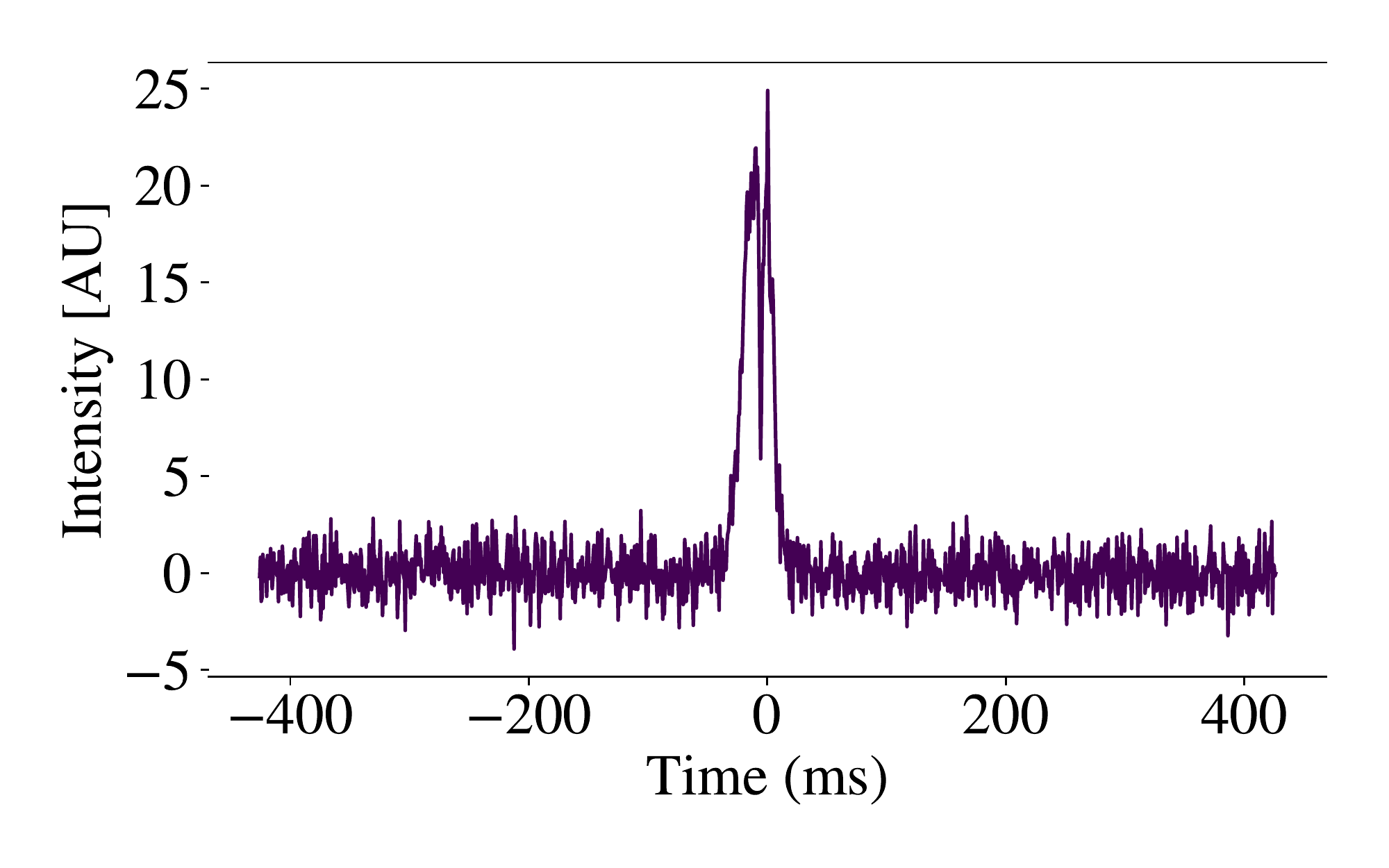} \\
\hphantom{3} (j) J0854+5449 &\hphantom{5} (k) J0939+45 &\hphantom{3} (l) J1006+3015 \\\\
\adjincludegraphics[height=0.14\textheight]{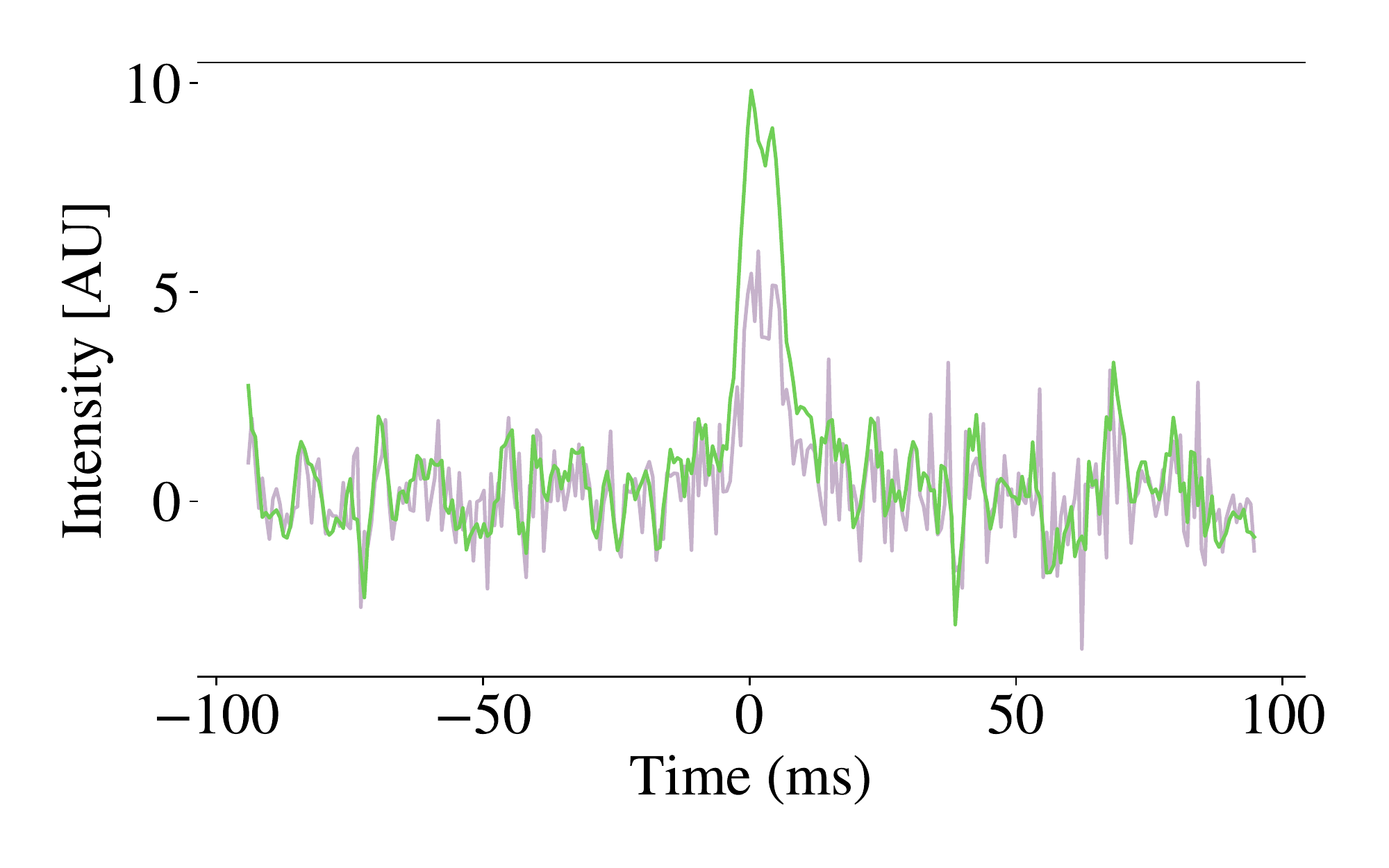} &\adjincludegraphics[height=0.14\textheight]{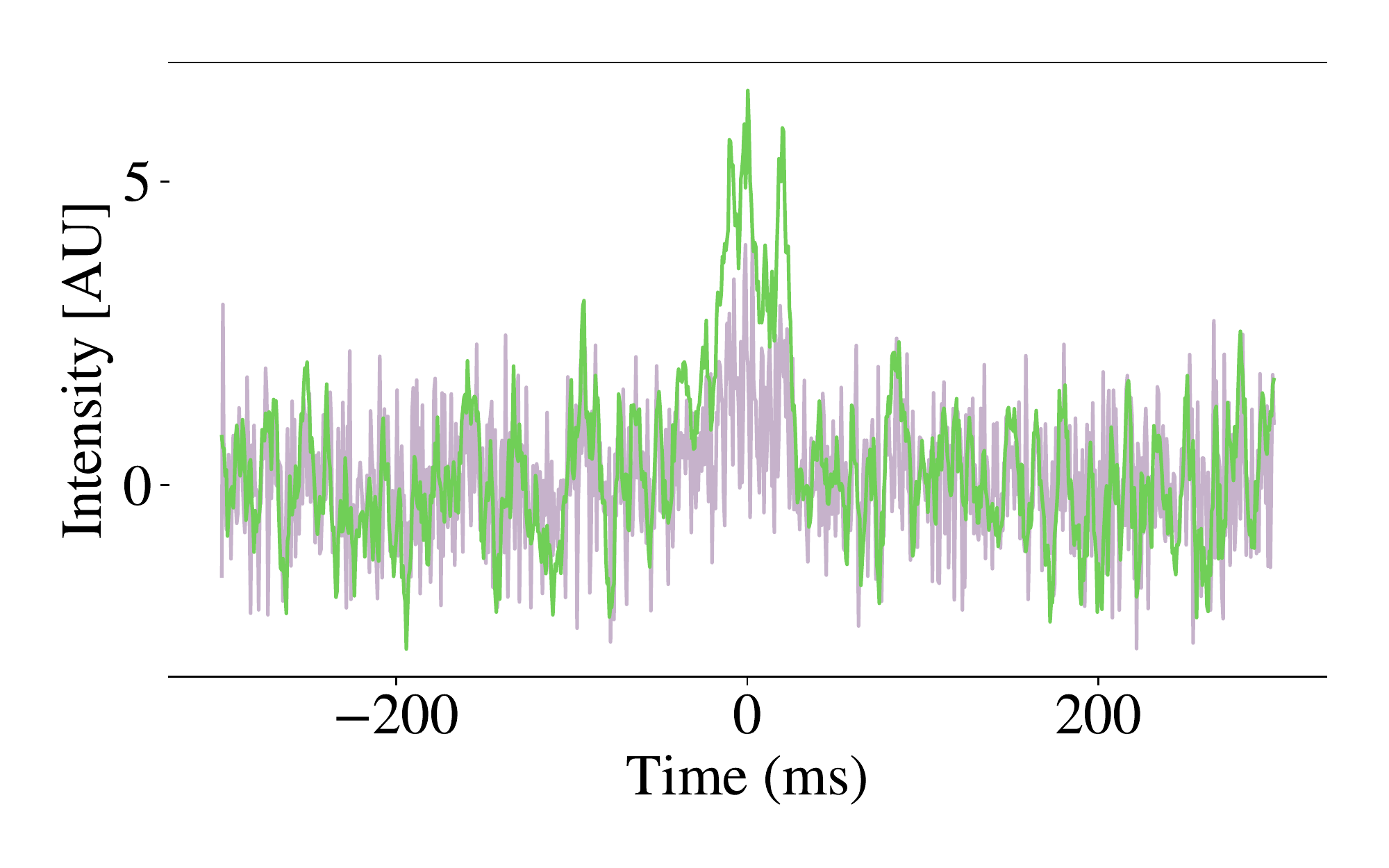} &\adjincludegraphics[height=0.14\textheight]{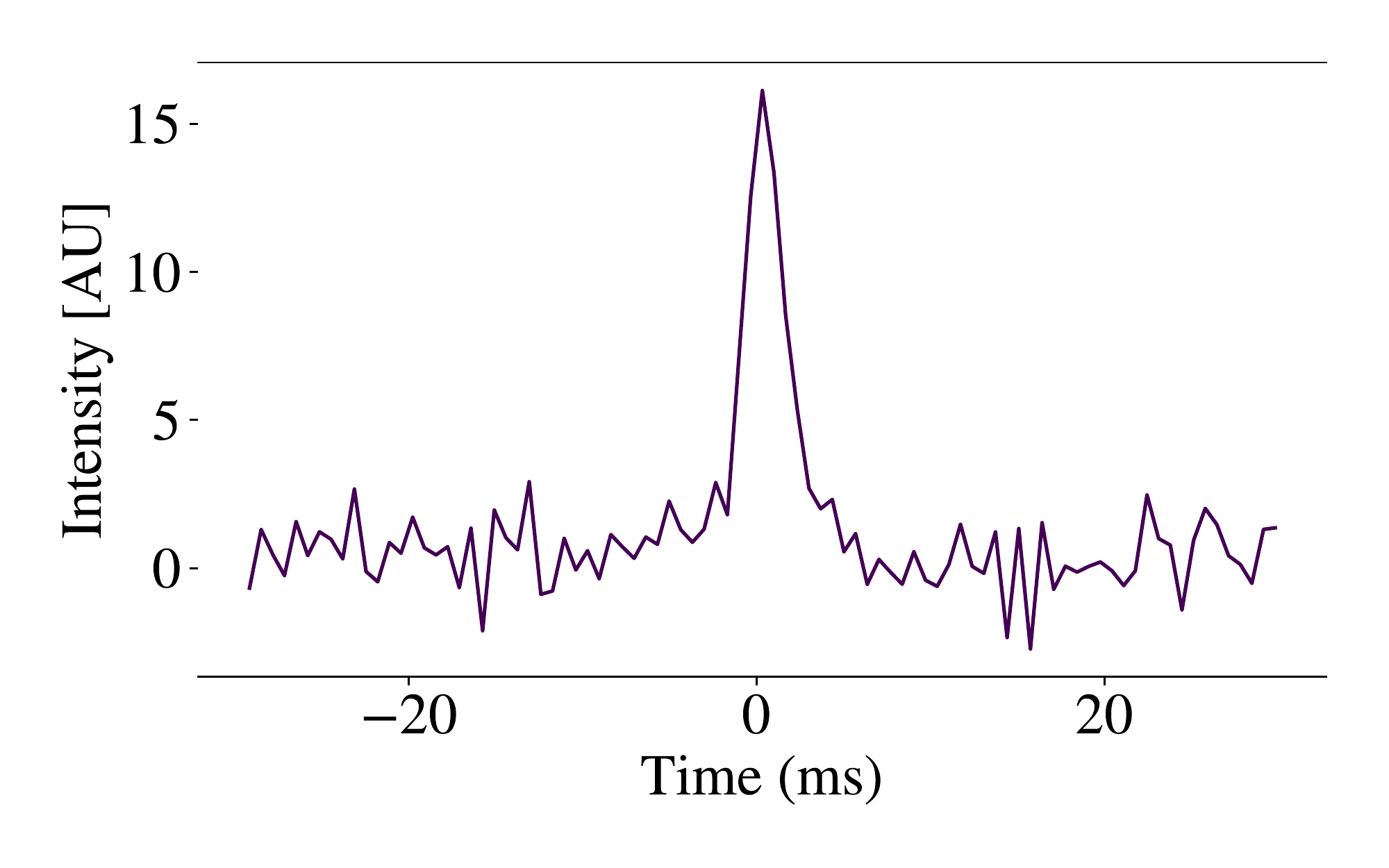} \\
\hphantom{5} (m) J1218+47 &\hphantom{3} (n) J1329+13 &\hphantom{3} (o) J1336+3414 \\

    \end{tabular}
    \caption{The brightest single-pulses for each source detected as a part of the census. For weak sources, a boxcar with a variable width (the larger choice of 4 bins and 1/8 of the pulse width) has been convolved with the data and plotted in light green. The time axis has been set to approximately 10 pulse widths. Continued on next page.}\label{fig:singleprofiles}
\end{figure*}

\begin{figure*}
    \centering
    \ContinuedFloat
    
    \begin{tabular}{ccc}
    
\adjincludegraphics[height=0.14\textheight]{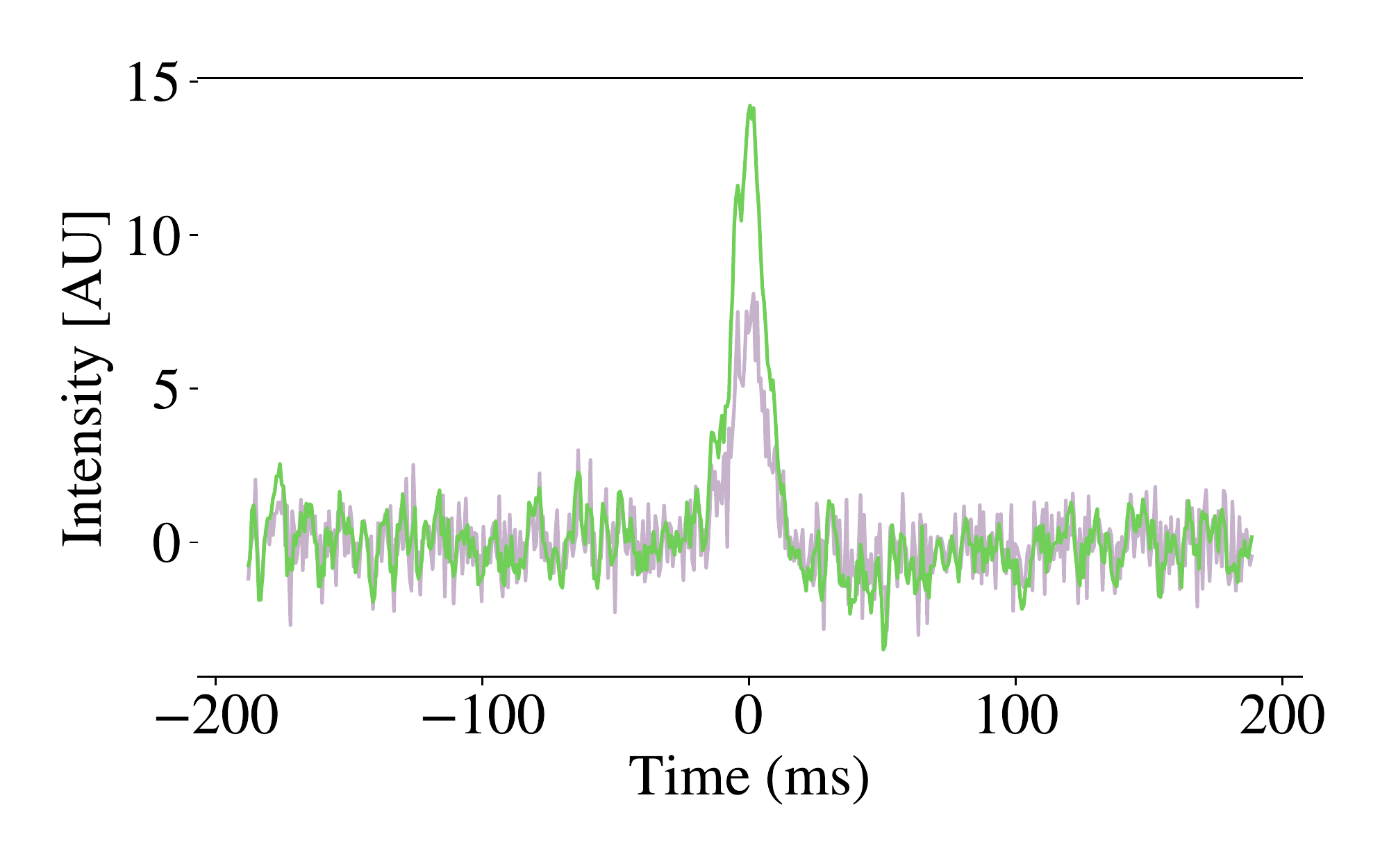} &\adjincludegraphics[height=0.14\textheight]{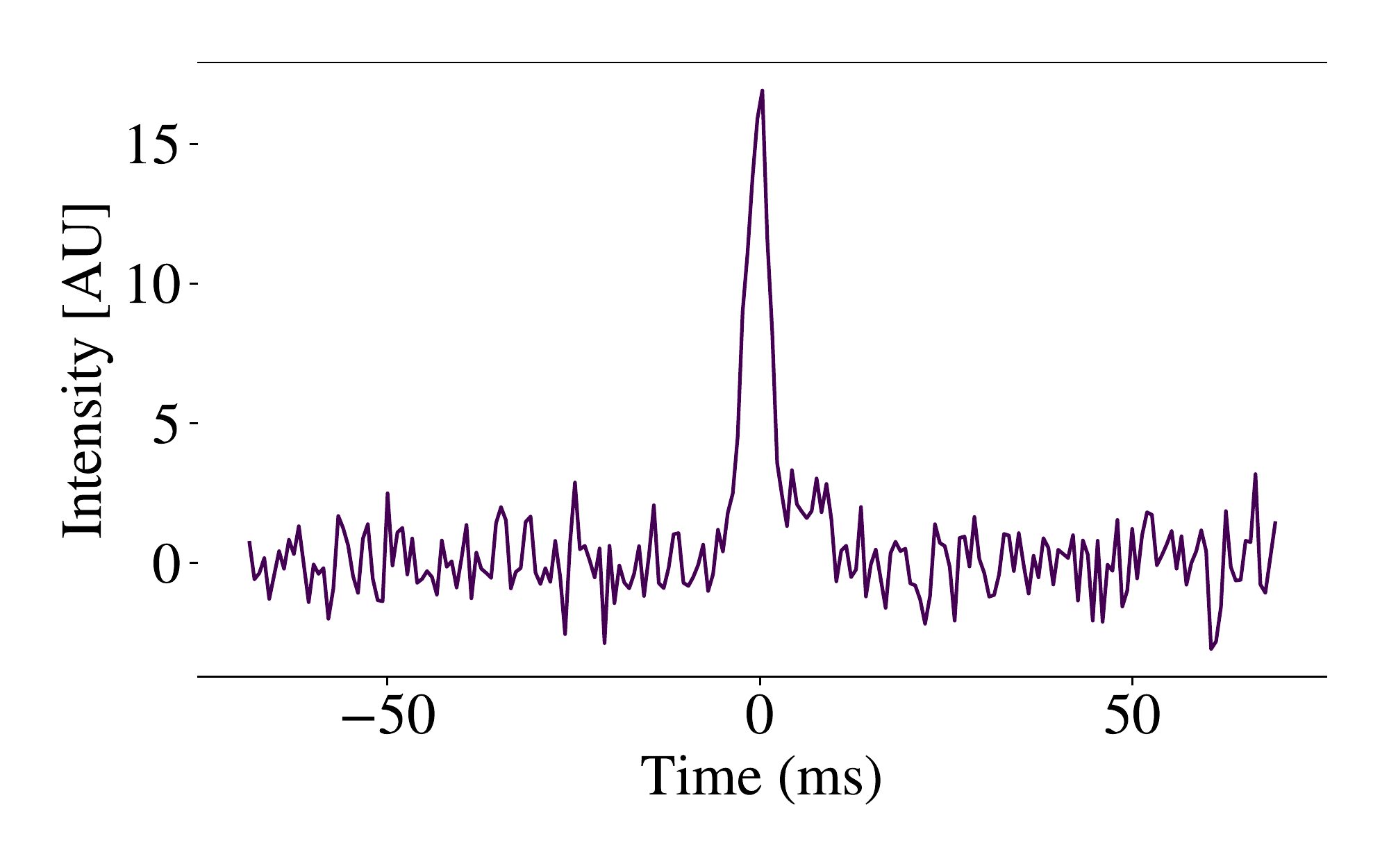} &\adjincludegraphics[height=0.14\textheight]{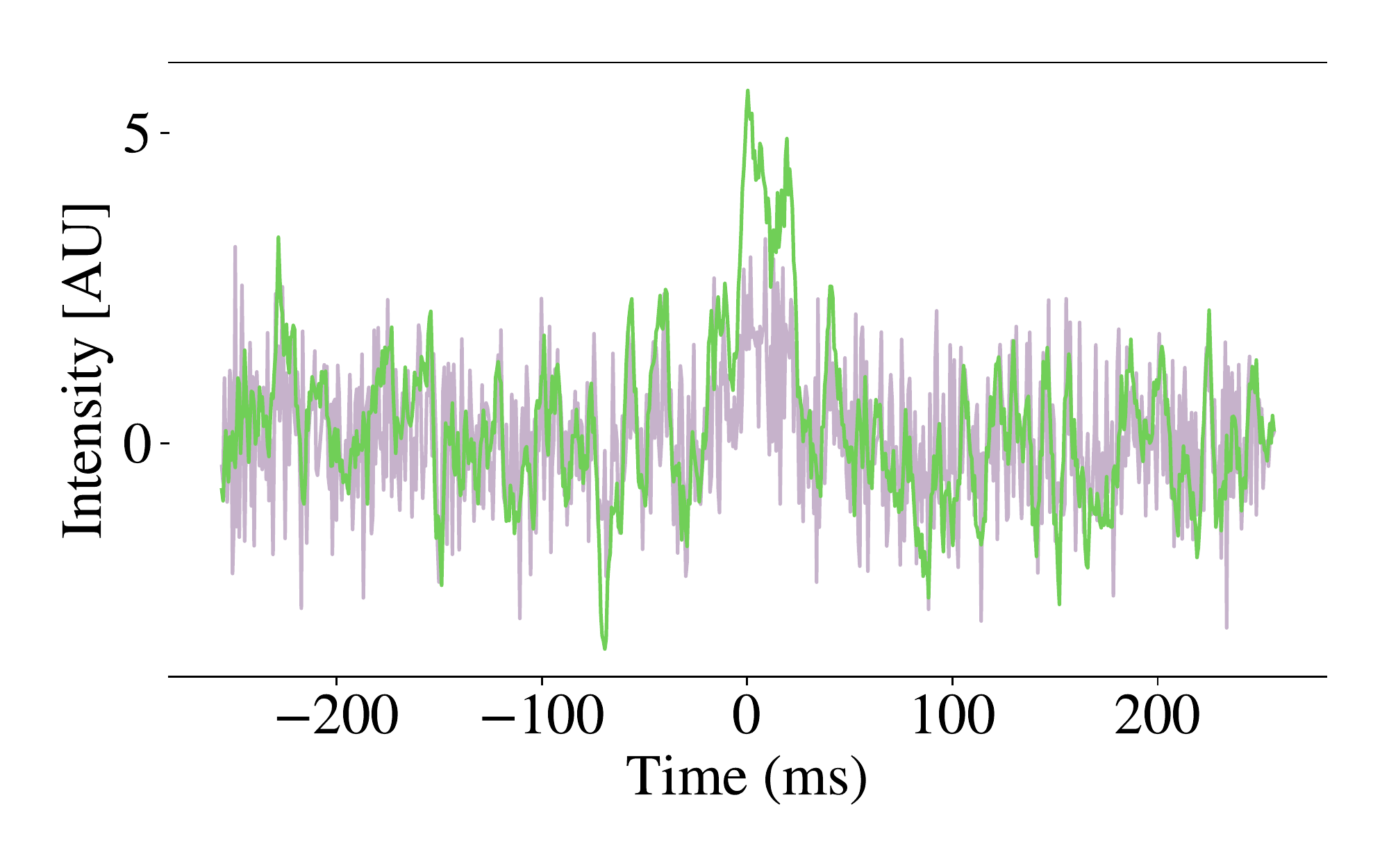} \\
\hphantom{3} (p) J1400+2125 &\hphantom{3} (q) J1538+2345 &\hphantom{3} (r) J1848+1516 \\\\
\adjincludegraphics[height=0.14\textheight]{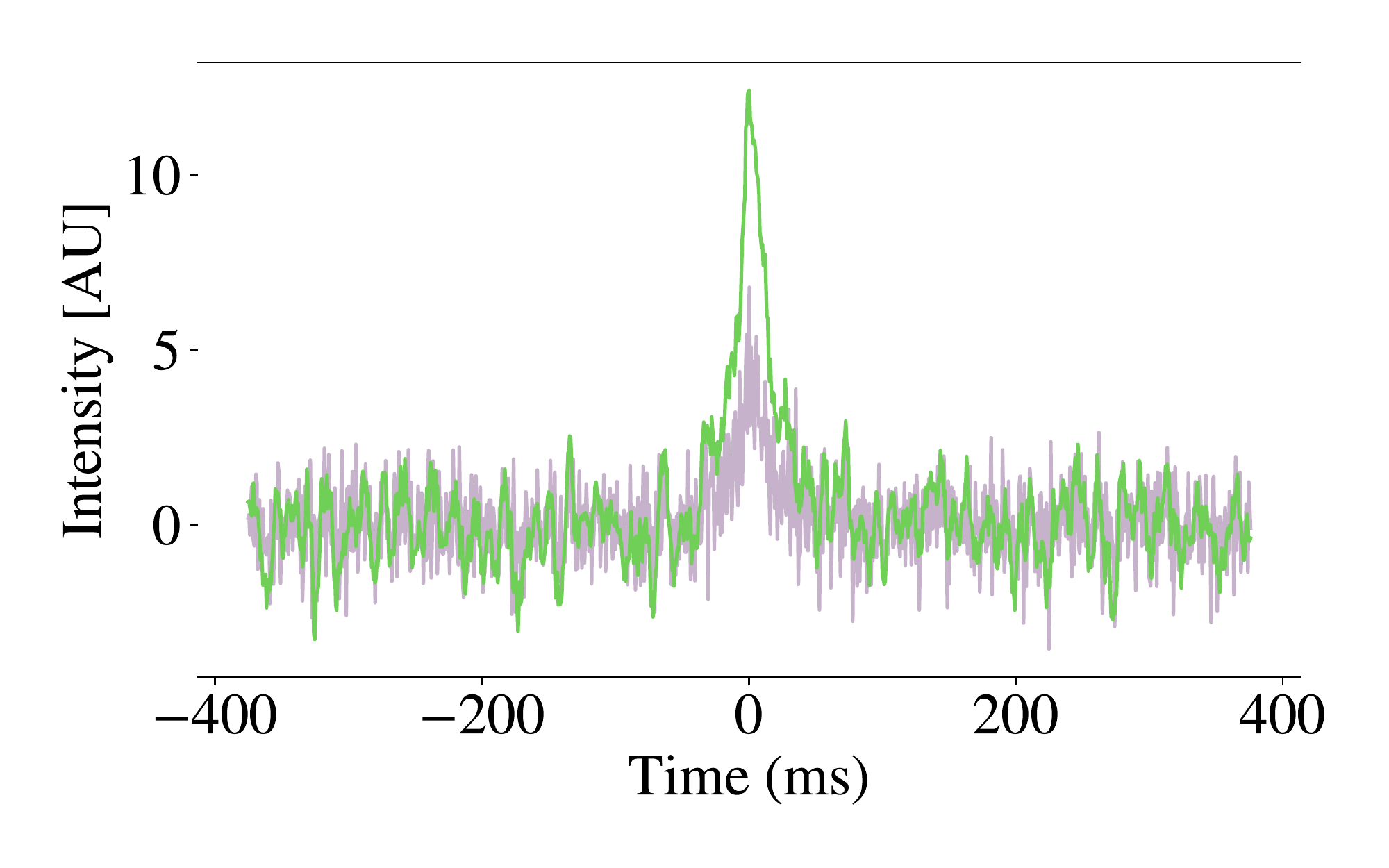} &\adjincludegraphics[height=0.14\textheight]{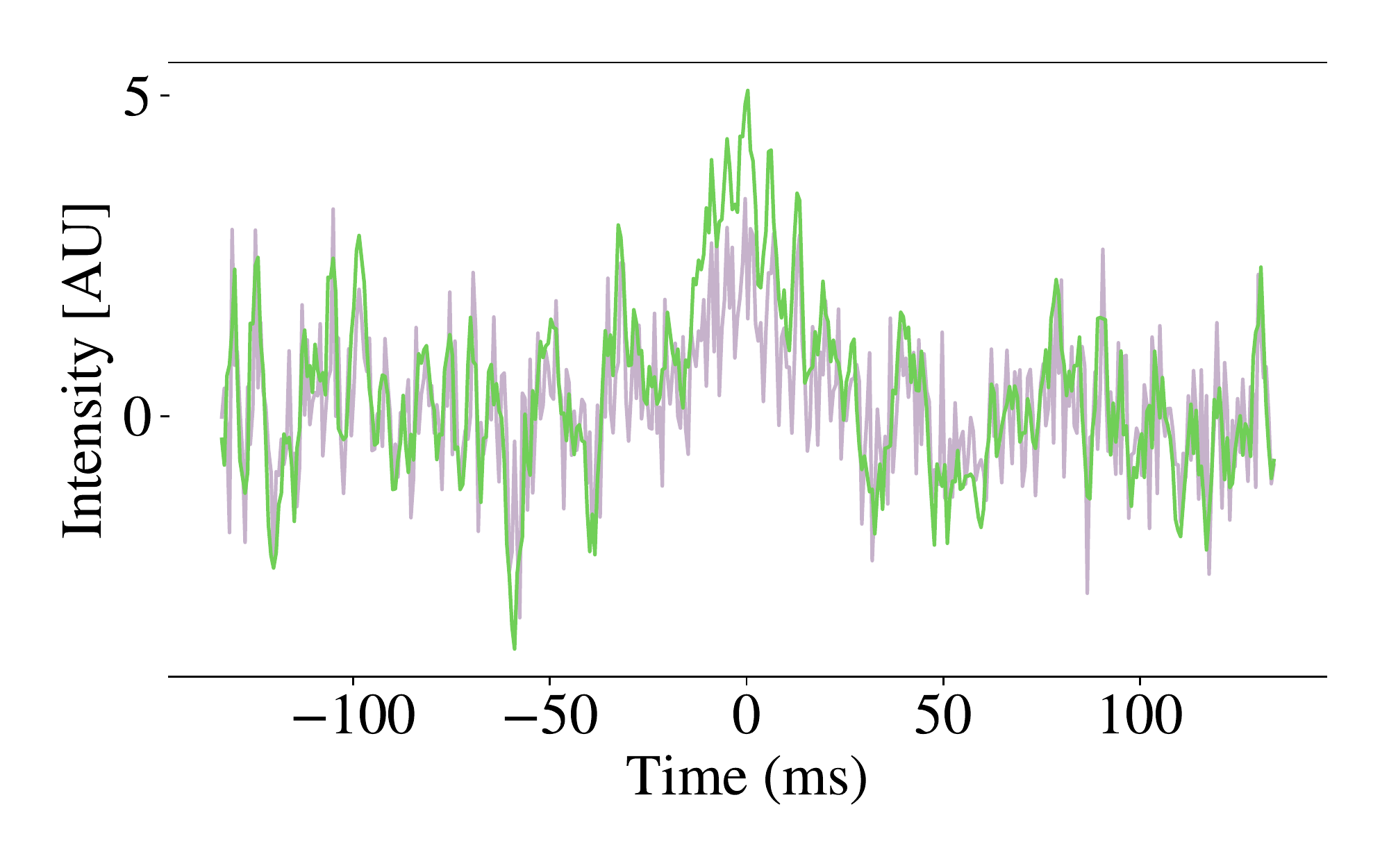} &\adjincludegraphics[height=0.14\textheight]{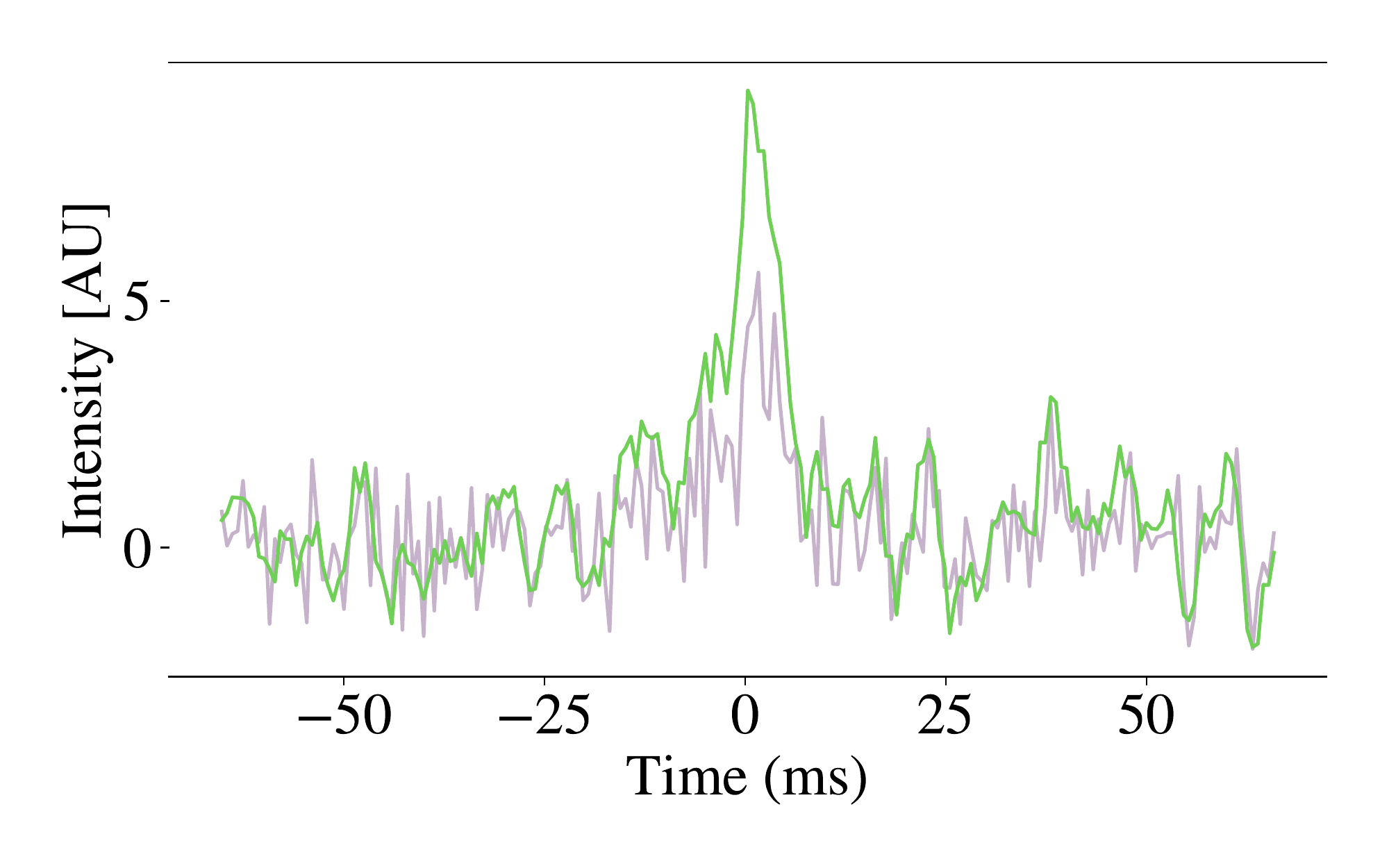} \\
\hphantom{3} (s) J1931+4229 &\hphantom{3} (t) J2108+4516 &\hphantom{5} (u) J2138+69 \\\\
\adjincludegraphics[height=0.14\textheight]{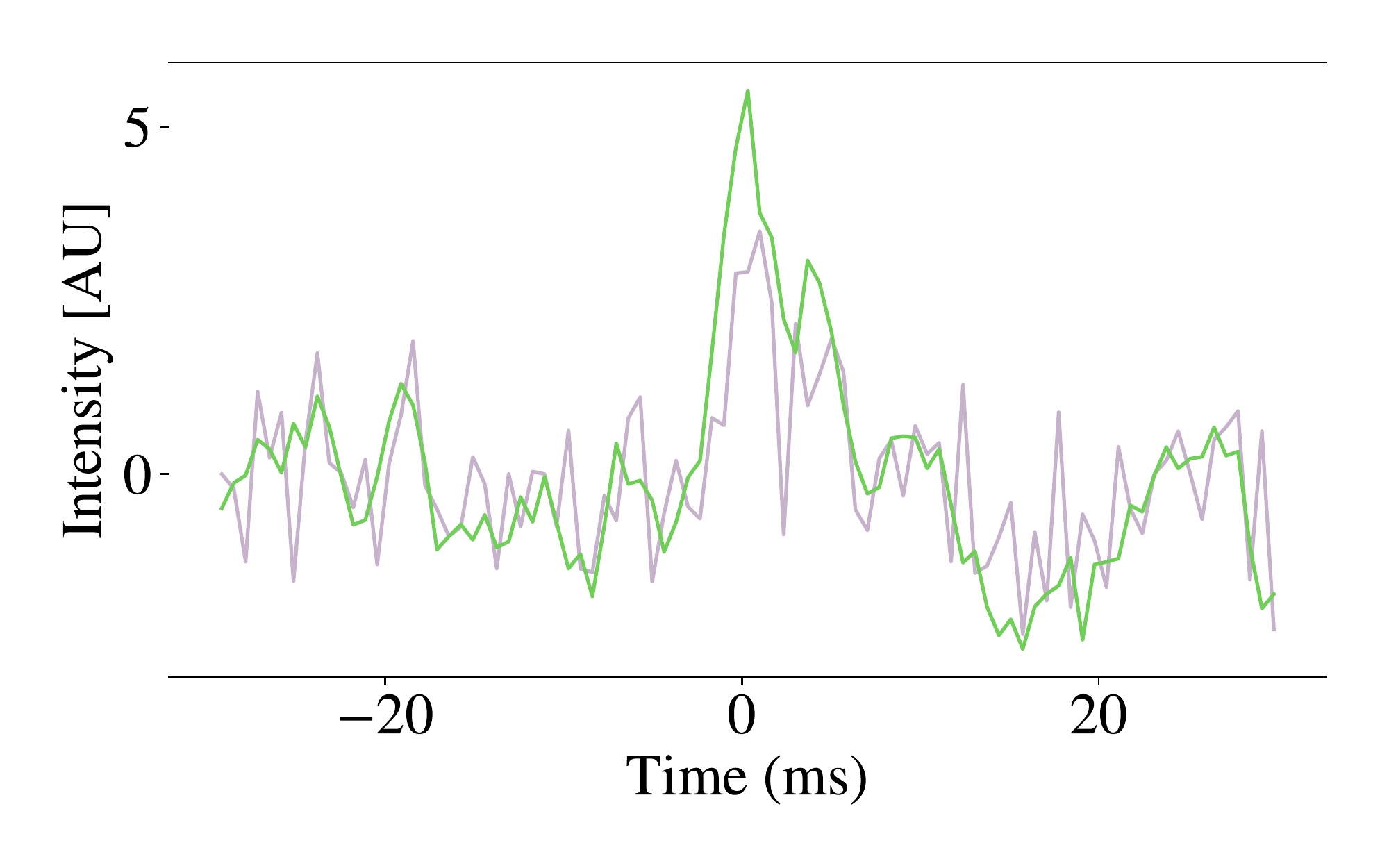} &\adjincludegraphics[height=0.14\textheight]{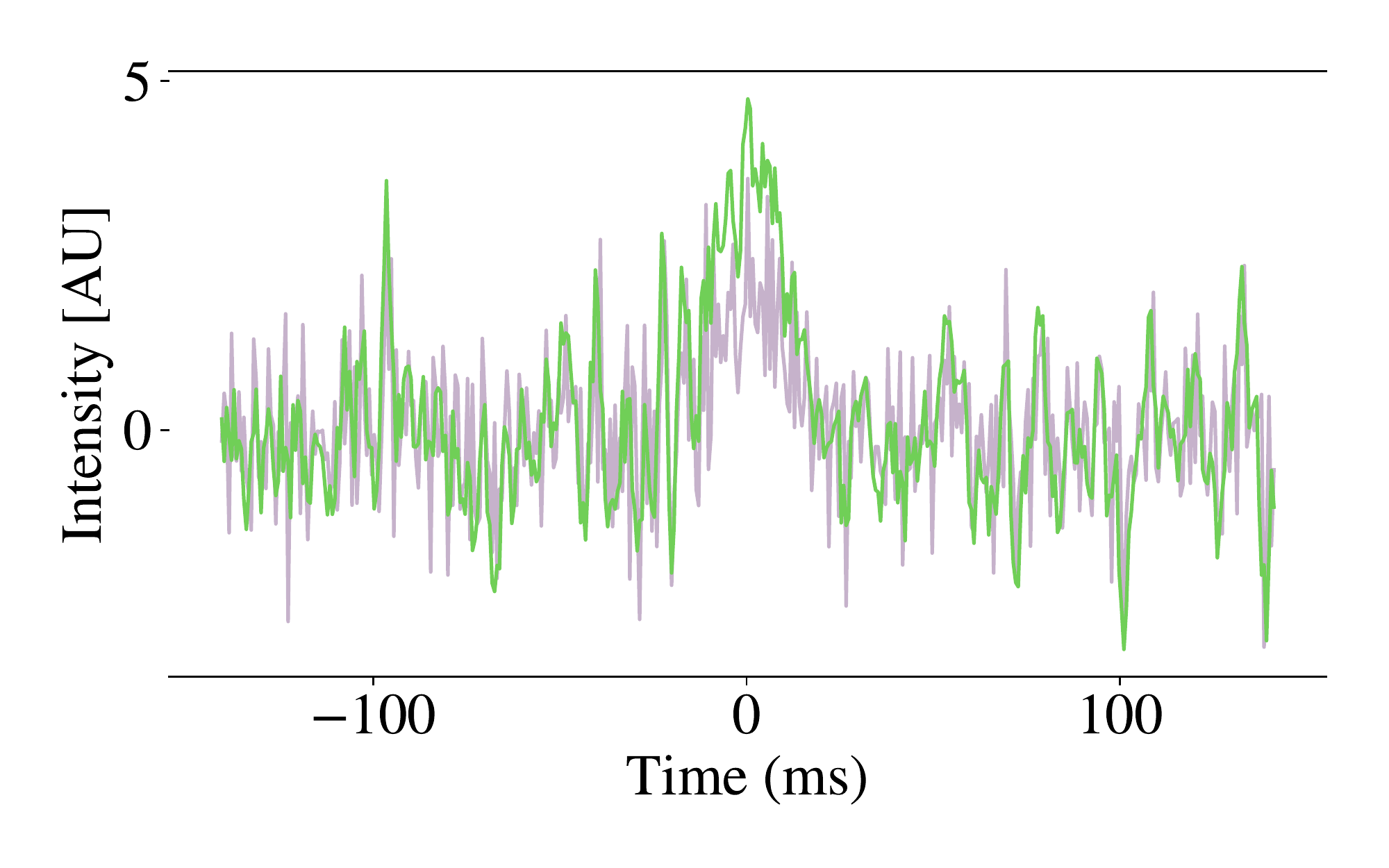} &\adjincludegraphics[height=0.14\textheight]{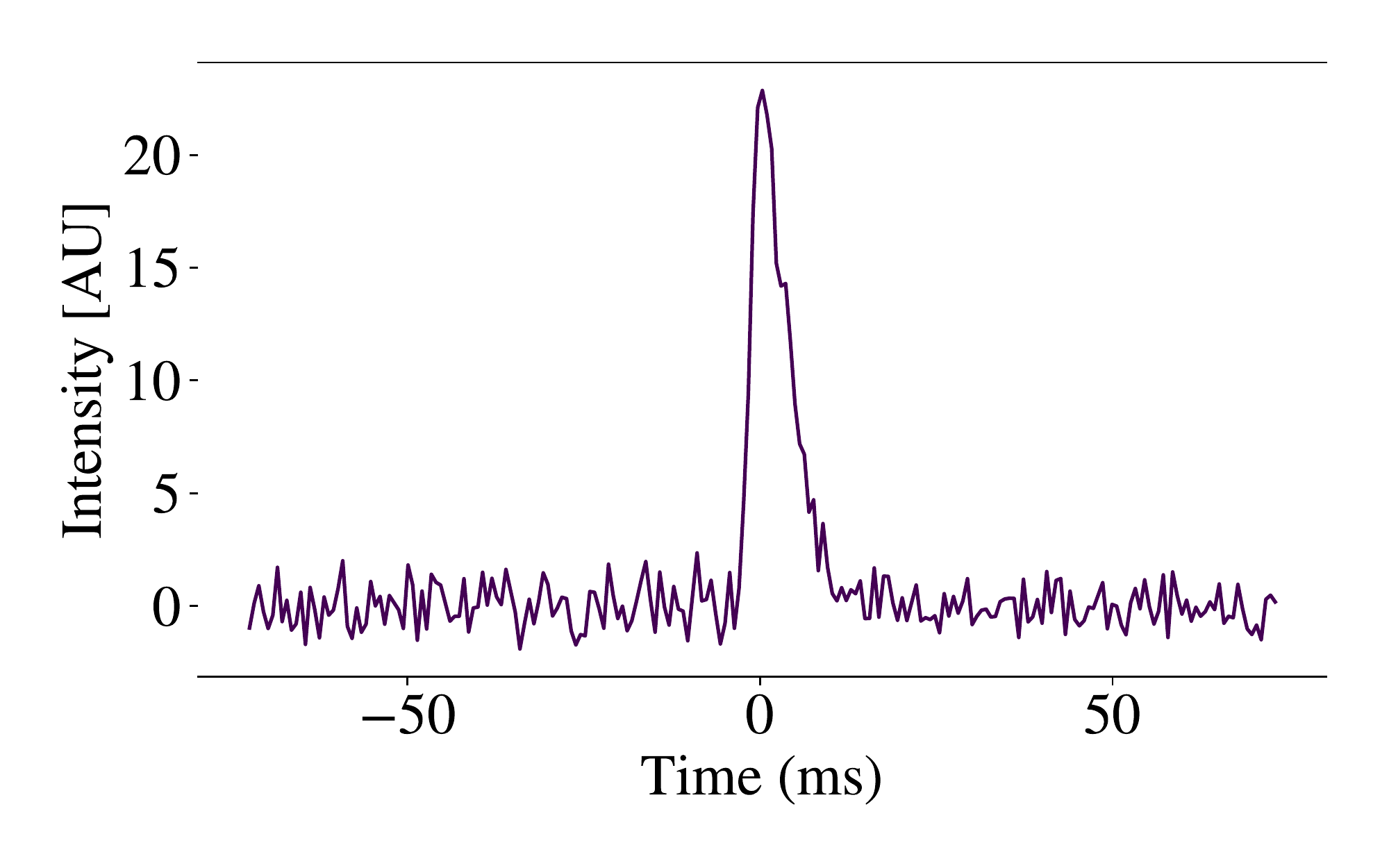} \\
\hphantom{3} (v) J2202+2134 &\hphantom{3} (w) J2215+4524 &\hphantom{3} (x) J2325$-$0530 \\\\
& \adjincludegraphics[height=0.14\textheight]{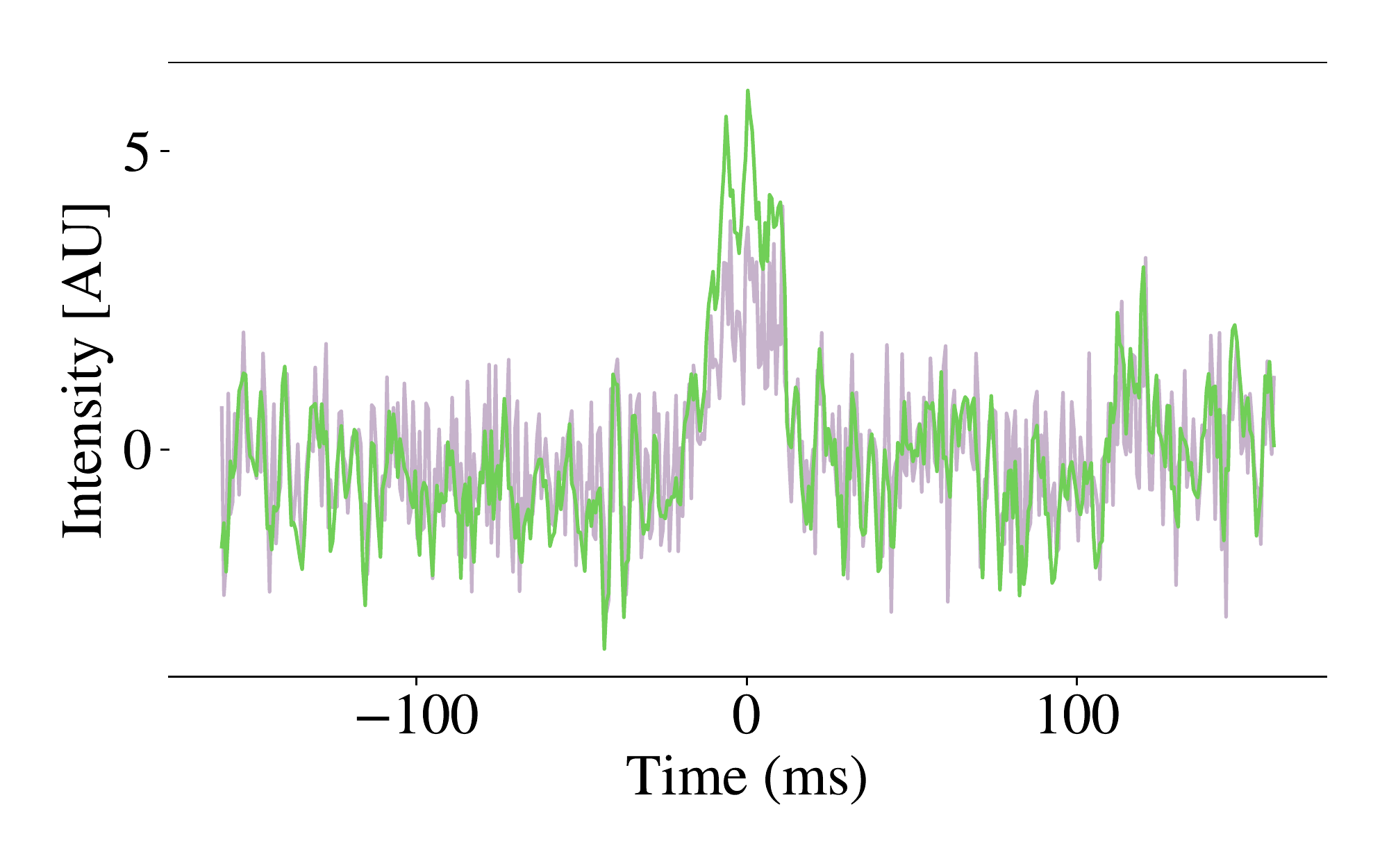} &\\
& \hphantom{3} (y) J2355+1523 &\\

    \end{tabular}
\captionsetup{list=off}
    \caption{A continuation of single-pulse profile plots from Fig.~\ref{fig:singleprofiles}.}
\end{figure*}

\section{Folded Profiles}
\begin{figure*}
    \centering
    \begin{tabular}{ccc}
\adjincludegraphics[height=0.14\textheight]{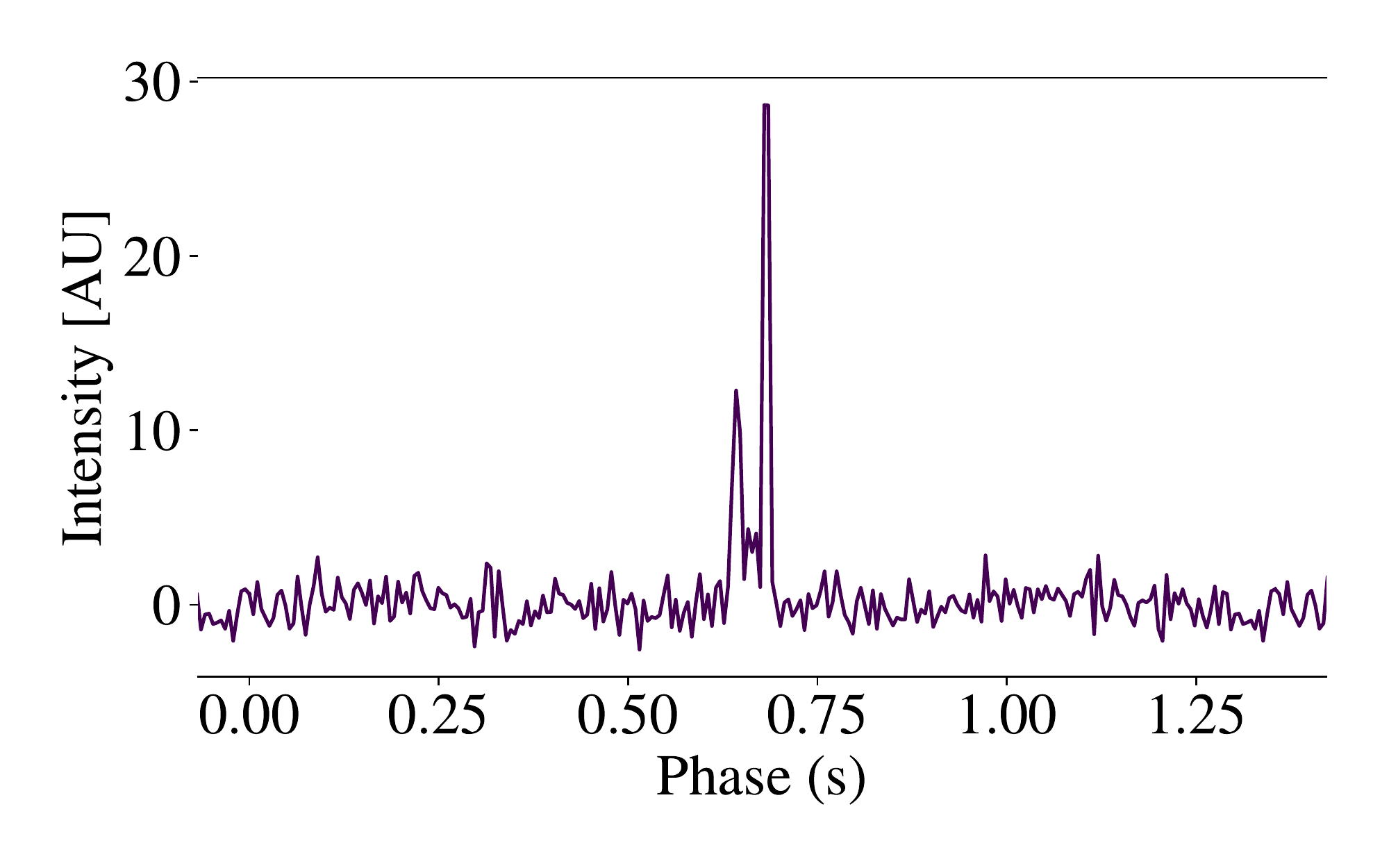} &\adjincludegraphics[height=0.14\textheight]{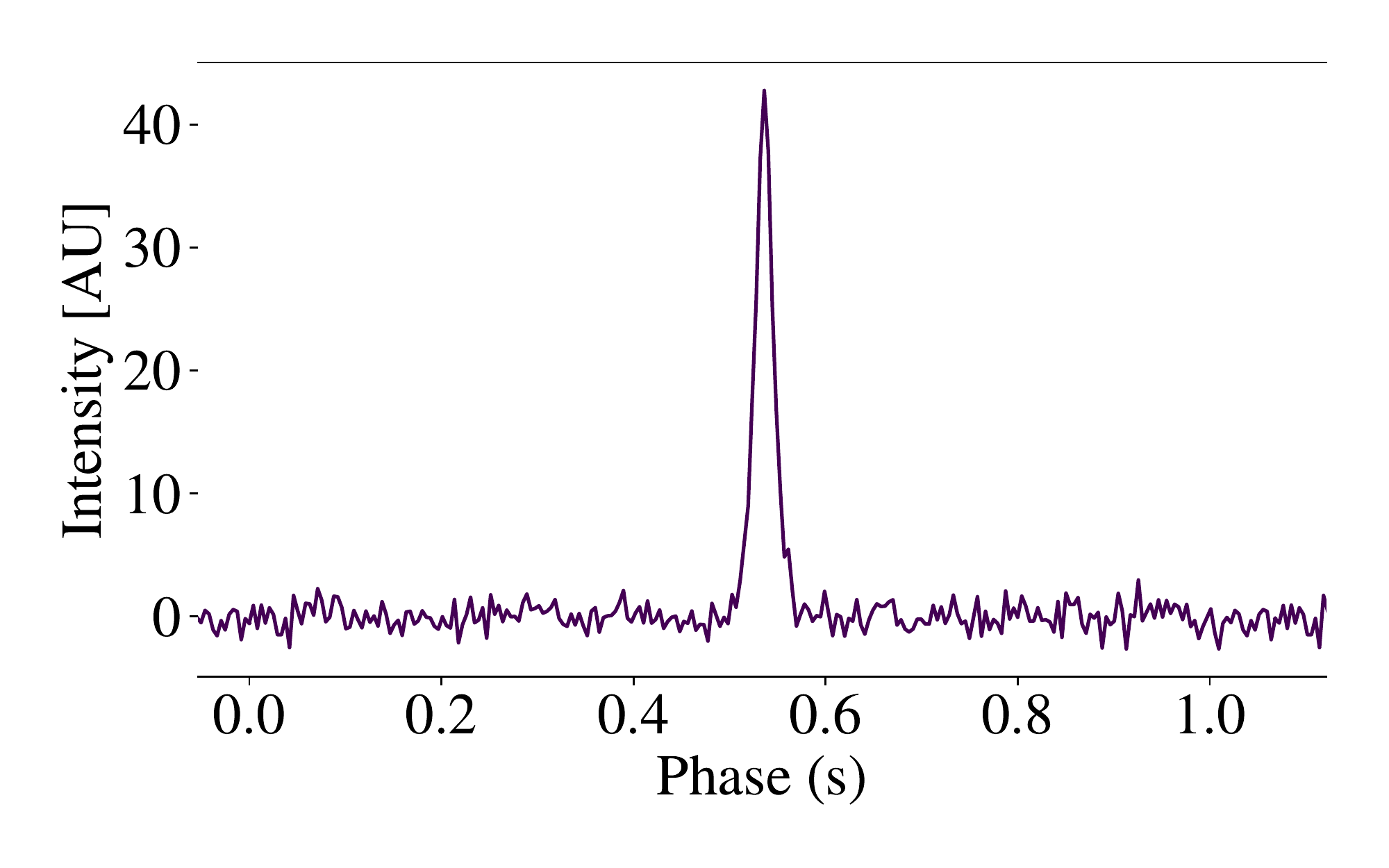} &\adjincludegraphics[height=0.14\textheight]{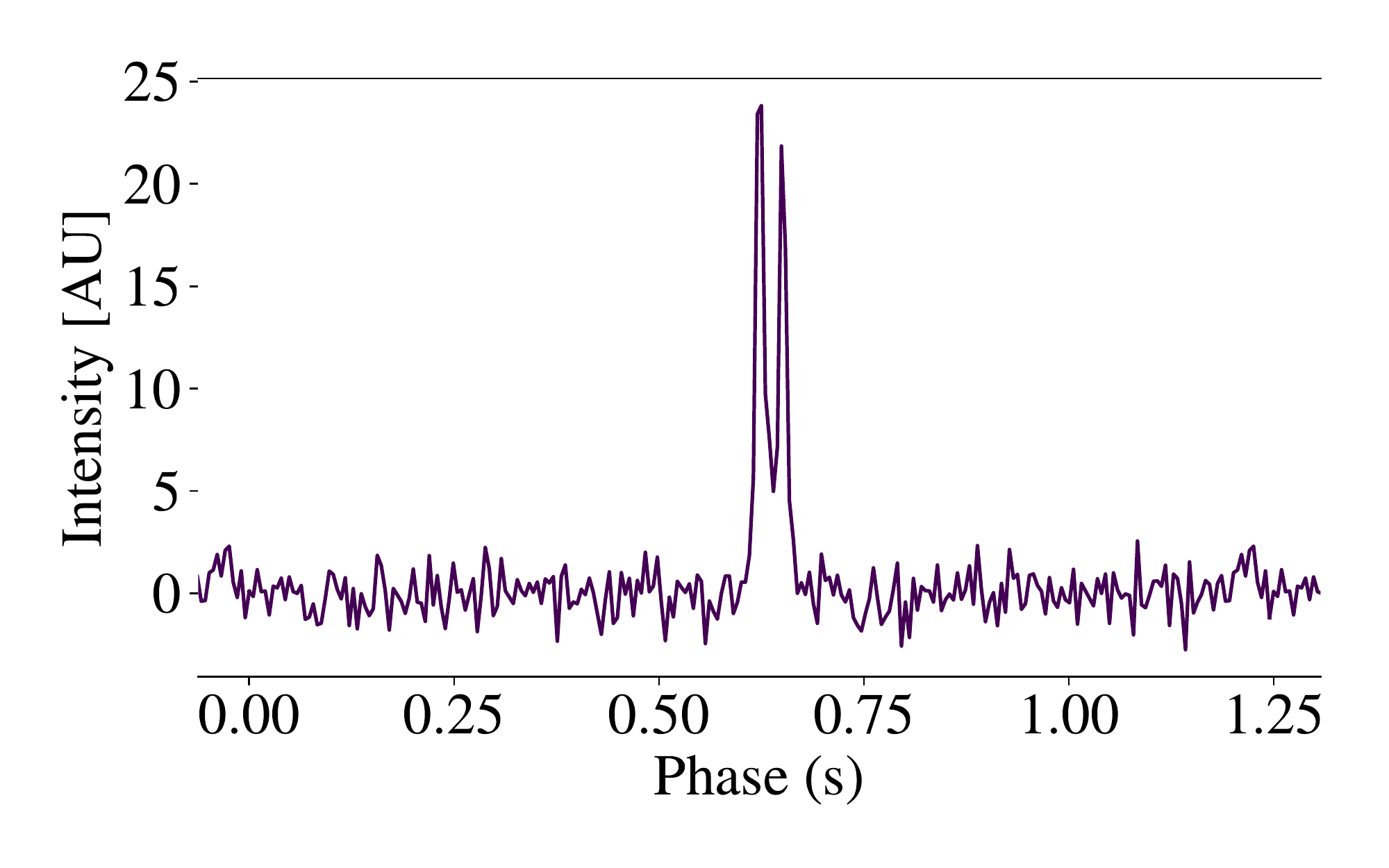} \\
\hphantom{3} (a) J0201+7005 &\hphantom{3} (b) J0209+5759 &\hphantom{3} (c) J0226+3356 \\\\
\adjincludegraphics[height=0.14\textheight]{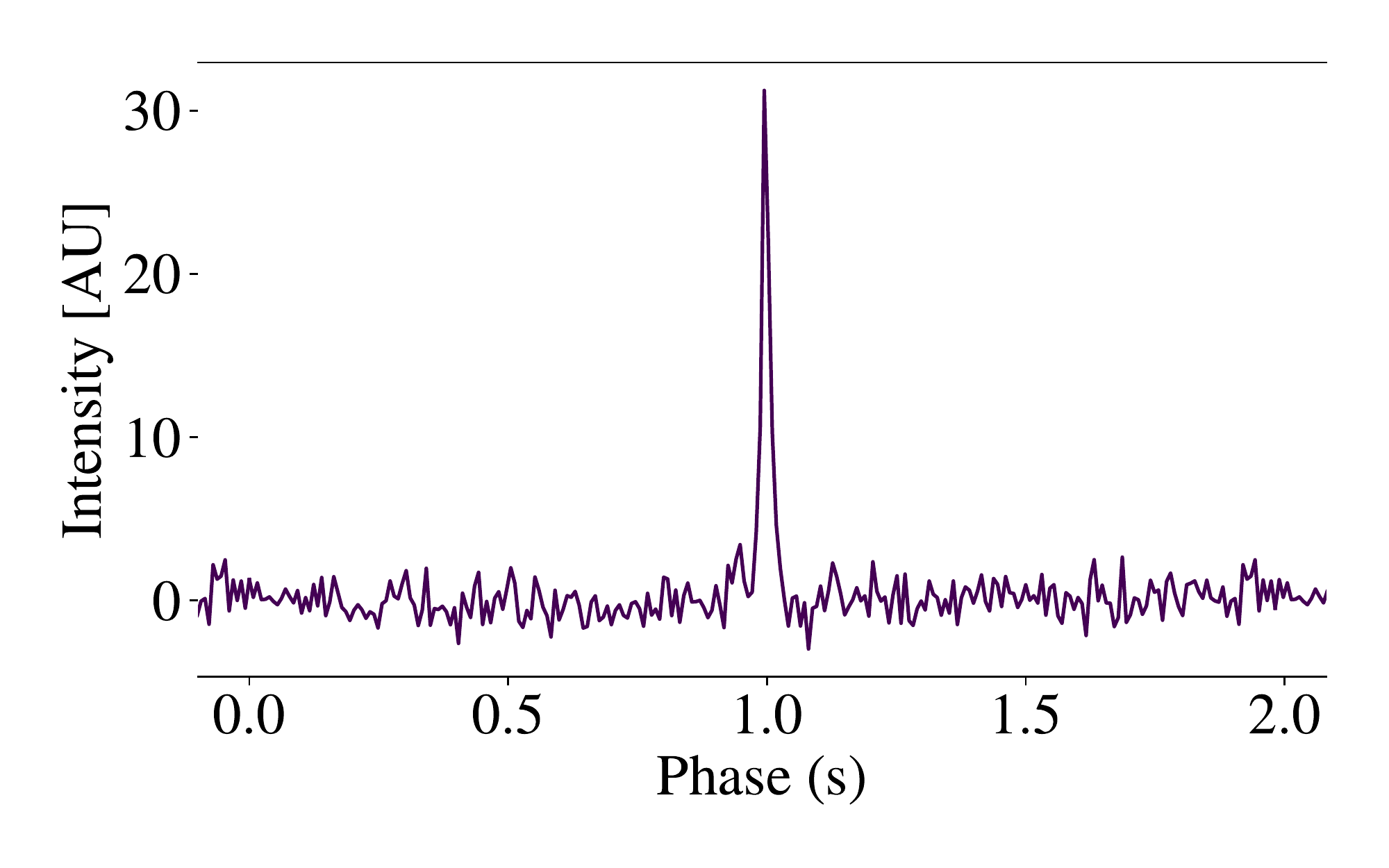} &\adjincludegraphics[height=0.14\textheight]{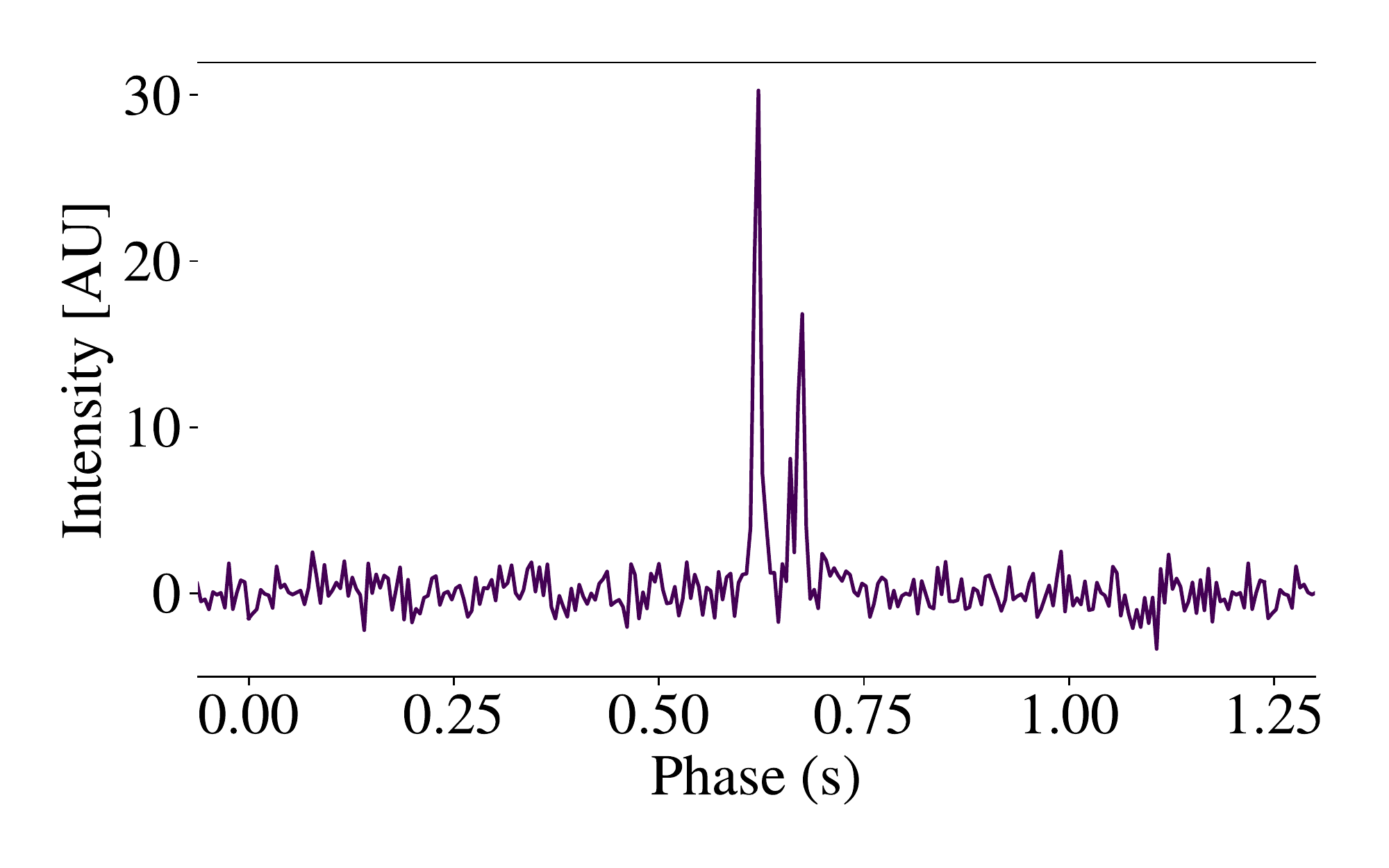} &\adjincludegraphics[height=0.14\textheight]{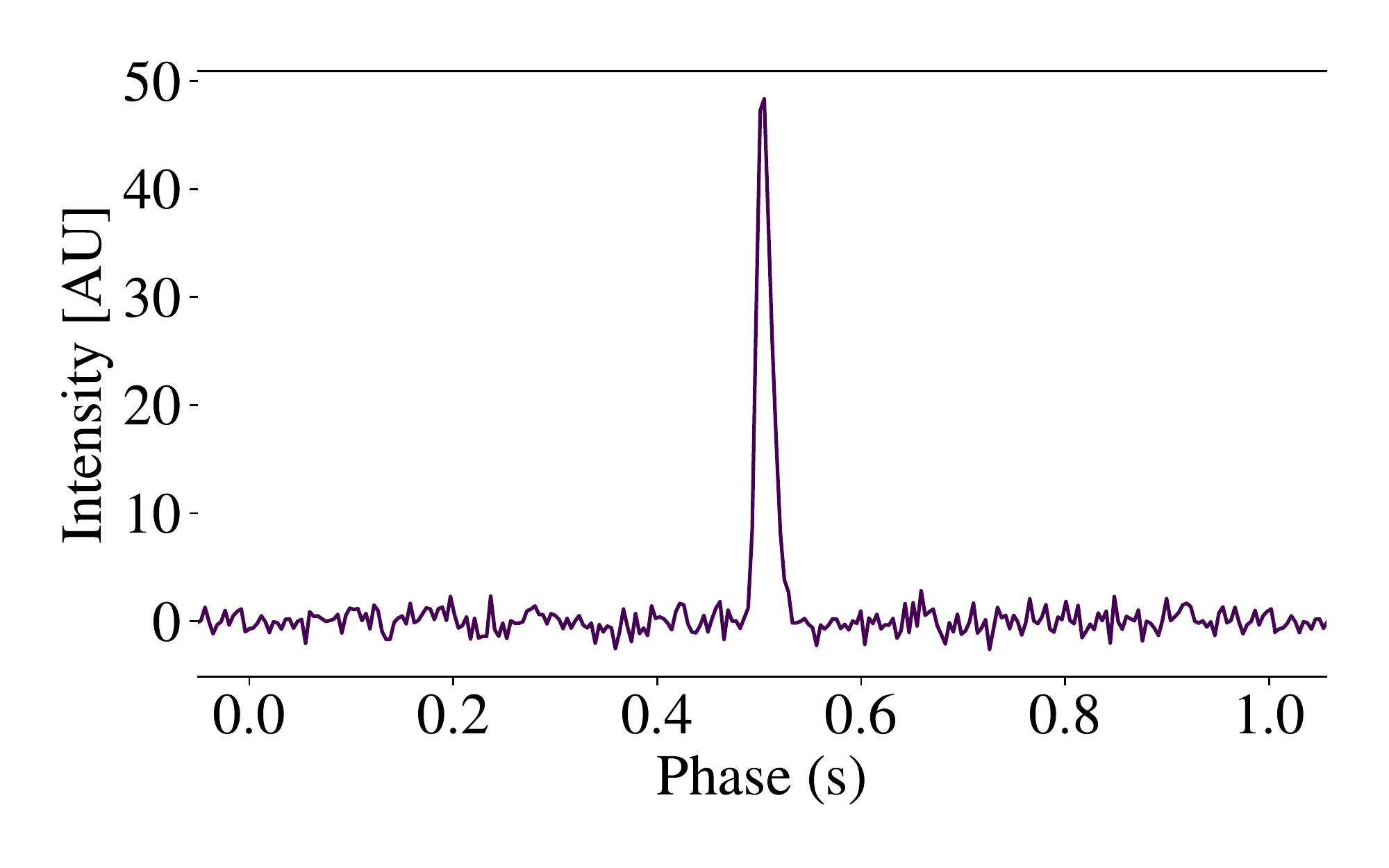} \\
\hphantom{3} (d) J0317+1328 &\hphantom{3} (e) J0854+5449 &\hphantom{3} (f) J1132+2513 \\\\
\adjincludegraphics[height=0.14\textheight]{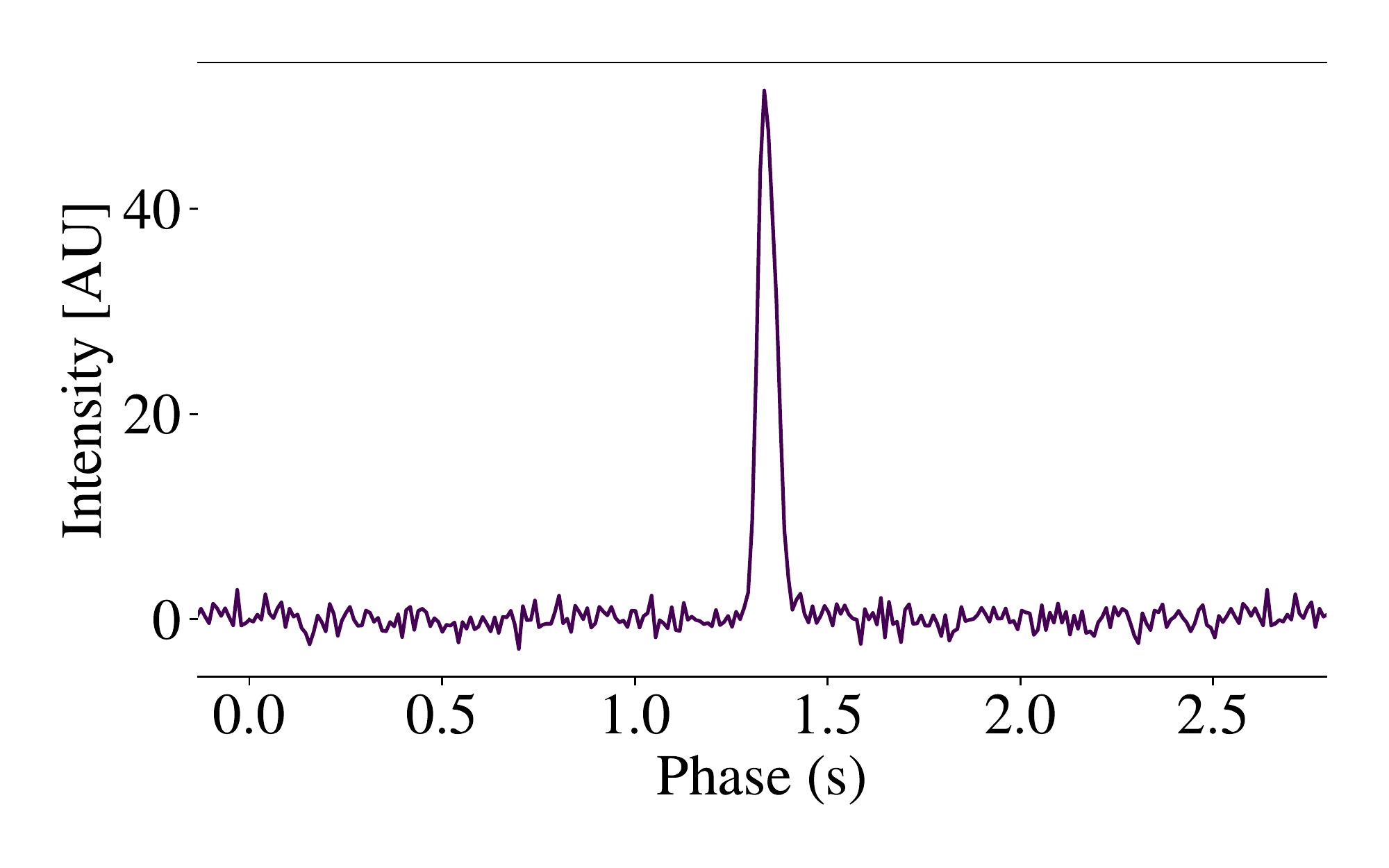} &\adjincludegraphics[height=0.14\textheight]{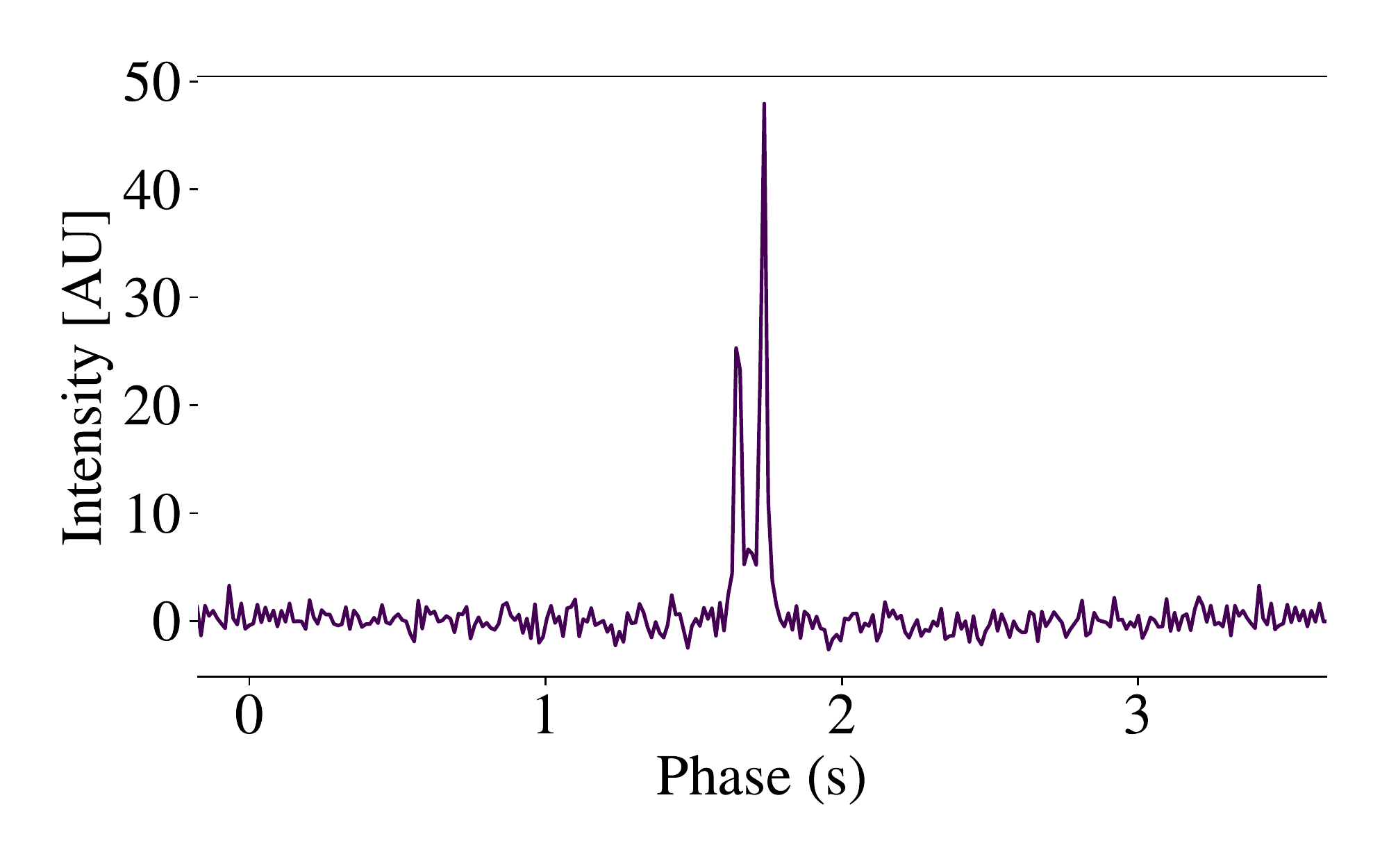} &\adjincludegraphics[height=0.14\textheight]{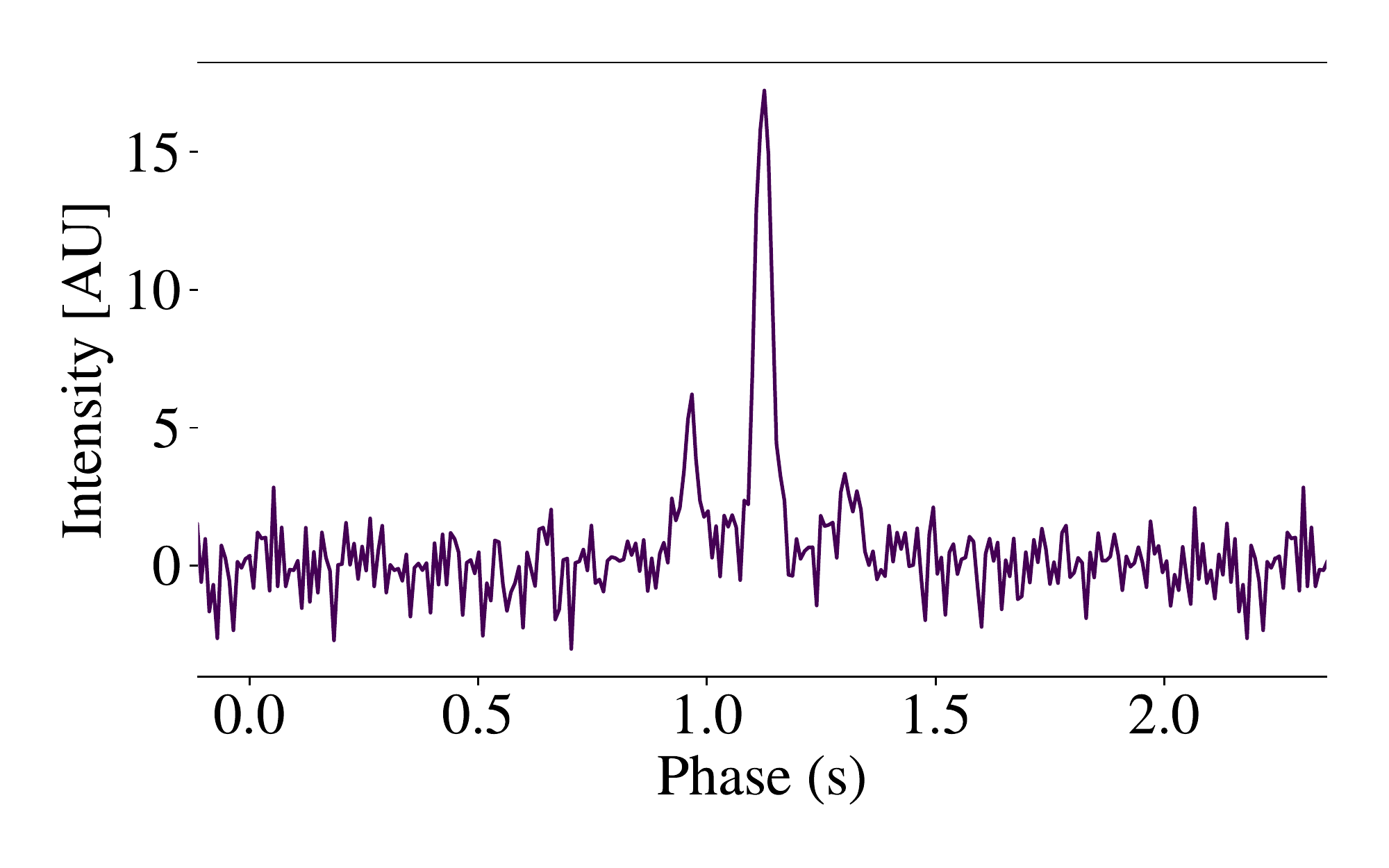} \\
\hphantom{3} (g) J1404+1159 &\hphantom{3} (h) J1538+2345 &\hphantom{3} (i) J1849+1518 \\\\
\adjincludegraphics[height=0.14\textheight]{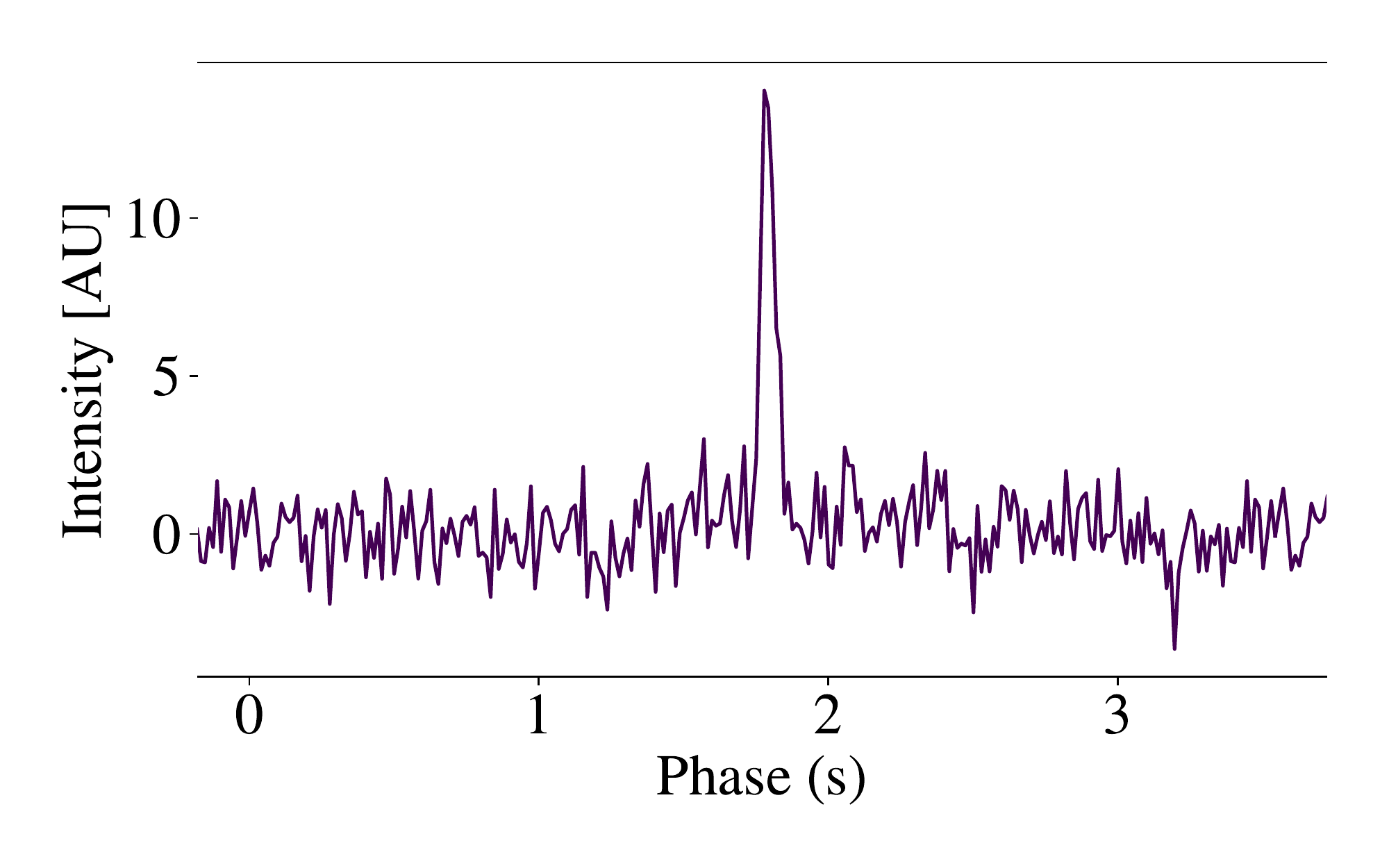} &\adjincludegraphics[height=0.14\textheight]{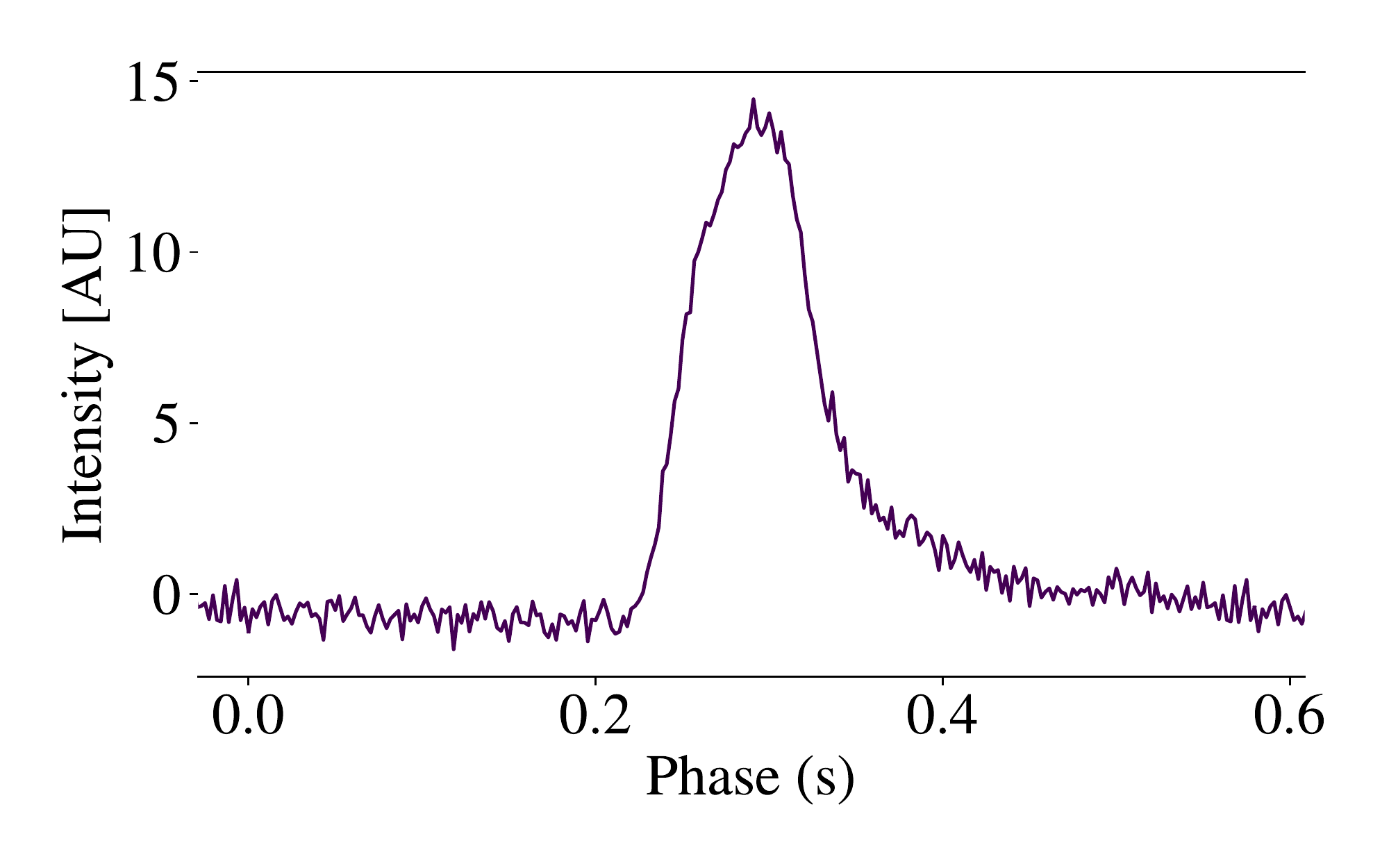} &\adjincludegraphics[height=0.14\textheight]{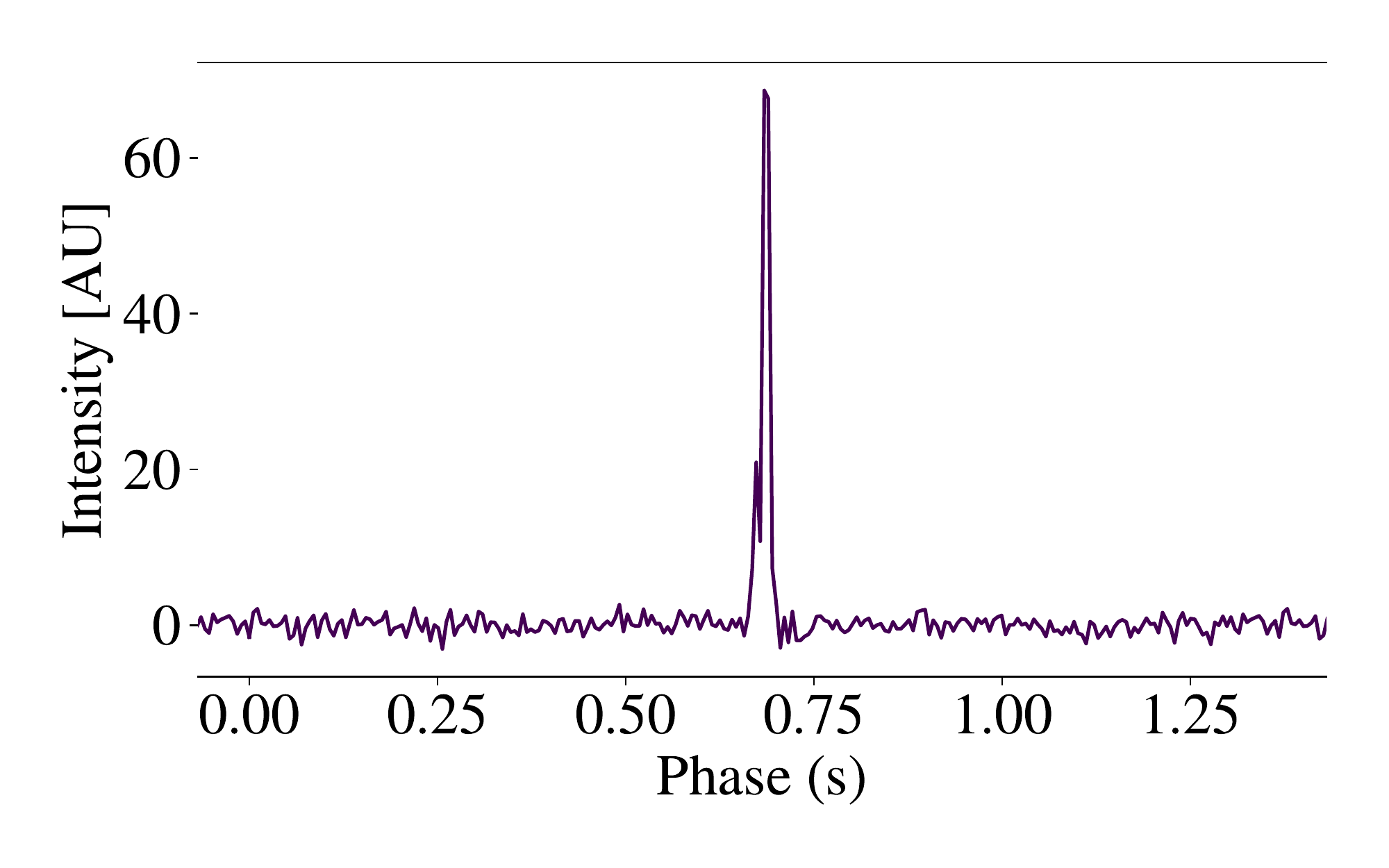} \\
\hphantom{5} (j) J2105+19 &\hphantom{3} (k) J2108+4516 &\hphantom{3} (l) J2108+4516 \\\\
\end{tabular}
\begin{tabular}{cc}
\adjincludegraphics[height=0.14\textheight]{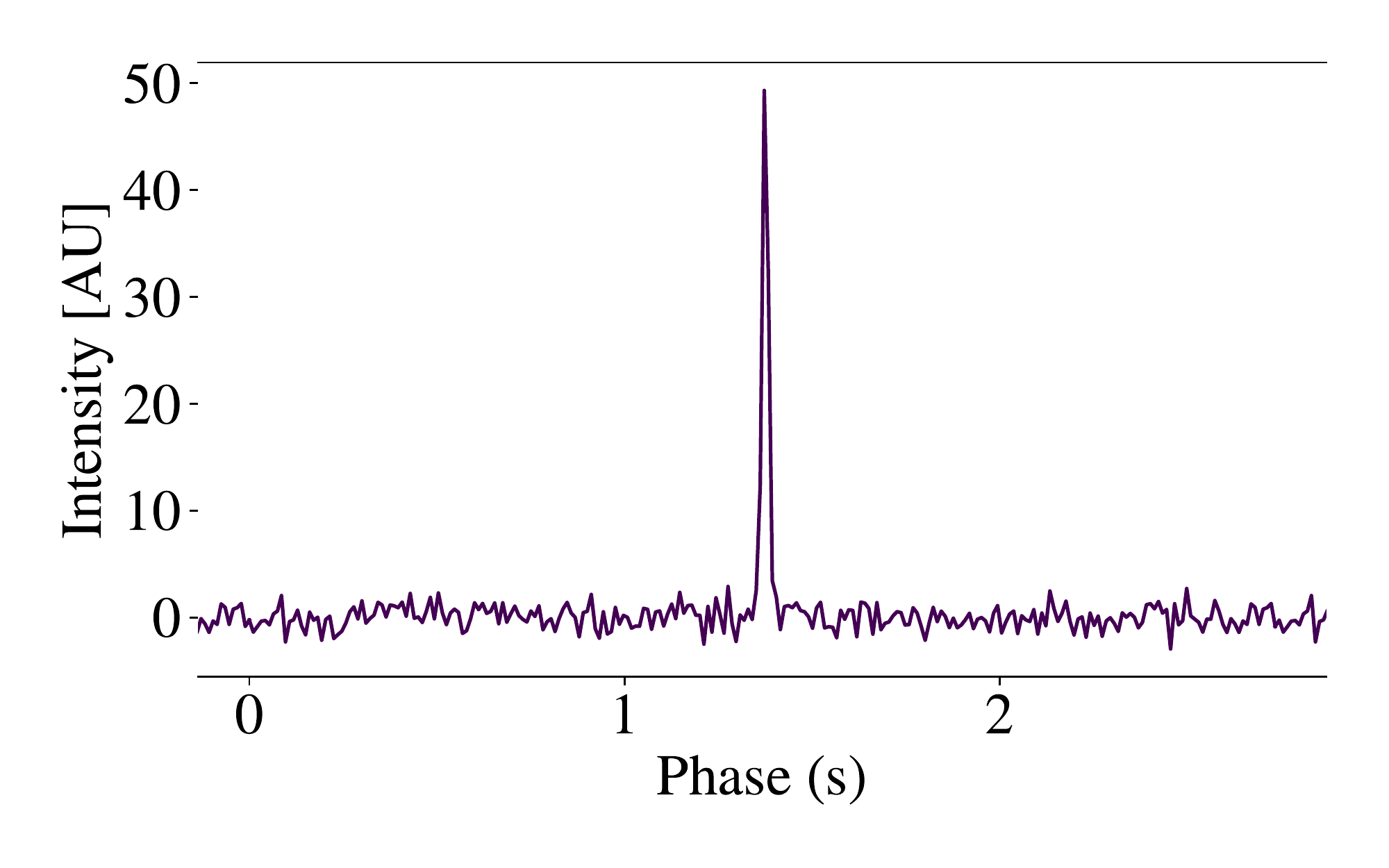} &\adjincludegraphics[height=0.14\textheight]{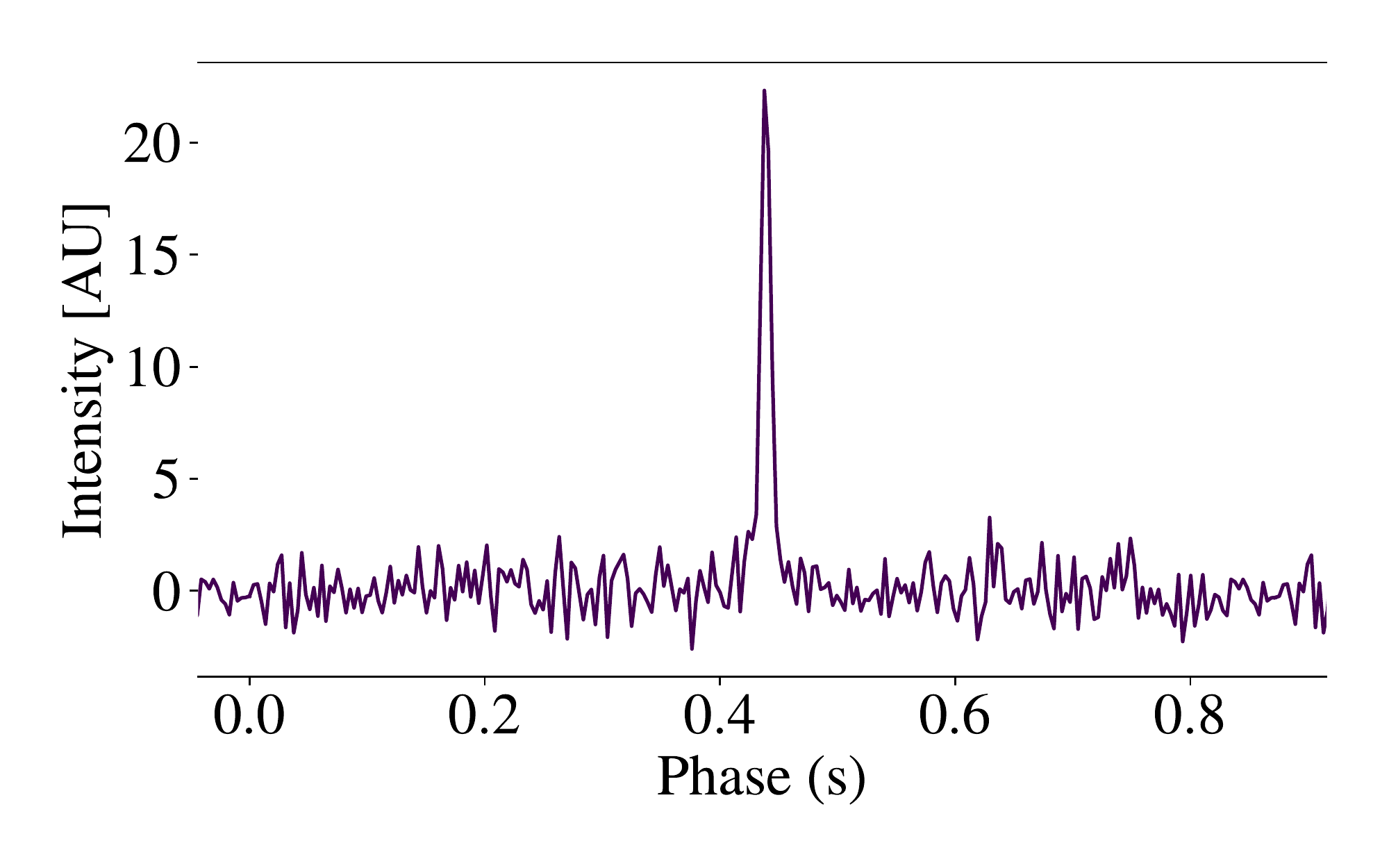} \\
\hphantom{3} (m) J2215+4524 &\hphantom{3} (n) J2325$-$0530 \\
\end{tabular}
    \caption{Figure of summed periodic emission profiles. Fig.~(b) only contains integrations where nulling pulsar J0208+5759 was active. Fig.~(k) only contains integrations from a single observation where the binary pulsar J2108+58 was active.}\label{fig:periodicprofiles}

\end{figure*}

Fig.~\ref{fig:periodicprofiles} provide plots of the time-average pulsar profile for each of the sources that had periodic emission, described in Table~\ref{tab:obs_summary}. In the case of nulling pulsar J0209+5759, this represents only the ``active'' time of the source, as determined by the 3-sigma criteria and methodology described in~\S\ref{sec:periodicres}.

%%%%%%%%%%%%%%%%%%%%%%%%%%%%%%%%%%%%%%%%%%%%%%%%%%

\section{Pulse Amplitude Distribution Parameters}\label{sec:pulseamplitudeparams}

Table~\ref{tab:sourcemappings} contains the best-fit parameters for each of the models used to generate the fits to the data in Fig.~\ref{fig:spectralmodfits}. Only the powerlaw, broken powerlaw and powerlaw--log--normal distribution fits have been provided, as the remaining models were not optimal for any of the observed data. The best model (as chosen by the AICc) has been emboldened.

The functions are provided in Equations~\ref{eq:pwl}, \ref{eq:pwlb}, and \ref{eq:pwllogn}, where the fit parameters match those provided in the table, $c_n$ refers to a fit constant, and for multi-model functions, $n(x_B)$ is a normalisation factor to ensure continuity, while $x_B$ references the ``break'' point between two models.

\begin{equation}\label{eq:pwl}
f_\text{PL}(x) = c_1 x^{n_F}    
\end{equation}
\begin{equation}
n_\text{BPL}(x_B) =  x_B^{\left(n_{F,1} - n_{F,2}\right)}
\end{equation}
\begin{equation}\label{eq:pwlb}
f_\text{BPL}(x) = \begin{cases}
c_1 x^{n_{F,1}} & x <= x_B \\
c_1 n_\text{BPL} x^{n_{F,2}} & x > x_B \\
\end{cases}
\end{equation}
\begin{equation} % (A / (np.sqrt(2* np.pi) * sigma)) * np.exp(-np.square(np.log(sep) - mu)/(2 * sigma ** 2) ) / sep
n_\text{PLLN}(x_B) = \frac{c_1}{\sqrt{2\pi}\sigma x_B} \exp{\left(-\frac{\left(\ln{x_B} - \mu\right)^2}{2\sigma^2}\right)}
\end{equation}
\begin{equation}\label{eq:pwllogn}
f_\text{PLLN}(x) = \begin{cases}
\left(n_\text{PLLN} * x_B^{-n_F}\right) x^{n_F}   & x <= x_B \\
\left(\frac{c_1}{\sqrt{2\pi}\sigma x}\right)\exp{\left(-\frac{\left(\ln{x} - \mu\right)^2}{2\sigma^2}\right)} & x > x_B \\
\end{cases}
\end{equation}

\begin{table*}
    \centering
    \caption{The best-fit parameters for a variety of models to the observed pulse amplitude distributions for detected sources with at least 40 pulses detected that were provided in Figure \ref{fig:spectralmodfits}, with the lowest AICc fit emboldened. Lognormal and Lognormal-Power-Law distribution parameters have not been presented, as they were not found to be the best fit for any of the observed distributions. ``Break'' refers to the separation point on the amplitude axis where the first model transitions into the second model.}
    \label{tab:pulseamplitudetab}
    \begin{tabular}{l|c|ccc|cccc}
    \hline\hline
Source & Power Law & \multicolumn{3}{c|}{Broken Power Law} & \multicolumn{4}{c}{Power Law-Log Normal} \\
& $n_F$ & Break & $n_{F,1}$ & $n_{F,2}$ & Break & $n_F$ & $\mu$ & $\sigma$ \\
    \hline
J0054+6650  & $-$0.1(5) & 16.5(1.2) & 2.3(4) & $-$3.6(1.0) & \textbf{8.7(1.5)} & \textbf{(3)} & \textbf{2.76(3)} & \textbf{0.40(3)}\\
J0102+5356  & $-$0.6(5) & \textbf{15.6(1.8)} & \textbf{2.5(1.1)} & \textbf{$-$2.6(9)} & 40(200) & (1) & (0.0052) & (48)\\
J0139+3336  & $-$1.0(4) & \textbf{14.3(8)} & \textbf{1.2(3)} & \textbf{$-$3.3(5)} & 11(4) & (1) & 2.61(7) & 0.47(6)\\
J0209+5759  & \textbf{$-$1.0(6)} & 9.6(4) & 5.5(1.8) & $-$2.4(6) & 22 & (1) & 83 & 36\\
J0317+1328  & \textbf{$-$1.3(6)} & 18(2) & 1.9(1.4) & $-$3.3(1.4) & 13 & 1 & 5.8e+03 & 3.1e+03\\
J0746+5514  & \textbf{$-$0.0(3)} & 9(3) & 1.1(9) & $-$0.8(7) & 6.1 & (1) & 2.92(97) & 1.1(1.2)\\
J1006+3015  & \textbf{$-$1.0(2)} & 11(3) & 0.1(8) & $-$1.7(5) & 5.4 & 1 & 2.5 & 0.86\\
J1336+3414  & 0.2(5) & 16.7(1.8) & 2.4(6) & $-$2.51(99) & \textbf{12} & \textbf{3.6} & \textbf{2.5} & \textbf{0.78}\\
J1400+2125  & \textbf{$-$1.0(6)} & 14.3(1.8) & 1.5(1.4) & $-$3.3(1.5) & 11 & 1 & 18 & 1.5e+02\\
J1538+2345  & 0.2(5) & 17.4(1.0) & 2.8(4) & $-$3.1(7) & \textbf{11.6(8)} & \textbf{(48)} & \textbf{2.86(8)} & \textbf{0.48(8)}\\
J1848+1516  & \textbf{$-$0.37(98)} & 11.9(6) & 2.8(7) & $-$7(3) & 8.3 & 0.99 & 1.7e$-$07 & 66\\
J1931+4229  & $-$1.2(3) & \textbf{6.12(16)} & \textbf{2.0(3)} & \textbf{$-$2.9(2)} & 5.8(1.1) & (1) & 1.6(6) & 0.6(3)\\
J2108+4516  & \textbf{$-$0.6(1.1)} & 8.9(1.2) & 1.8(1.0) & $-$14(19) & 8.6 & 7.8 & 2 & 0.16\\
J2215+4524  & $-$1.3(6) & \textbf{10.0(4)} & \textbf{1.8(4)} & \textbf{$-$4.9(1.0)} & 5.9 & 1 & 1e+02 & 91\\
J2325$$-$$0530  & $-$0.3(4) & \textbf{41(4)} & \textbf{1.7(5)} & \textbf{$-$2.4(7)} & 23 & 1 & 4.7 & 1.2e+03 \\
\hline\hline

    \end{tabular}
\end{table*}

\section{Adapted RRATalog}\label{ap:extendedrratalog}
An updated version of the RRATalog to contain sources announced by surveys that were a part of the 2016 release, and additional sources from the catalogues mentioned in~\S\ref{sec:sources}, is distributed alongside the machine--readable versions of the tables of this work \citep{dmckennRRATCat2023}.

Several of the cited survey webpages have gone offline in the time between the last update of the RRATalog in 2016 and the publishing of this work, preventing the catalogue from being fully rebuilt from scratch. As a result, the updated version has been derived from the last version, published in September 2016.

While distributed as both a text file and a webpage on the WVU website, we opted to use the webpage for updating the catalogue as it contains additional sources, and separated parameters into their respective frequency bands.

The overall changes include, but are not limited to:
\begin{itemize}
    \item Updating source names to match \texttt{psrcat} entries and recent publications, fixing incorrect name labels
    \item Including data from the CHIME and Pushchino catalogues, alongside adding new candidates from previously referenced surveys
    \item Fixing incorrect survey references
    \item Updating entries where newer data is available through \texttt{psrcat} or other surveys
    \item Updating derived property columns
\end{itemize}

% Don't change these lines
\bsp	% typesetting comment
\label{lastpage}
\end{document}